\documentclass[a4paper]{article}

\usepackage{times}
\usepackage{amsmath, amssymb, textcomp}
\usepackage[latin1]{inputenc}
\usepackage[english]{babel}
\usepackage{acronym}
\usepackage{graphicx}
\usepackage{url}
\usepackage[numbers]{natbib}

\usepackage[margin=3cm]{geometry}

\newcommand{\captionfonts}{\small\it}

\makeatletter  
\long\def\@makecaption#1#2{%
  \vskip\abovecaptionskip
  \sbox\@tempboxa{{\captionfonts #1: #2}}%
  \ifdim \wd\@tempboxa >\hsize
    {\captionfonts #1: #2\par}
  \else
    \hbox to\hsize{\hfil\box\@tempboxa\hfil}%
  \fi
  \vskip\belowcaptionskip}
\makeatother   

\newenvironment{example}[1][Example]{\noindent\rule{\textwidth}{1pt}\begin{trivlist}
\item[\hskip \labelsep {\bfseries #1}]}{\end{trivlist}\noindent\rule{\textwidth}{1pt}}

\begin{document}

\renewcommand{\thepage}{\roman{page}}

\thispagestyle{empty}

\setlength{\voffset}{-0.0cm}
\begin{minipage}[t c]{.9\linewidth}
\vspace{-1cm}
\begin{center}
UU-NF 07\#02 (February 2007)\\
\begin{large}
\sc{Uppsala University Neutron Physics Report}\\
ISSN 1401-6269\\
\end{large}
\footnotesize{Editor: J. Blomgren}
\end{center}
\hspace{-6cm}
\end{minipage}

\begin{center}
\rule[6pt]{.9\linewidth}{1pt}
\end{center}
\vspace{.5cm}
\begin{center}
{\LARGE \sc Energy spectra for protons}
\end{center}
\begin{center}
{\LARGE \sc emitted in the reaction}
\end{center}
\begin{center}
{\LARGE $^{\textrm{nat}}{\textrm{Ca}}(\textrm{n},x\textrm{p})$ {\sc
at 94 MeV}}
\end{center}
\begin{center}
    {\sc P\"{a}r-Anders S\"{o}derstr\"{o}m}
\end{center}
\begin{center}
    \footnotesize \textit{Department of Neutron Research, Uppsala
    University,}

    \textit{Box 525, SE-75120, Uppsala, Sweden}
\end{center}
\vspace{12pt}

\begin{center}
    {\bf \sc Abstract}
\end{center}

\noindent The MEDLEY setup based at The Svedberg Laboratory (TSL),
Uppsala, Sweden has previously been used to measure
double-differential cross sections for elastic $nd$ scattering, as
well as light ion production reactions for various nuclei in the
interaction with neutrons around 95~MeV. When moved to the new beam
line, the first experimental campaign was on light-ion production
from Ca at 94~MeV in February 2005.

These data sets have been analyzed for proton production in forward
and backward angles. The  $\Delta E - \Delta E - E$ technique have
been used to identify protons, and a cutoff as low as 2.5~MeV is
achieved. Suppression of events induced by neutrons in the
low-energy tail of the neutron field is achieved by time-of-flight
techniques. The data are normalized relative to elastic $np$
scattering measured in the 20-degree telescope. Results from an
accepted neutron spectrum are presented and some methods to correct
for events from low energy neutrons are presented and evaluated. The
data are compared with calculations using the nuclear code TALYS. It
was found that TALYS systematically overestimates the compound part,
and underestimates the pre-equilibrium part of the cross section.

\newpage

\tableofcontents 

\clearpage \addcontentsline{toc}{section}{List of acronyms}
\section*{List of acronyms}

\begin{acronym}
  \acro{ADC}{analog-to-digital converter}
  \acro{BD}{beam dump}
  \acro{CFD}{constant fraction discriminator}
  \acro{CM}{center of mass}
  \acro{CsI(Tl)}{thallium activated cesium iodide}
  \acro{DWBA}{distorted wave Born approximation}
  \acro{FIFO}{fan-in/fan-out}
  \acro{FWHM}{full width at half maximum}
  \acro{ICM}{ionization chamber monitor}
  \acro{INF}{Department of Neutron Physics, Uppsala University}
  \acro{LET}{linear-energy-transfer}
  \acro{NN}{Nucleon-Nucleon}
  \acro{OMP}{optical model potential}
  \acro{PWA}{partial wave analysis}
  \acro{QDC}{charge-to-digital converter}
  \acro{RF}{radio-frequency}
  \acro{SRIM}{The Stopping and Range of Ions in Matter}
  \acro{T1}{Telescope 1}
  \acro{T2}{Telescope 2}
  \acro{T7}{Telescope 7}
  \acro{T8}{Telescope 8}
  \acro{TDC}{time-to-digital converter}
  \acro{TFBC}{thin-film breakdown counter}
  \acro{TOF}{time-of-flight}
  \acro{TSL}{The Svedberg Laboratory}
\end{acronym}

\clearpage \addcontentsline{toc}{section}{List of figures}
\listoffigures \clearpage \addcontentsline{toc}{section}{List of
tables} \listoftables \clearpage

\renewcommand{\thepage}{\arabic{page}}
\setcounter{page}{1}

\setcounter{figure}{0} \setcounter{table}{0}
\setcounter{equation}{0}

\section{Introduction}

In the last years, fast neutrons have gathered more and more
interest worldwide. Lots of applications involving fast neutrons in
some way are predicted to be available within 50 years, for example
one can mention fusion power reactors and transmutation of spent
nuclear fuel. Even today, fast neutrons play an important role in
the field of electronics, where it can cause so called single event
effects, doing damage to the hardware and software. Fast neutrons
are also used for hadronic radiotherapy of cancer in several places
in the world, for exampel at iTemba LABS in South Africa. So
understanding of fast neutron interactions is a large and important
part of applied nuclear physics.

In this work the interaction between 94 MeV neutrons with calcium
will be examined, in particular the productions of protons at
forward and backward angles. A first rough estimation of the angular
distributions will also be carried out. In this introduction to the
work, some general properties of the neutron, and experiments with
the neutron, will be discussed, as well as a short introduction to
the nucleus under study. This is followed by a few sections about
what is actually measured and some possible applications and
motivations for this work in the form of a biological introduction.

\subsection{The neutron}

\begin{figure}[!htb]
 \begin{center}
\includegraphics[width=0.4\textwidth,bb=0 0 405 241]{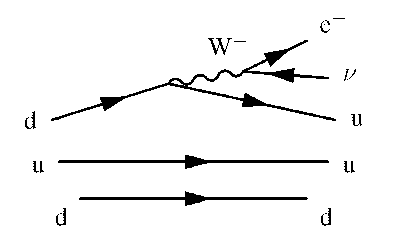}
  \caption[$\beta$-decay of a free neutron]{$\beta$-decay of a free neutron into a proton, an electron and an antineutrino.}\label{fig:neutronbetadecay}
 \end{center}
\end{figure}

The second lightest (known) baryon in the universe is the uncharged
neutron, discovered in 1932 by James Chadwick
\cite{existneutron}, when he was examining mysterious radiation
emitted from beryllium, boron and lithium that had been discovered a
few years earlier by \citet{1930ZPhy...66..289B}. Since the neutron
is the second lightest, the slight mass difference between it and
the lightest, the proton, makes the neutron unstable when it is in
an unbound state and it decays according to figure
\ref{fig:neutronbetadecay} with a lifetime of 887 seconds
\cite{physics}.

The neutron interacts with its surroundings through all four of the
fundamental forces: the strong nuclear force\footnote{\noindent More
or less all nuclear reactions.}, the weak nuclear force\footnote{In
the $\beta$-decay showed in figure \ref{fig:neutronbetadecay}}, the
electromagnetic force\footnote{Even though the neutron is
electrically neutral it still holds a magnetic dipole moment of
about -1.91 nuclear magnetons, $\mu_\textrm{N}$. This is due to the
fact that a baryon has a magnetic moment equal to the sums of the
quarks magnetic moments, which in the case of the neutron gives
$\mu_\textrm{n} = \frac{4}{3}\mu_\textrm{d} -
\frac{1}{3}\mu_\textrm{u} = -
\frac{2}{3}\frac{M_\textrm{p}}{m_{\textrm{u},\textrm{d}}}$
. The electrical dipole moment is measured to be less than $2.9
\cdot 10^{-26}$ $e\cdot$cm \cite{2006PhRvL..97m1801B}.} and the
gravitational force\footnote{For example in neutron stars.}. In
experiments with neutrons there exist a few difficulties that do
occur due to its special properties. Acceleration of neutrons is
impossible today since accelerators use the charge of the particles.
Instead, one has to rely on a neutron beam as a secondary beam. This
means that one uses a primary beam of charged particles, for example
protons, and let them hit some target that can induce a neutron
production nuclear reaction. Similarly the focusing of particle
beams is done by large magnets, while the beam shaping for a beam of
uncharged particles must be done with collimators. Dealing with
neutrons also gives rise to background and radiation protection
issues. This since high-energy neutrons has a lower probability than
charged particles for interaction and higher possibility for passing
through the shielding material.

\subsection{Calcium}

The nucleus under study in this work builds up a quite hard, silvery
white, alkaline earth metal. The name calcium origins from the Latin
word \emph{calcis} that means \emph{lime}, referring to the group of
minerals dominated by calcium-carbonates, -oxides and -hydroxides
often used as building material, for example in limestone, concrete
and mortar between bricks in masonry \cite{encyclo}.

In the human body, calcium is the most common mineral where almost
all of it is located in the skeleton in the form of hydroxide
apatite crystals. The skeleton of a normal adult contains about 1 kg
of calcium. Some of it can also be found in the soft tissue in the
form of Ca$^{2+}$ ions \cite{biology}. There are six stable
isotopes of calcium where the doubly magic\footnote{See section
\ref{sec:magicmodel}} isotope $^{40}$Ca is the absolutely most
common one, with a natural abundance of 96.941~\%\ \cite{physics}.

\subsection{Cross-section measurements}

One of the most fundamental needs in nuclear physics, as well as in
all other physics and even all other sciences, is theoretical models
to explain nature. Of course, another fundamental need in the field
of science is also the to go the other way, by obtaining
experimental data to
verify theoretical models.

In nuclear and particle physics the cross section is in a way the
connection between everyday life and the often complicated and
abstract theoretical models. It is a quantity that both can be
measured in experiments, as well as calculated from simple or
complex nuclear models. The unit is a non-SI, although accepted for
use with the SI, areal unit named barn\footnote{Rumors claim that
\citet{barn} coined the term barn in December 1942. Trying to come
up with a name for the unit 10$^{-28}$ m$^2$ they rejected both
bethe, oppenheimer, and manley before they decided to use barn,
since they thought that the probability of a reaction in the
experiment was as high as the probability of hitting a barn when
firing a gun.} that can be visualized as the effective area a
projectile sees when approaching a target. This makes the
interpretation of a cross section as a reaction probability a little
more intuitive; the larger the effective area of the target -- the
larger the probability of hitting it. It is important to keep in
mind though that the reaction cross section is not the same as the
geometric cross section.

\subsection{Ionizing radiation and humans\label{sec:ionizing_humans}}

\begin{figure}[!htb]
  \centering
\includegraphics[width=0.48\textwidth,bb=0 0 375 324]{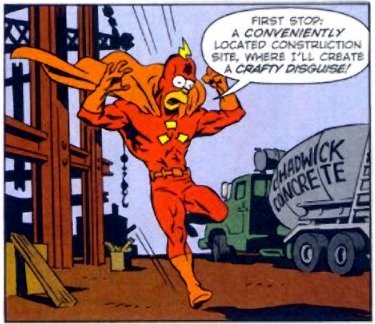}
  \caption[Man as a source for radiation]{In some sense, everyone of us is a Radioactive Man \protect \cite{simpsons}.}\label{fig:radioman}
\end{figure}

The human body is constantly exposed to radioactivity in terms of
ionizing radiation. The sources are partly due to human activities
like rest products from atmospheric nuclear weapons tests, nuclear
power accidents, dental x-rays and many other things. But there are
also various natural sources of radioactivity. For example radon gas
from earths crust and cosmic radiation. Even man itself contains
various radioactive isotopes and is a source for radiation, as
illustrated in figure \ref{fig:radioman}.

The damage that ionizing radiation does on human tissue is mainly on
the DNA and a majority of the DNA damage is indirect by ionizing the
water in the cell and creating very reactive free radicals.
Following \cite{radiobio}, there are three possible damages that
can occur, namely base pair damage, single strand break and double
strand break. Base pair damage and single strand break are the most
common of these three, but causes generally no problems for the cell
and are fixed fast and easy. Double strand break on the other hand
can create some more severe problems.

\begin{figure}[!htb]
  \centering
\includegraphics[width=0.6\textwidth,bb=0 0 545 286]{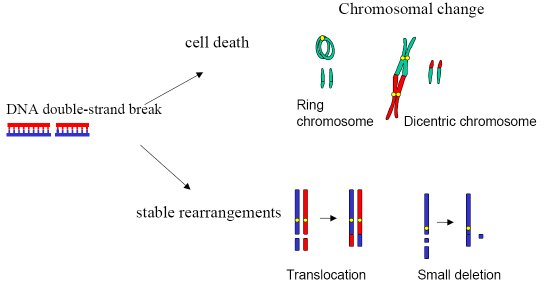}
  \caption[Reparation errors and DNA-damage]{Possible reparation errors when fixing a double strand break and the resulting DNA-damage \protect \cite{radioslides}.}\label{fig:dnadam}
\end{figure}

When the DNA, in the form of chromosomes, gets exposed to a double
strand break and the cell tries to repair it there are three
possible outcomes. The first is that the reparation succeeds and
everything is back to normal, but the reparation can also go wrong
in two different ways. One way is that we get an abnormal
chromosome, like a ring chromosome where the end of each arm has
been lost and repaired into a ring formation. Another abnormal
chromosome, that will appear in about 60 \% of chromosome
aberrations, is the di-centric chromosome where the chromosome has
been repaired into one with two centromeres, the region where the
two the two identical parts of the chromosome touch during cell
division, instead of just one. Both these examples of abnormal
chromosomes are, although fatal to the cell, quite harmless to
humans since the damaged cell never can survive mitosis, cell
division. This is illustrated in figure \ref{fig:dnadam}.

The third outcome, also illustrated in figure \ref{fig:dnadam}, of a
double strand break is the worst one. Here the chromosomes undergo
stable rearrangements instead of getting abnormal shapes. The stable
rearrangements could consist of translocation of two of the arms or
that a small portion of an arm is removed from the chromosome. This
mutation does not lead to mitotic cell death. Instead the damage
will multiply at the same rate as the cell, something that very well
can be a part of inducing a cancer tumor. This is the reason why
cancer-inducing effects are often refereed to as stochastic effects.
The probability of the statistically rare event of mutation
increases with increasing exposure, but the effect of an injury is
independent of the received dose.

\subsection{Cancer therapy}

In principle there are three ways to treat cancer. One can use
surgery, chemotherapy or radiation therapy. The basics behind
radiation therapy of cancer are to damage the cancer cells DNA
strands and in that way prevent the cell to divide. In this case one
actually wants a double strand break that ends up in an abnormal
chromosome and prevents the mitosis, as mentioned in section
\ref{sec:ionizing_humans}. The most common type of radiation therapy
is using $\gamma$-rays from $^{60}$Co, x-rays or electron beams.
However some tumors, called radioresistant tumors, seem to respond
quite poorly to this treatment and therefore other methods are
needed.

One way is to use hadrons such as protons, neutrons, helium ions or
pions instead of photons or electrons. Since hadrons have a larger
\ac{LET}, that is the stopping power in the medium, than electrons
and photons, they has a higher chance to do biological damage to the
cell. Actually there are three things that make a high \ac{LET}
particle more effective than a low \ac{LET} particle. Firstly it is
harder for the cells to repair the larger damage, secondly high
\ac{LET} particles is more efficient in damaging the oxygen-poor
tissue in  a tumor. And finally high \ac{LET} radiation is not as
sensitive to whether or not the cell is currently dividing as low
\ac{LET} radiation \cite{radiobio,hadrons}. Another property of
charged hadrons is the interaction depth. While photons and neutrons
deposit most of their energy in the surface, charged hadrons deposit
most of their energy in a well-defined depth in the patient. This
known as the Bragg peak.

Using fast neutrons for cancer therapy is a little more complex
method than using photons, electrons or hadrons, since neutrons do
not damage the tissue via direct interactions. Instead, nuclear
reactions with neutrons damage the tissue indirectly by production
of various types of light ions. This means that in order to make
accurate dose calculations, typically via Monte Carlo based
radiation transport codes, it is fundamental that the
double-differential cross sections for light ion production are well
known \cite{tippawan}.

\subsection{TALYS\label{sec:talys}}

TALYS is a joint project between NRG Petten in the Netherlands and
CEA Bruy\'{e}res-le-Ch\'{e}tel in France, and is a nuclear reactions
modelling program working with nuclear reactions of energies 1
keV~-~200 MeV. It is written completely from scratch in Fortran77,
with the only exception of the implemented coupled-channels code
ECIS-97 \cite{2005AIPC..769.1154K}.

The two main purposes of TALYS are to be used as a physics tool
during experimental analysis, and as a data tool to generate nuclear
data for use in various applications. The program is also very
flexible, since one can get a reasonable cross section by a minimal
input like the four lines:
\begin{verbatim}
projectile n
element    Ca
mass       40
energy     94
\end{verbatim}
But it also reacts to more than 150 keywords, making it possible to
adjust the nuclear models and the output \cite{talys}.

\clearpage
\setcounter{figure}{0} \setcounter{table}{0}
\setcounter{equation}{0}

\section{Theory}

As mentioned in section \ref{sec:talys} the theoretical predictions
are made using a nuclear reactions code called TALYS. This is an
outline of how TALYS does the calculations in this particular case,
after a brief introduction to the basics of nuclear reactions.

One should remember that this is only a crude simplification of the
calculation process valid only for incident high-energy neutrons on
a spherical nucleus and emission of protons. When dealing with more
complex particles like deuterons, tritons, helium isotopes and so on
effects like continuum stripping, pick-up and knock-out also needs
to be taken into account. Deformed nuclei also introduce a number of
corrections not needed here. An example of the different reactions
contribution to the total cross section can be found in figure \ref{fig:talplot}.

\subsection{The shell model and magic nuclei\label{sec:magicmodel}}

In nuclear physics there is not a single simple model to describe
the atomic nucleus completely. Instead several different models that
fit different kinds of parameters or properties for the nuclei are
used. For example there are the liquid drop model and the shell
model.

The shell model for nuclei is quite analogous to the shell model for
atoms. In the atomic shell model the electrons populate shells
outside the nucleus and when a shell is filled they start filling
the next. In a nucleus the principle is the same, only here it is
protons and neutrons that occupy shells and not electrons. Since the
neutron and the proton have individual shells two layers of shells
are filled in parallel \cite{krane}.

Since the shell model deals with particles populating different
energy levels there are some regions of extra interest and these are
the nuclei with filled, or close to filled, shells. The nuclei with
certain number of nucleons, corresponding to filled shells in the
shell model, are found to be more tightly bound than others. These
numbers of nucleons are commonly referred to as magic numbers and
are: 2, 8, 20, 28, 50, 82 and 126. Nuclei containing filled shells
are consequently called magic nuclei and nuclei with both protons
and neutrons in filled shells are called double magic nuclei.
Examples of double magic nuclei are $^{16}$O, $^{40}$Ca, $^{48}$Ca
and $^{208}$Pb.

There are theories that also higher numbers of neutrons and of
protons are magic numbers which would give rise to stable exotic
nuclei such as $^{298}_{114}$Uuq, $^{304}_{120}$Ubn and
$^{310}_{126}$Ubh in an island of stability among the otherwise
highly unstable elements at the end of the periodic table
\cite{island}.

\subsection{Fundamental nuclear reactions}

One important field of research to understand nuclear physics is the
study of nuclear reactions. The general nuclear reaction is when an
incoming particle, $a$, hits a target, $X$, causing the emission of
a particle, $b$, and a recoil nucleus, $Y$. This can be written on
the same form as a chemical reaction
\begin{equation}\label{eq:nuclreact}
    a + X \rightarrow Y + b
\end{equation}
or in a more compact form $X(a,b)Y$ where the $Y$ often is dropped.
An example\footnote{This example is actually quite historically
interesting since it was the first accelerator induced nuclear
reaction \cite{krane}.} of a reaction could be
\begin{equation}\label{eq:nuclreact2}
    \textrm{p} + {^7\textrm{Li}} \rightarrow {^4\textrm{He}} + \alpha
\end{equation}
or correspondingly ${^7\textrm{Li}}(\textrm{p},\alpha)$. One can now
divide (\ref{eq:nuclreact}) into two subgroups, elastic and
inelastic nuclear reactions. An elastic reaction is when $a=b$ and
$X=Y$ in (\ref{eq:nuclreact}) and no energy is lost due to
excitation of any of the components, while in an inelastic process
the target gets excited and/or $a \neq b$.

\begin{figure}[!htb]
  \centering
\includegraphics[width=0.48\textwidth,bb=0 0 450 250]{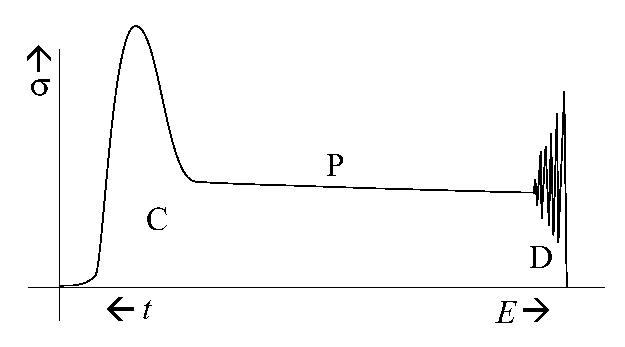}
  \caption[The different types of reactions]{The different types of reactions, adapted from \protect \cite{talys}: Compound reactions are indicated by C, pre-equilibrium reactions by P and direct reactions by D. The x-axis represent reaction time and particle energy and the y-axis the cross section.}\label{fig:cpdspectra}
\end{figure}

When studying high-energy reactions, another way to classify nuclear
reactions is in terms of the time scales at which they occur, or
equivalently in terms of the number of collisions within the
nucleus. The fastest reactions, with the fewest collisions, are
called direct reactions and occur on time scales of about
$10^{-22}$~s. Here the incoming projectile primarily interacts with
the surface of the target. At the opposite end there is the compound
reaction at time scales of about $10^{-16}-10^{-18}$ s
\cite{krane}, and in between these types pre-equilibrium reactions
may occur. The spectrum of the different reactions is illustrated in
figure \ref{fig:cpdspectra}, where also the corresponding shape of
the cross section is sketched.

\subsection{The optical model}

Before diving into details in the different nuclear reaction it is
suitable to have a closer look at the basis of nuclear reaction
calculations, something that is called the \ac{OMP}. An analogy to
this nuclear model is a murky glass sphere with some incoming light.
Some of the light will be refracted, or elastically scattered,
through the sphere, while other rays will be partially absorbed. In
the nuclear model, this effect is mathematically described by
defining a scattering potential as
\begin{equation}\label{eq:genomp}
    \mathcal{U}(r,E) = \mathcal{V}(r,E) + i\mathcal{W}(r,E)
\end{equation}
that can be inserted into the Schr\"{o}dinger equation. Following the
parametrisation in \cite{talys, 2003NuPhA.713..231K}, these
potentials can in turn be described as a sum of potentials caused by
different properties of the nucleus as
\begin{subequations}
    \begin{equation}
        \mathcal{V}(r,E) = - \mathcal{V}_V(r,E) + \mathcal{V}_{SO}(r,E) \cdot \mathbf{L} \cdot \mathbf{\sigma} +
        \mathcal{V}_C(r)\label{eq:genvop}
    \end{equation}
    \begin{equation}
        \mathcal{W}(r,E) = - \mathcal{W}_V(r,E) - \mathcal{W}_D(r,E) + \mathcal{W}_{SO}(r,E) \cdot \mathbf{L} \cdot \mathbf{\sigma}\label{eq:genwop}
    \end{equation}
\end{subequations}
where the indices denote the volume part, $V$, the spin-orbit
coupling part, $SO$, the surface derivative part, $D$, and the
Coulomb part, $C$. Finally, all of these can be further broken down
into an energy dependent part and a radial dependent form function
part as follows
\begin{subequations}
    \begin{equation}
        \mathcal{V}_V(r,E) = V_V(E) f(r,R_V,a_V)\label{eq:genvv}
    \end{equation}
    \begin{equation}
        \mathcal{V}_{SO}(r,E) = V_{SO}(E) \left( \frac{\hbar}{m_\pi c} \right)^2\frac{1}{r}\frac{\textrm{d}}{\textrm{d}r}f(r,R_{SO},a_{SO})\label{eq:genvso}
    \end{equation}
    \begin{equation}
        \mathcal{V}_C(r) = \left\{ \begin{array}{cl}
                                  \frac{Zze^2}{2R_C}\left( 3-\frac{r^2}{R_C^2} \right) & \text{for $r \leq R_C$}\\
                                  \frac{Zze^2}{r} & \text{for $r \geq
                                  R_C$}
                                \end{array}
        \right. \label{eq:genvc}
    \end{equation}
    \begin{equation}
        \mathcal{W}_V(r,E) = W_V(E) f(r,R_V,a_V)\label{eq:genwv}
    \end{equation}
    \begin{equation}
        \mathcal{W}_D(r,E) = -4 a_D W_D(E)\frac{\textrm{d}}{\textrm{d}r}f(r,R_D,a_D)\label{eq:genwd}
    \end{equation}
    \begin{equation}
        \mathcal{W}_{SO}(r,E) = W_{SO}(E) \left( \frac{\hbar}{m_\pi c} \right)^2\frac{1}{r}\frac{\textrm{d}}{\textrm{d}r}f(r,R_{SO},a_{SO}).\label{eq:genwso}
    \end{equation}
\end{subequations}
The parameters $R_i = r_iA^{1/3}$ and $a_i$ are the radius and
diffuseness parameters for the nucleus and the form function often
is assumed to be of Woods-Saxon type
\begin{equation}\label{eq:wodds-saxon}
    f(r,R_i,a_i) = \frac{1}{1+e^{\frac{r-R_i}{a_i}}}.
\end{equation}

\begin{figure}[!htb]
  \centering
  \begin{minipage}{0.48\textwidth}
\includegraphics[width=\textwidth,bb=0 0 567 567]{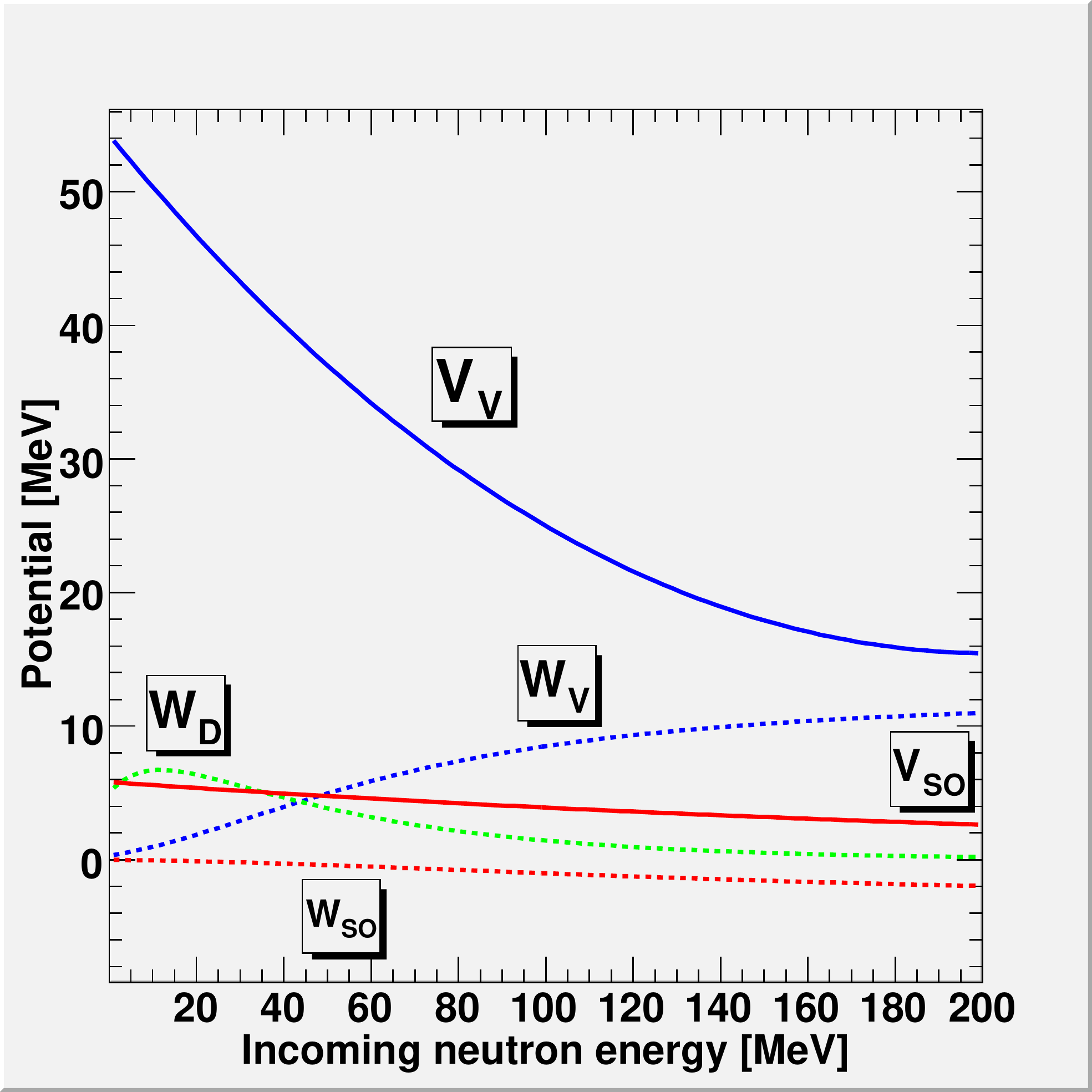}
  \end{minipage}
  \begin{minipage}{0.48\textwidth}
\includegraphics[width=\textwidth,bb=0 0 567 567]{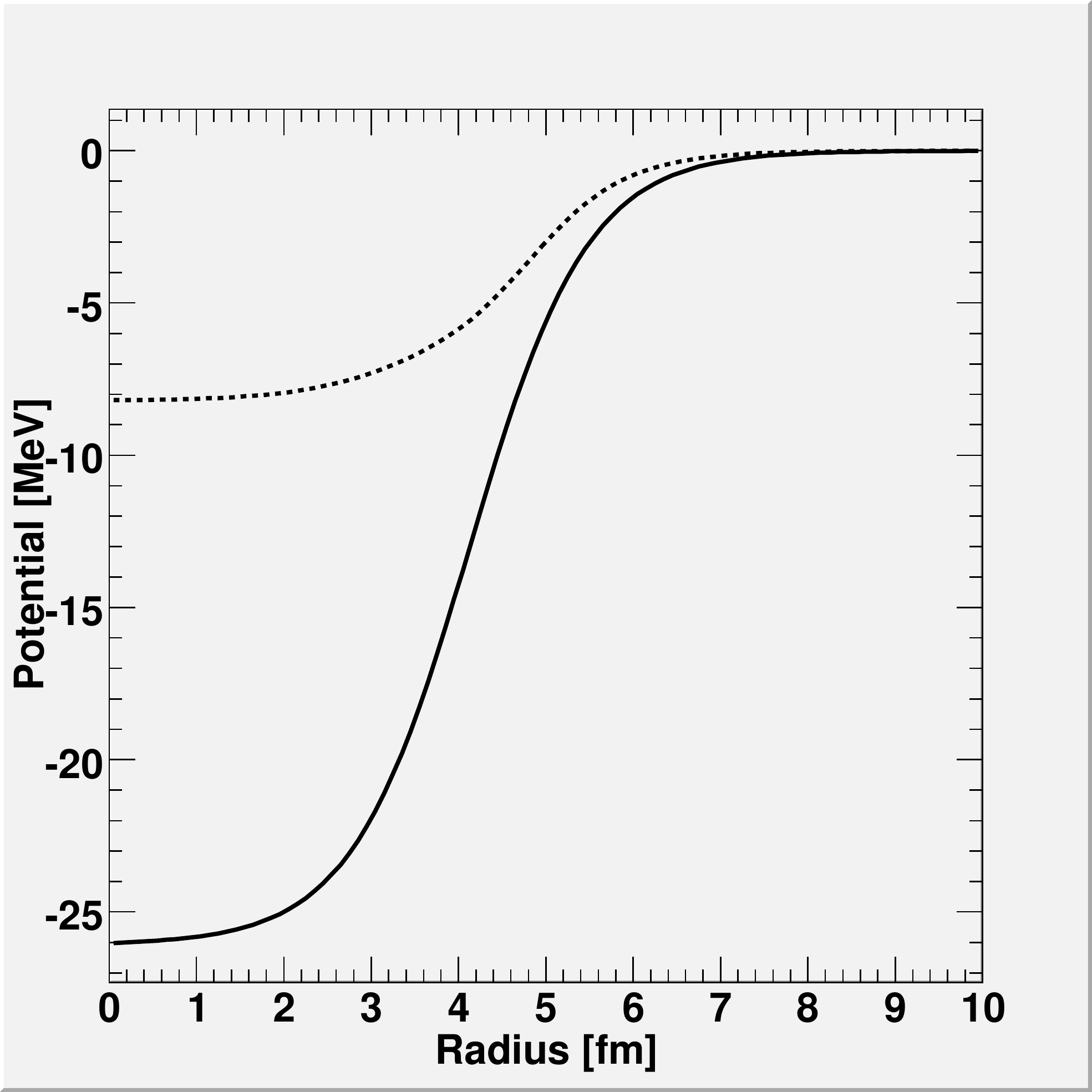}
  \end{minipage}
  \caption[Optical potentials plotted for $^{40}$Ca and incoming neutrons.]{Optical potentials plotted for $^{40}$Ca and incoming neutrons. The left panel shows $V_V(E)$, $V_{SO}(E)$, $W_V(E)$, $W_D(E)$ and $W_{SO}(E)$ for incident neutron energies between 0 and 200 MeV while the right panel shows $\mathcal{V}(r)$ (solid line) and $\mathcal{W}(r)$ (dashed line) for an energy of 94.5 MeV. Parametrized via \protect \cite{2003NuPhA.713..231K}.}\label{fig:Ca-potentials}
\end{figure}

These parameters are listed by \citet{2003NuPhA.713..231K} for a
vast amount of nuclei and for  $^{40}$Ca  and incoming neutrons they
are given by table \ref{tab:ca40omp}. The change in the potentials
with energy of the incoming neutron and distance from the center of
the nucleus is plotted in figure \ref{fig:Ca-potentials}.

\begin{table}[!htb]
  \centering
  \begin{tabular}{cccccc}
    \hline
    $r_V$ & $a_V$ & $r_D$ & $a_D$ & $r_{SO}$ & $a_{SO}$ \\
    1.206 & 0.676 & 1.295 & 0.543 & 1.01 & 0.60 \\
    \hline
    \hline
    $V_V$ & $V_{SO}$ & $V_C$ & $W_V$ & $W_D$ & $W_{SO}$\\
    26.06 & $3.984$ & 0 & $8.216$ & $1.597$ & -0.9518 \\
    \hline
  \end{tabular}
  \caption[OMP parameters for $^{40}$Ca]{Numerical values for the \ac{OMP} parameters for $^{40}$Ca with an incident neutron.}\label{tab:ca40omp}
\end{table}

\subsection{Direct reactions}

To calculate the cross sections from the direct formation processes
TALYS has several models incorporated, like the weak coupling model
or giant resonances. For near-spherical nuclides TALYS calculates
the cross sections for the direct formation processes with the
\ac{DWBA}. This basically means that one uses first order
perturbation theory on the regular Born approximation, that the
incoming and outgoing wave functions are just plane waves.
This will give a result where the simple plane waves of the Born
approximation is disturbed by the scattering potential of Eq.
(\ref{eq:genomp}). This first order model for small deformations is
valid since $^{40}$Ca is a double magic nucleus, and hence near
spherical. For a reaction like (\ref{eq:nuclreact}), the cross
section is obtained from the relation
\begin{equation}\label{eq:dwbaxsec}
    \frac{\textrm{d}\sigma^{\textrm{D}}}{\textrm{d}\Omega} = \frac{E_a E_b}{(2\pi)^2(\hbar
    c)^4} \frac{p_b}{p_a}\frac{1}{(2s_a+1)(2s_X+1)}\sum_{m_a, m_X, m_b,
    m_Y}\left|\left\langle f \left| H_\textrm{int} \right| i
    \right\rangle \right|^2
\end{equation}
where $E_i$ is the total energy, $p_i$ the momentum, $s_i$ the spin
of the particle, $i$, and $m_i$ the mass. The initial and final
states $\left\langle f \right|$ and $\left| i \right\rangle$ are
given by
the \ac{DWBA}
and $H_\textrm{int}$ is the
interaction potential \cite{lecturenotes}.

\subsection{Pre-equilibrium reactions}

\begin{figure}[!htb]
  \centering
\includegraphics[width=0.58\textwidth,bb=0 0 866 444]{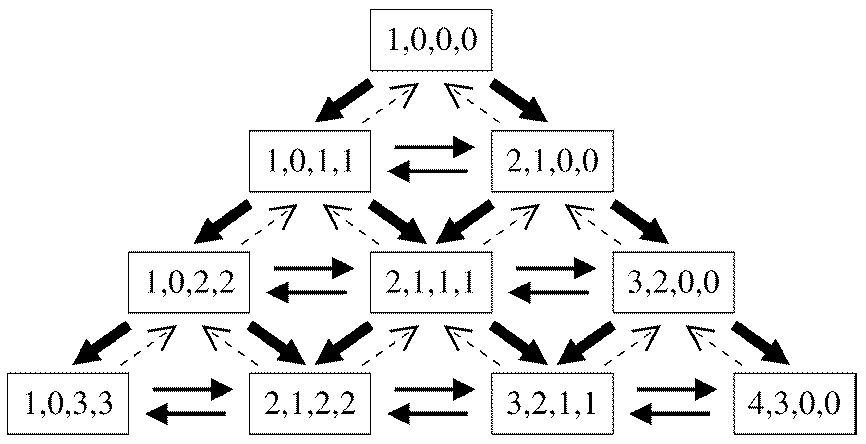}
  \caption[Different reaction paths in the two component exciton model]{Different possible reaction paths in the two component exciton model for an incoming proton \protect \cite{2004NuPhA.744...15K}. The numbers in the boxes correspond to $p_{\pi}$, $h_{\pi}$, $p_{\nu}$ and $h_{\nu}$ of some population $\mathcal{P}\left( p_{\pi}, h_{\pi}, p_{\nu}, h_{\nu} \right)$. The thickness of the
arrows is proportional to the transition rate, and as the arrows
pointing backwards are really small compared to the others, they are
neglected by TALYS. This is the so called never-come-back
approximation.}\label{fig:exciton}
\end{figure}

After the direct reaction and before the nucleus reaches equilibrium
some pre-equilibrium processes takes place. In TALYS the
pre-equilibrium stage is calculated using a model called the
two-component exciton model, following \cite{talys,
2004NuPhA.744...15K}. In analogy to solid state physics one can
define particle-hole pairs, also called pairs of excitons, above the
Fermi surface\footnote{The Fermi surface is the highest energy level
when all the nucleons in a nucleus, or electrons in an atom, are at
their lowest lying states - the Fermi sea.} as \newpage
\begin{subequations}
    \begin{equation}
        n_{\pi} = p_{\pi} + h_{\pi} \label{eq:exi1}
    \end{equation}
    \begin{equation}
        n_{\nu} = p_{\nu} + h_{\nu} \label{eq:exi2}
    \end{equation}
    \begin{equation}
        p = p_{\pi} + p_{\nu} \label{eq:exi3}
    \end{equation}
    \begin{equation}
        h = h_{\pi} + h_{\nu} \label{eq:exi4}
    \end{equation}
    \begin{equation}
        n = n_{\pi} + n_{\nu} \label{eq:exi5}
    \end{equation}
\end{subequations}
where $p$ is the number of particles, $h$ the number of holes and
$n$ the number of excitons while $\pi$ corresponds to protons and
$\nu$ to neutrons. When the incoming projectile interacts with the
nucleus it can excite a particle-hole pair. On its path through the
nucleus it can either excite new particle-hole pairs, annihilate
existing ones or change configuration of the existing pairs as
illustrated in figure \ref{fig:exciton}, as well as release some
excitation energy by emitting a particle.
The more interactions the particle undergoes on its way through the
nucleus, the more it looses its memory of incoming direction and
energy until it reaches some equilibrium state and the compound
model takes over. The cross section for this pre-equilibrium state
is described by
\begin{equation}\label{eq:exixsec}
    \frac{\textrm{d}\sigma_{k}^{\textrm{P}}}{\textrm{d} E_k} =
    \sigma^{\textrm{CF}}
    \sum_{p_{\pi}=p_{\pi}^0}^{p_{\pi}^\textrm{max}}
    \sum_{p_{\nu}=p_{\nu}^0}^{p_{\nu}^\textrm{max}}
    W_k\left( p_{\pi} , h_{\pi} , p_{\nu} , h_{\nu} , E_k \right)
    \tau\left( p_{\pi} , h_{\pi} , p_{\nu} , h_{\nu} \right) P\left( p_{\pi} , h_{\pi} , p_{\nu} , h_{\nu} \right)
\end{equation}
where $\sigma^{\textrm{CF}}$ is the total compound formation cross
section, $W_k$ is the emission rate, $\tau$ the lifetime of the
particle-hole states and $P$ the states that survived and are left
from the last process.

The total compound-formation cross section is just the reaction
cross section, that can be calculated from the \ac{OMP}, with the
sum of the cross sections for direct reactions calculated from the
\ac{DWBA} subtracted, $\sigma^{\textrm{CF}} =
\sigma^{\textrm{reac}}-\sigma^{\textrm{direct}}$. The emission rate
depends on known parameters, the inverse reaction cross section
calculated from the \ac{OMP} and the particle-hole state density.
Both the lifetime and the survived population are calculated from
the internal transition rates for particle-hole pair creation,
conversion and annihilation \cite{talys}.

\subsection{Compound formation}

For the compound formation one can separate two different processes.
At low energies it is a binary reaction where the projectile is
captured by the target, followed by emission of another particle.
But at high incident energies the nucleus will still be left in a
highly excited state after the binary reaction and other process
take over. Although the two processes at first sight seem identical,
there are two major differences. One of them being that the first
process is non-isotropic, and the other the presence of a width
fluctuation factor in the first process that correlates the incoming
and outgoing waves \cite{2005AIPC..769.1154K}. An analogy is often
made with the evaporation of a hot fluid, that works in basically
the same way, where particles evaporate from the nucleus until it is
cooled down.

Also for compound reactions TALYS includes several models for
different situations. At high energies, as in this case, the default
model used is multiple pre-equilibrium emission within the exciton
model. This is an extension of the pre-equilibrium model and
includes the pre-equilibrium population $\mathcal{P}^{\textrm{pre}}
\left( Z,N,p_\pi,h_\pi,p_\nu,h_\nu,E_x(i) \right)$ that needs to be
included in (\ref{eq:exixsec}) and summed over all possible
combinations of parameters. The populations that are not used for a
new $\mathcal{P}^{\textrm{pre}}$ are used to feed another multiple
emission model called multiple Hauser-Feshbach decay \cite{talys}.

\subsection{Angular distribution}

Continuum angular distributions can be achieved using the
systematics by \citet{1988PhRvC..37.2350K}. The systematics is
based on that the angular distribution, to a first approximation, is
independent of the energy of the projectile, as well as the type of
the target, projectile and ejectile. Instead it depends on the
emitted particles energy and the fraction of multi-step direct and
multi-step compound emissions  for the given energy
\cite{1988PhRvC..37.2350K}. In this phenomenological approach
the angular distribution is described as
\begin{equation}\label{eq:kalbach}
    \frac{\textrm{d}^2 \sigma}{\textrm{d} \Omega \textrm{d} E} = \frac{1}{4 \pi} \frac{\textrm{d} \sigma}{\textrm{d}
    E} \frac{a}{\sinh(a)}\left( \cosh(a\cos\theta) +
    f_{\textrm{MSD}} \sinh(a\cos\theta) \right)
\end{equation}
with $a=a(E)$ as a free parameter to be fitted and $f_{\textrm{MSD}}$ as
the fraction of emissions that comes from multi-step direct
processes. This parameter, as well as the energy-differential cross
section, $\frac{d \sigma}{d E}$, is assumed to be known. This result
is physically motivated by \citet{1994PhRvC..50.2490C}. The form in
Eq. (\ref{eq:kalbach}) is the form that TALYS uses, but for fitting
purposes the generalized form

\begin{equation}\label{eq:gen_kalbach}
    \frac{\textrm{d}^2 \sigma}{\textrm{d} \Omega \textrm{d} E} = b e^{a\cos\theta}
\end{equation}
is used. Here $a=a(E)$ is still a free parameter and another free
parameter, $b=b(E)$, is introduced.

\clearpage
\setcounter{figure}{0} \setcounter{table}{0}
\setcounter{equation}{0}

\section{Experimental setup\label{sec:setup}}

The experiment was carried out at \ac{TSL} in Uppsala in February
2005 using a setup called MEDLEY. The MEDLEY setup is designed to
measure light ion reactions like (n,p), (n,d), (n,t), (n,$^3$He) and
(n,$\alpha$). Besides calcium, MEDLEY has also performed cross
section measurements at 96 MeV on other nuclei of biological or
technical relevance like silicon \citep{2004PhRvC..69f4609T}, oxygen
\citep{2006PhRvC..73c4611T}, carbon \cite{tippcomm}, iron, lead and
uranium \citep{2004PhRvC..70a4607B}. This chapter contains a brief
description of \ac{TSL}, the neutron beam, and a more in depth
description of the MEDLEY spectrometer setup. The last two sections
are devoted to the electronics and data acquisition system involved.

\subsection{The Svedberg Laboratory}

The protons for the primary beam was produced by the Gustaf Werner
isochronous cyclotron that can accelerate protons up to energies of
180 MeV and heavy ions with charge $Q$ and mass number $A$ up to 192
$Q^2/A$ MeV. It consists of two D-shaped electrodes with a
perpendicular magnetic field that makes the particles go in a
circular pattern, and a high-voltage field between the electrodes
that make the particles accelerate in the gap between the
electrodes. By using a \ac{RF} alternating electric field the
particle will get a spiral trajectory outwards until extracted. The
Gustaf Werner cyclotron is operated with a \ac{RF} of 58 ns for 98
MeV protons. However, for protons of energies higher than 100 MeV,
as well as ${^3}$He at their highest energies, relativistic effects
need to be compensated for and the cyclotron switches to a frequency
modulated synchrocyclotron mode where the driving \ac{RF} is varied.
This switching between modes is something that makes the Gustaf
Werner cyclotron unique.

\begin{figure}[!htb]
  \centering
\includegraphics[width=\textwidth,bb=0 0 576 756]{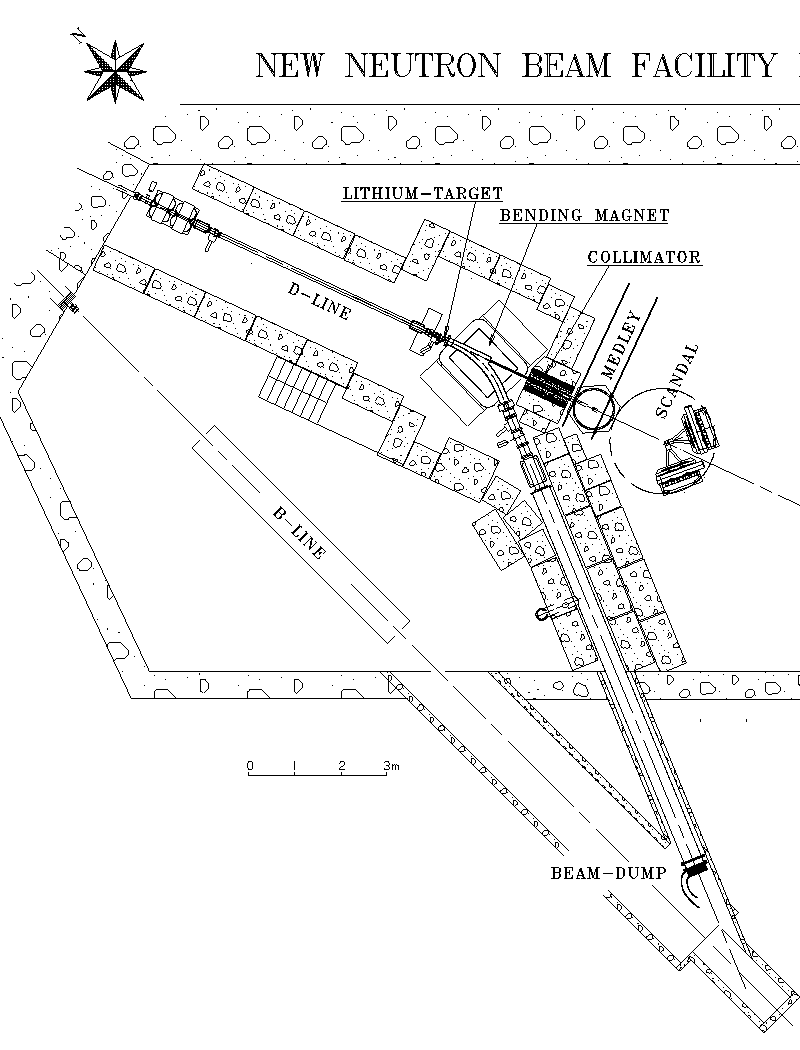}
  \caption[Drawing of the new neutron beam facility in the Blue Hall]{Drawing of the new neutron beam facility in the Blue Hall at \ac{TSL}}\label{fig:beam}
\end{figure}

From the cyclotron the proton beam is transported to the Blue Hall,
as seen in the drawing in figure \ref{fig:beam}. Here it hits a
highly enriched target of $^7$Li, where it induces the nuclear
reaction $^7$Li(p,n)$^7$Be with a $Q$ value of 1.64 MeV
\cite{physics}, resulting in a quasi-monoenergetic neutron beam in
the range of 20-180 MeV. There are different thicknesses of the
lithium target available, ranging from 2 to 24 mm
\citep{2005AIPC..769..780P}, in this experiment the thickness used
was 8 mm. The remaining proton beam is bent away using a bending
magnet, actually recycled from the former LISA spectrometer, into a
well-shielded \ac{BD} and integrated in a Faraday cup. The charge
deposited in the \ac{BD} is used as one of three independent
monitors for the neutron flux.

\begin{figure}[!htb]
  \centering
  \begin{minipage}{0.48\textwidth}
\includegraphics[width=\textwidth,bb=0 0 72 54]{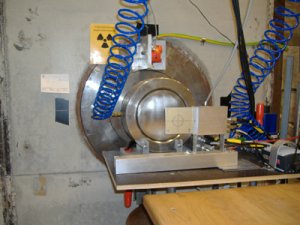}
  \end{minipage}
  \begin{minipage}{0.30\textwidth}
\includegraphics[width=\textwidth,bb=0 0 238 337]{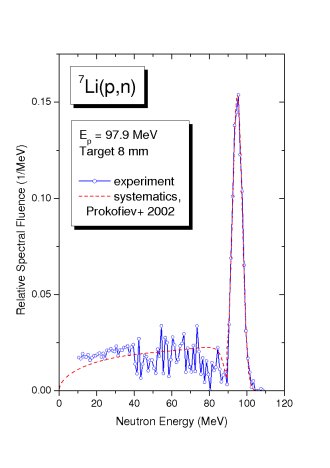}
  \end{minipage}
  \caption[The exit of the collimator]{In the left panel, the exit of the collimator with the larger cylindrical \ac{ICM} and the smaller rectangular \ac{TFBC} in front. The right panel shows the neutron spectrum emerging from the collimator \protect \citep{pompcomm}.}\label{fig:monitors}
\end{figure}

The actual beam is defined by a collimator, consisting of 100 cm
long iron cylinders of different diameter, embedded in a wall of
concrete. These cylinders can give a beam of, more or less, any
shape and diameter up to 30 cm, even though the standard forms
available are circular beams in steps of about 5 cm. Right after the
collimator is the experimental area with the MEDLEY setup, that is
used for experiments concerning neutron-induced charged particle
production and will be described in more detail in section
\ref{sec:medley}. It is centered at a distance of 3.74 m from the
lithium production target. After MEDLEY there is another setup
called SCANDAL \citep{2002NIMPA.489..282K} used for experiments on
elastic neutron scattering. Running several experiments at once like
this is possible since almost the entire neutron beam passes the
first setup without interacting. Two more independent neutron
monitors are available namely the \ac{ICM} that uses the ionization
of the gas inside the detector and via an electric field collects
the ion pairs created \citep{knoll}, and the \ac{TFBC} \citep{tfbc}.
The exit area from the collimator as well as the \ac{ICM} and the
\ac{TFBC} can be seen in figure \ref{fig:monitors}. However, in this
experiment there was a problem with the \ac{ICM} so that it could
not be used as an absolute monitor of the neutron flux, but instead
as a relative monitor calibrated to the \ac{TFBC}.

\subsection{MEDLEY spectrometer\label{sec:medley}}

\begin{figure}[!htb]
  \centering
  \begin{minipage}{0.48\textwidth}
\includegraphics[width=\textwidth,bb=0 0 253 179]{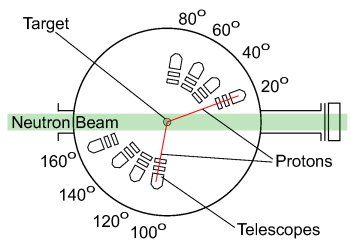}
  \end{minipage}
  \begin{minipage}{0.48\textwidth}
\includegraphics[width=\textwidth,bb=0 0 72 54]{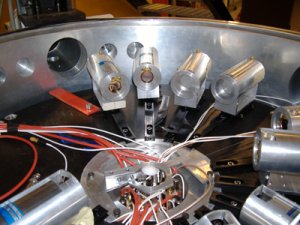}
  \end{minipage}
  \caption[The MEDLEY spectrometer]{The MEDLEY spectrometer. The left panel shows a schematic sketch of the scattering chamber and in the right panel is a photography of a similar setup.}\label{fig:medley}
\end{figure}

The MEDLEY spectrometer, as seen in figure \ref{fig:medley},
consists of a 24 cm high cylindrical vacuum chamber with 90 cm
diameter. Inside this chamber there are eight telescopes mounted at
even intervals covering angles from 20 to 160 degrees, four on each
side, and can be changed by rotating the table inside the chamber.
The distance from the target to the different telescopes can also be
varied.

\begin{figure}[!htb]
  \centering
\includegraphics[width=0.48\textwidth,bb=0 0 72 54]{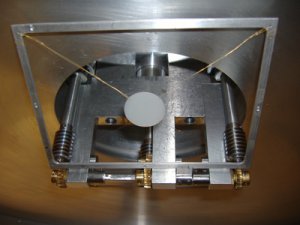}
  \caption[One of the aluminum frames and the CH$_{2}$ target]{One of the aluminum frames and the CH$_{2}$ target stinged up inside the MEDLEY spectrometer.}\label{fig:medleytarget}
\end{figure}

To get a clean positioning of the target within the beam without
getting too much background signal from the target holder, a small
aluminum frame is used that is well off the beam. The target itself
is strung up in the frame using thin wires. This setup can be seen
in figure \ref{fig:medleytarget}. There are three of these frames so
one can easily switch between targets without breaking the vacuum.
For calibration purposes an ${^{241}}$Am source is also available,
though this was not used in this experiment.

\subsection{Detector telescopes}

\begin{figure}[!htb]
  \centering
  \begin{minipage}{0.48\textwidth}
\includegraphics[width=\textwidth,bb=0 0 253 179]{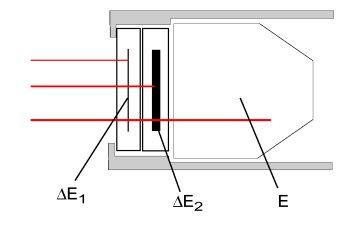}
  \end{minipage}
  \begin{minipage}{0.48\textwidth}
\includegraphics[width=\textwidth,bb=0 0 72 54]{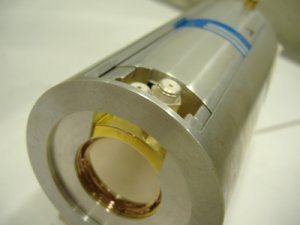}
  \end{minipage}
  \caption[One of the detector telescopes]{One of the detector telescopes. The left panel shows a schematic sketch of the telescope as well as the different kinds of events mentioned in section \ref{sec:silicon} and in the right panel is a photography of a similar telescope.}\label{fig:telescope}
\end{figure}

Each of the telescopes, as can be seen in figure
\ref{fig:telescope}, in the MEDLEY spectrometer consists of three
individual detectors. First there are two silicon detectors, one
thin of about 50 - 60~\textmu m and one thick of about 400 -
500~\textmu m thickness respectively. Finally there is a 5 cm long
crystal of \ac{CsI(Tl)}. These detectors are often referred to as
the $\Delta E_1$, the $\Delta E_2$, and the $E$ detector, or
sometimes A, B, and C detector. The detectors are mounted in an
aluminum housing. Since the configuration of the detectors in this
experiment was not the usual for MEDLEY it is listed in table
\ref{tab:decset}.

There also exists a possibility to use plastic scintillators mounted
as active anti-coincidence collimators to discard the signals from
particles that did not pass straight into the detector. Attempts
have actually been made in earlier experiments, but due to problems
with them they were not used in this experiment.

\subsubsection{Silicon detectors\label{sec:silicon}}

The two most common semiconductor detectors are silicon and
germanium detectors, and are in principle a large diode in reverse
bias. The primary way of energy loss is via Coulomb interaction in
the material where the atoms are either ionized or excited. This
will create electron-hole pairs that can be collected by two
electrodes and measured as a small current pulse when a particle is
passing through.

\begin{figure}[!htb]
  \centering
\includegraphics[width=0.48\textwidth,bb=0 0 567 567]{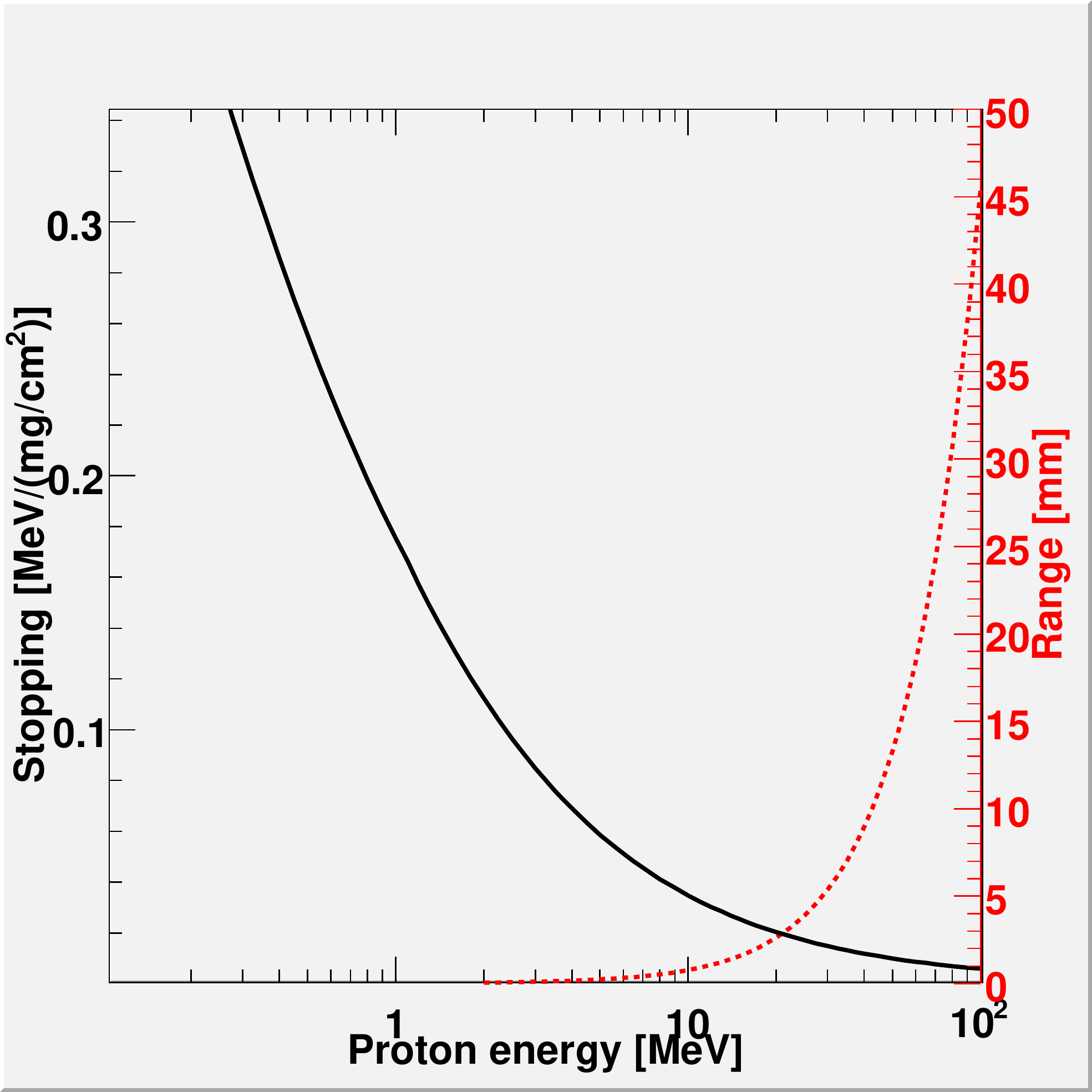}
  \caption[Stopping and range of protons in silicon]{Stopping power (solid line) and range (dashed line) of protons in silicon calculated using SRIM.}\label{fig:stopping}
\end{figure}

Another feature that is fundamental for this experiment is that the
amount of energy lost per unit length is not constant. Actually it
decreases with increasing energy. In figure \ref{fig:stopping} the
stopping power and the distance a proton can travel before it is
completely stopped, the so called range, is plotted for the range of
proton energies relevant in this experiment. This property will be
used for particle identification in the analysis.

These detectors are standard fully depleted surface-barrier
detectors from ORTEC \citep{2000NIMPA.452..484D} were the silicon
wafer is glued to a metal ring. One side is covered with a thin
layer of gold and the other with aluminum \citep{ortec}. The $\Delta
E_1$ detector is used as a transmission detector, that is the
particles pass through the detector leaving only a fraction of its
energy. This is also the case for the $\Delta E_2$ detector at
energies above 10~MeV, but for low energy events the proton is
stopped in this detector. Protons that stop in $\Delta E_1$ are
below the cutoff and are not included in the analysis. For an
illustration of the different kinds of events, see
figure~\ref{fig:telescope}.

\subsubsection{CsI(Tl) detectors\label{sec:CsI(Ti)}}

An inorganic scintillator is used as $E$ detector in this setup.
Here the incoming radiation creates electron-hole pairs in the
crystal by moving an electron from the valence band to the
conduction band on its way through the crystal. In the crystal, some
impurities are often added. These impurities will have lower
ionization energy than the crystal, which is why the holes will
drift here and ionize the impurities. The ionized impurities will
then pick up an electron from the conduction band and decay to the
ground state and emit a scintillation photon \citep{knoll}.

A quite popular inorganic scintillator is the cesium iodide,
available with both thallium and sodium activation. The high density
of the \ac{CsI(Tl)} gives it large stopping power and it also has a
really high light yield of about 65000 photons/MeV \citep{knoll}.
The wavelength of the emission light, 540 nm, is quite bad suited
for a photomultiplier tube so instead a Hamamatsu photodiode is used
\citep{2000NIMPA.452..484D}. To suite this, the last 2 cm of the 4
cm diameter cylinder is tapered off to 1.8 cm to fit the surface of
the diode. Another useful feature of the \ac{CsI(Tl)} worth
mentioning is that it has a variable decay time for different kinds
of particles \citep{knoll} that makes it suitable to use with pulse
shape analysis.

\subsection{Readout and electronics\label{sec:readelectr}}

\begin{figure}[!htb]
  \centering
\includegraphics[angle=90, width=\textwidth, bb=0 0 864 612]{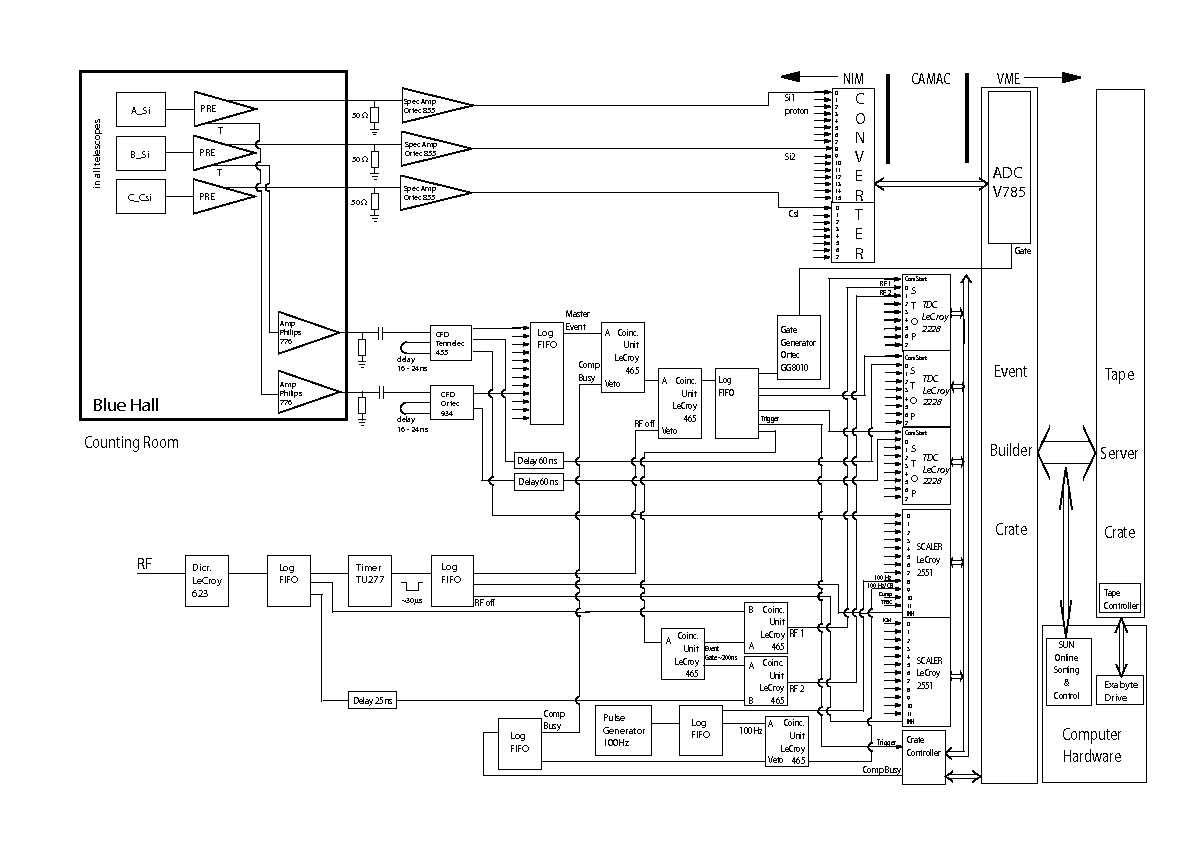}
  \caption[MEDLEY electronics]{Scheme of the MEDLEY electronics as of week 9, 2004}\label{fig:electro}
\end{figure}

Both the $\Delta E_1$ and the $\Delta E_2$ detector can generate a
trigger signal, in order to allow triggers from both low- and
high-energy particles. The signal from the detector is split up in
two branches, a timing branch and an energy branch. A complete
scheme is found in figure \ref{fig:electro}.

A signal from a detector is first amplified and then, if the pulse
height is enough, converted to a logical signal using a \ac{CFD}.
The timing signal is split up into two branches. One of them is
connected via a \ac{FIFO} logic to the other detector and telescopes
as a master event. The other branch is delayed and sent as a stop
signal for the timing channel in the \ac{TDC}. If the computer is
busy processing earlier event the current event is discarded,
otherwise it, through a logic \ac{FIFO}, generates an energy gate
signal for the \ac{ADC} and \ac{QDC} units, a start signal for the
\ac{TDC} and finally a trigger for the data acquisition.

The energy signal is recorded from all three detectors and all eight
telescopes as soon as one detector triggers. It is sent into a
\ac{ADC}, with the previously named master signal as gate, after
amplifying and shaping. From the cyclotron one also gets a \ac{RF}
signal that indicates a pulse from the neutron beam. The \ac{RF} is
split up through a \ac{FIFO} into three branches. Two of the \ac{RF}
signals, where one is delayed, are used as a stop signal for the
\ac{TOF} channels in the \ac{TDC}\footnote{It may sound backwards to
use the registration of a particle as a start signal and the next
beam pulse as stop for the \ac{TDC}, but this is simply a matter of
not overloading the electronics with unnecessary start signals. The
reason to store the \ac{RF} two times where one is delayed may not
be intuitively clear either, but this is explained in more detail in
section \ref{sec:timing}, and also in \citep{aboutRF}.}. It is also
possible to use this \ac{RF} signal as a veto for the master signal
if the cyclotron is switched off.

\subsection{Data acquisition}

For the data acquisition a system called SVEDAQ is used, and it is
summarized here from \citep{svedaqdescr}. The system is taken from
the EUROGAM experiment and has been modified to be suitable for use
at the \ac{TSL}. It is divided into three main
blocks, as can be seen in figure
\ref{fig:electro}, connected via two independent networks. The three
blocks are the event builder, the tape server and the control and
monitoring workstation.

The event builder is the block that reads out the data from the
\ac{ADC} units in the VME crate as well as the \ac{TDC} units and
scalers in the CAMAC crate. The event builder itself is also located
in the VME. It mainly consists of a Motorola CPU that runs the
readout program and sends packages to the data network when the
readout buffers are full. It also sends the computer busy signals
for vetoing of the master event and the 100 Hz clock so that events
occurring during processing of other events are discarded.

The tape server is an optional part of the acquisition system. Of
course if one wants to use the data after the experiment one needs
to write it to the hard drive, but if one wants to conduct different
kinds of test runs it is not needed by the event builder.

To control all this a SUN workstation is used. It is via this
workstation that the system is started and stopped, but since the
three blocks are more or less independent of each other the event
builder and tape server will continue to write data from the CAMAC
and VME even if the workstation is completely shut down. This
workstation is connected to the local area network at \ac{TSL} and
can be remotely controlled from there. But it is also connected to
the data network via a sort-spy daemon process so it can listen to
the traffic without formally being a part of the network with its
own IP number. This makes it possible to display a large number of
one- and two-dimensional spectra from the CAMAC signals and conduct
some on-line sorting and analysis.

\clearpage
\setcounter{figure}{0} \setcounter{table}{0}
\setcounter{equation}{0}

\section{Data analysis}

The analysis of the data is split up into several parts. Since the
SVEDAQ system writes the events as binary data, the raw files first
have to be properly decoded. After the decoding, the information
needs to be calibrated so that it corresponds to correct energy and
timing information, before the particle selection procedure can take
place, where events corresponding to the right particle type and
incoming neutron energy are selected. Finally the relevant
corrections for background and other effects can be carried out
before extracting the final cross section.

During the analysis, three different data sets were analyzed
simultaneously. Two of them were the calcium data and the background
data, while the third was a reference data set of polyethene for
calibration and normalization purposes. The complete set of runs is
listed in tables \ref{tab:runlist180} and \ref{tab:runlist0}.

\subsection{ROOT}

The main tool used for the analysis is ROOT. ROOT is an object
oriented data analysis framework initiated at CERN by Ren\'{e} Brun and
Fons Rademakers\footnote{These two have also been a part of creating
other popular data analysis and simulation tools like PAW, PIAF, and
GEANT.}. The framework is freely distributed as open source, where
everyone is free to further distribute and modify it. This is
something that has contributed, and still is contributing, to its
rapid growth. It contains a large amount of packages concerning
different part of data analysis like histogramming, curve fitting,
minimization, statistics tools and much more \citep{ROOT}.

The framework of ROOT is closely bound to a C++ interpreter called
CINT that is written by Masahuru Goto. This means that one actually
can use C++ as a scripting language for rapid prototype development
of programs. Then the same code can be compiled to take advantage of
the fast running of a machine language executable, unlike if one had
used a normal scripting language like Ruby, Perl or Python for the
prototyping.

\subsection{Decoding}

The raw data from SVEDAQ are written into binary data files in the
EUROGAM standards with a 24 byte header followed by event-by-event
records \citep{svedaqdescr}. A typical example of a data sequence
can be found in figure \ref{tab:hexraw}. The events are recorded as
a series of shorts\footnote{One short is 2 bytes}, with the first
one being the number of bytes that are used to register one event.
Between the first short and the telescope events there are 20 scaler
channels with numbers 1-13 containing information about the
telescope triggers, the 100 Hz clocks and the neutron monitors. The
last seven channels are left empty. Next is the telescope events,
each of them written in five shorts corresponding to three energy
signals, one from each detector, and timing signals from the two
silicon detectors. Finally the two \ac{RF} signals are written
before the event is ended. To translate the SVEDAQ files into ROOT
files a small computer code called SV2ROOT \citep{terucomm} was
used.

\subsection{Calibration}

\subsubsection{Energy calibration}

Since the data from the runfile are given in the unit of channel
numbers from the \ac{ADC} an energy calibration is needed to convert
these to incoming energies. The silicon detectors are assumed to
have a linear response to the energy, that is the correlation looks
like
\begin{equation}\label{eq:linear}
    E=kx+m
\end{equation}
where $E$ is the energy output, $x$ is the channel number, and $k$
and $m$ the constants to be fitted.

Unfortunately the \ac{CsI(Tl)} is not that simple since it has a
non-linear response function between light output, $L$, and incoming
energy, $E$, which makes the calibration more complicated. The light
output is assumed to be given by a three-parameter formula that
follows
\begin{equation}\label{eq:scintresponse}
    L = a_0 + a_1 \left( E - a_2 A z^2 \ln \left[ \frac{E + a_2 A z^2}{a_2 A z^2} \right]
    \right),
\end{equation}
where $a_i$ is the fitting constants, and $A$ and $z$ is the mass
number and charge of the incident particle
\citep{2000NIMPA.452..484D}. However, to conduct the calibration we
need the inverse of expression (\ref{eq:scintresponse}). Since this
inverse is analytically complicated to calculate, the approximation
according to \citet{2000NIMPA.452..484D}
\begin{equation}\label{eq:energulight}
    E \approx a + b L + c (bL)^2
\end{equation}
is used for the hydrogen isotopes. The parameter $c$ depends only on
particle type and is found by \citet{tippawan} to be $c = 0.0032$
for protons, which leaves only $a$ and $b$ as constants to be
fitted.

Here the calibration of the silicon detectors is done by calculating
the incoming energies needed to break through each of the silicon
detectors - the punch through energies. These are calculated using a
program called \ac{SRIM}, based on the work of \citet{srim}. The
punch through values for \ac{T1}, \ac{T2}, \ac{T7} and \ac{T8} are
found in table \ref{tab:Sidecs}.

\begin{table}
  \centering
  \begin{tabular}{lcccccccc}
    \hline
    & \multicolumn{2}{c}{T1} & \multicolumn{2}{c}{T2} & \multicolumn{2}{c}{T7} & \multicolumn{2}{c}{T8}\\
    & $\Delta E_1$ & $\Delta E_2$ & $\Delta E_1$ & $\Delta E_2$ & $\Delta E_1$ & $\Delta E_2$ & $\Delta E_1$ & $\Delta E_2$\\
    \hline
    \hline
    Thickness [\textmu m] & 64.9 & 549 & 60.5 & 538 & 52.9 & 424 & 61.6 & 550\\
    \hline
    p [MeV]& 2.42 & 8.62 & 2.32 & 8.51 & 2.34 & 8.63 & 2.13 & 7.41\\
    d [MeV]& 3.12 & 11.5 & 2.99 & 11.4 & 3.02 & 11.5 & 2.74 & 9.87\\
    t [MeV]& 3.60 & 13.5 & 3.43 & 13.4 & 3.47 & 13.6 & 3.14 & 11.6\\
    $^{3}$He [MeV]& 8.60 & 30.6 & 8.23 & 30.2 & 8.32 & 30.6 & 7.56 & 26.3\\
    $^{4}$He [MeV]& 9.56 & 34.4 & 9.14 & 34.0 & 9.24 & 34.5 & 8.38 & 29.6\\
    $^{6}$Li [MeV]& 18.1 & 65.3 & 17.3 & 64.6 & 17.5 & 65.4 & 15.9 & 56.3\\
    $^{7}$Li [MeV]& 19.1 & 69.7 & 18.3 & 68.9 & 18.5 & 69.8 & 16.7 & 60.0\\
    \hline
  \end{tabular}
  \caption{Thickness and punch through energies for the different silicon detectors used.}\label{tab:Sidecs}
\end{table}

\begin{figure}[!htb]
  \centering
  \begin{minipage}{0.48\textwidth}
\includegraphics[width=\textwidth,bb=0 0 350 350]{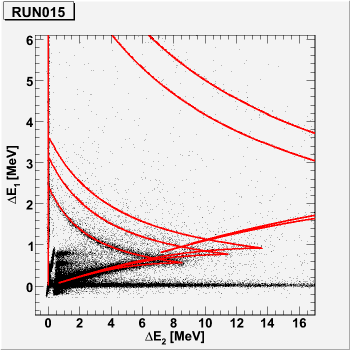}
  \end{minipage}
  \begin{minipage}{0.48\textwidth}
\includegraphics[width=\textwidth,bb=0 0 350 350]{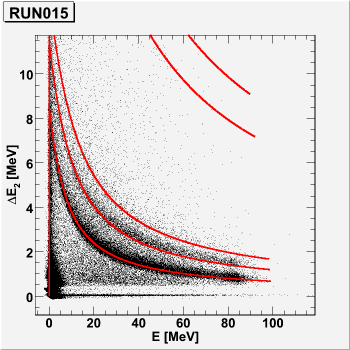}
  \end{minipage}
  \caption[Calibrated $\Delta E_1$-$\Delta E_2$ and $\Delta E_2$-$E$ plots for T1]{Calibrated $\Delta E_1$-$\Delta E_2$ and $\Delta E_2$-$E$ plots for \ac{T1}. The red lines correspond to tabulated energy loss values for protons, deutrons, tritons, $^3$He and alpha particles in ascending order. The apparent mismatch for some lines in the right panel is due to the non-linear effects in the \ac{CsI(Tl)}.}\label{fig:rawplots}
\end{figure}

If one plots the $\Delta E_1$ versus $\Delta E_2$, these punch
though points manifest themselves as turning points in the particle
bands as can be seen in the left panel of figure \ref{fig:rawplots}.
It is now quite straightforward to compare the channel numbers of
these areas to the values in table \ref{tab:Sidecs} and get the
coefficients to (\ref{eq:linear}).

But once again, the \ac{CsI(Tl)} poses a bit of a problem. As can be
seen in the right panel of figure \ref{fig:rawplots}, there are no
turning points available for calibration. Instead, at least in
forward angles, one can use the quite strong peak of elastic p(n,p)
scattering in the CH$_2$ runs as one calibration point. In the lower
part of the $E$ scale the proton band is quite bent, so here a
couple of points are selected and the energy deposited in the $E$
detector is calculated from the energy lost in $\Delta E_2$. 
Numerical values obtained for the calibration constants are listed
in table \ref{tab:fitparamet}.

\subsubsection{Time calibration\label{sec:timing}}

For the time calibration a couple of special runs devoted were made.
During these runs cables of different lengths, giving different
delay times, were put between the \ac{RF} signal and the \ac{TDC},
displacing the peak in the recorded \ac{RF} signal. The result from
these runs can be fitted with a linear calibration formula

\begin{figure}[!htb]
  \centering
\includegraphics[width=0.48\textwidth,bb=0 0 283 283]{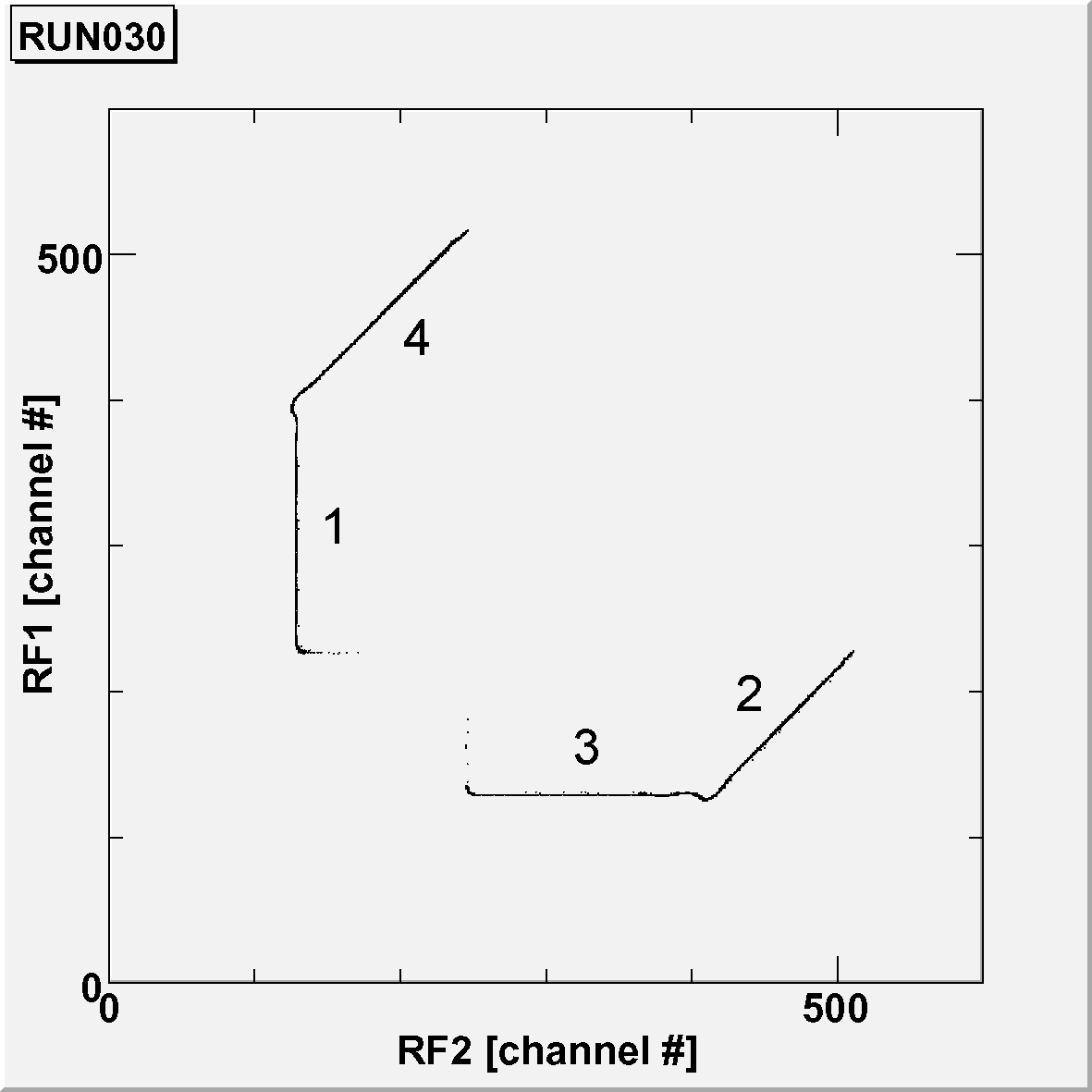}
  \caption[RF 1 versus RF 2]{RF 1 versus RF 2. Each point in the 8-shaped trajectory correspond to a time between detection and production of the next burst. One lap around is 58 ns.}\label{fig:rf1rf2}
\end{figure}

\begin{figure}[!htb]
  \centering
\includegraphics[width=0.6\textwidth,bb=0 0 1015 303]{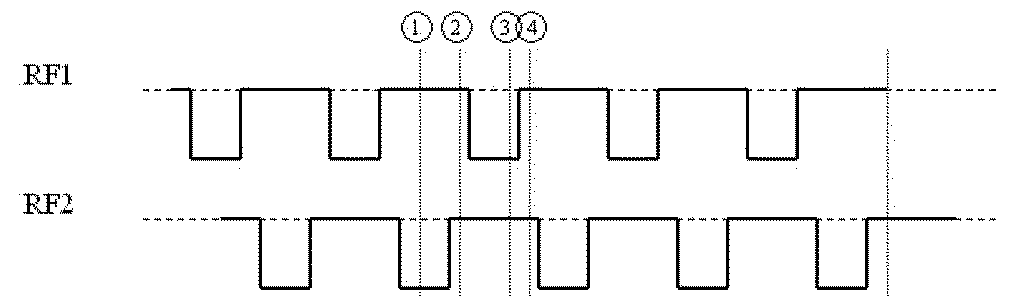}
  \caption[Scematic illustration of the two RF signals]{Scematic illustration of the two \ac{RF} signals and the different possible events that can occur \protect \citep{aboutRF}. The \ac{RF} is measured as the from the trigger to the next
signal from the cyclotron.}\label{fig:rfareas}
\end{figure}

When each of the \ac{RF} signals are time calibrated they need to be
combined into a single time line. The two \ac{RF} signals can be
plotted against each other in figure \ref{fig:rf1rf2}, where four
areas can be identified, schematically illustrated in figure
\ref{fig:rfareas}. Events in area 1 and area 3 are events that occur
during one of the \ac{RF} signal pulses, and hence get start and
stop signals simultaneously. So in these areas we only get signal
from one of the \ac{RF}. In area 2 and area 4 there is a signal from
both \ac{RF} 1 and \ac{RF} 2.

To translate the two \ac{RF} signals into a single time scale, one
of the \ac{RF} signals is chosen as base, in this case \ac{RF} 1. In
the areas where no information exists for the chosen base \ac{RF}
signal, the information from the other \ac{RF} signal is transformed
to the chosen base according to the algorithm in \citep{aboutRF}.

\subsection{Particle identification\label{sec:partid}}

\begin{figure}[!htb]
  \centering
  \begin{minipage}{0.48\textwidth}
\includegraphics[width=\textwidth,bb=0 0 350 350]{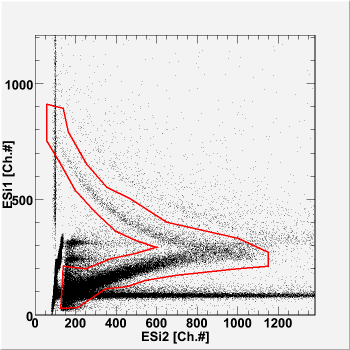}
  \end{minipage}
  \begin{minipage}{0.48\textwidth}
\includegraphics[width=\textwidth,bb=0 0 350 350]{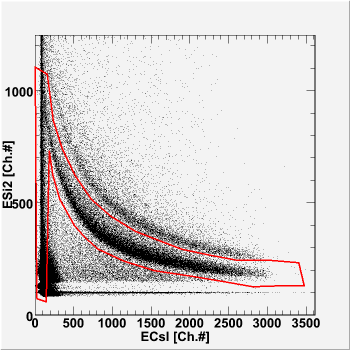}
  \end{minipage}
  \caption[First graphical selection cuts for protons]{First graphical selection cuts for protons in \ac{T1}. In the left panel the cut for the $\Delta E_1$-$\Delta E_2$ plot is shown and in the right panel the $\Delta E_2$-$E$ plot is shown.}\label{fig:firstcuts}
\end{figure}

As mentioned earlier, the bands that are shown in figure
\ref{fig:rawplots} correspond to different kinds of charged
particles. This makes the particle identification procedure quite
straightforward by taking some rough cuts around the structures of
the desired particle type. These cuts should be very generous, since
it is always easier to remove unwanted events than to recreate
wanted ones\footnote{At least concerning background events.
Misidentified events from, for example, deuterons might still be
tricky to get rid of if the cuts are too generous.} and is shown for
\ac{T1} in figure \ref{fig:firstcuts}. The criterion here is that an
event must be in both of these to be selected.

\begin{figure}[!htb]
  \centering
  \begin{minipage}{0.48\textwidth}
\includegraphics[width=\textwidth,bb=0 0 350 350]{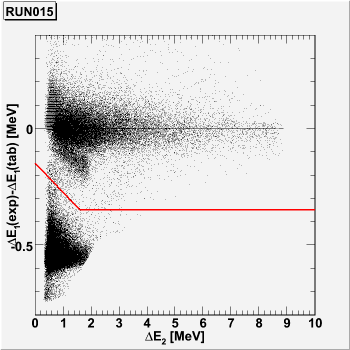}
  \end{minipage}
  \begin{minipage}{0.48\textwidth}
\includegraphics[width=\textwidth,bb=0 0 350 350]{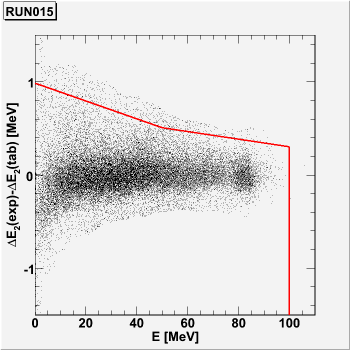}
  \end{minipage}\\
  \begin{minipage}{0.48\textwidth}
\includegraphics[width=\textwidth,bb=0 0 350 350]{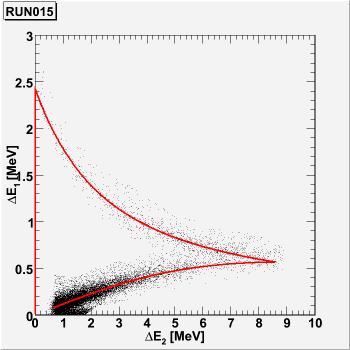}
  \end{minipage}
  \begin{minipage}{0.48\textwidth}
\includegraphics[width=\textwidth,bb=0 0 350 350]{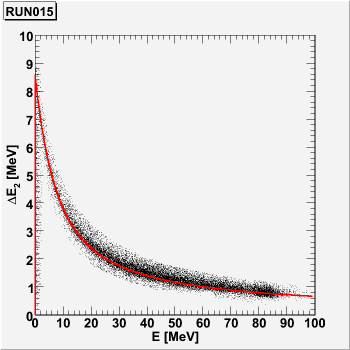}
  \end{minipage}
  \caption[Refined selection cuts for protons]{Refined selection cuts for protons in \ac{T1}. In the upper left panel the cut for the $\Delta E_{1}(\textrm{exp})-\Delta E_{1}(\textrm{tab})$ plot is shown and in the upper right panel the $\Delta
E_{2}(\textrm{exp})-\Delta E_{2}(\textrm{tab})$ plot is shown. The
lower panels contains the final proton cuts as in figure \protect
\ref{fig:firstcuts} for \ac{T1} together with the tabulated curve
from \ac{SRIM}}\label{fig:diff2d}
\end{figure}

When these first rough cuts are done, the particle bands are
straightened for a more refined selection of events. To do this, the
energy recorded in the $E$ detector is used together with \ac{SRIM}
to calculate the corresponding energies in the $\Delta E_{1}$ and
$\Delta E_{2}$ detectors. By subtracting this value from the energy
actually recorded the particle bands are straightened out around
zero. The $\Delta E_{1}(\textrm{exp})-\Delta E_{1}(\textrm{tab})$
selection is mainly used to remove noise in the low energy region in
the $\Delta E_1$-$\Delta E_2$ plot, and the $\Delta
E_{2}(\textrm{exp})-\Delta E_{2}(\textrm{tab})$ selection is used to
get a cleaner separation between protons and deuterons. These two
are shown in figure~\ref{fig:diff2d}.

\subsection{Time-of-Flight measurements\label{sec:tof}}

When only protons are selected, the next restriction is to only
select protons that are produced by neutrons of the correct energy.
Since the neutron beam is not completely monoenergetic, but contains
about 60 \% lower energy neutrons \citep{2000NIMPA.452..484D}, the
contribution from these has to be removed. To do this the \ac{TOF}
information obtained in section \ref{sec:timing} is used. The
\ac{TOF}, $t$,  for a particle with energy $E_k$ and velocity $v$ to
travel a distance $s$ can be calculated as
\begin{equation}\label{eq:simpletof}
    t = \frac{s}{v}.
\end{equation}
The velocity of the particle is given in its rest mass, $m_0$, and
its momentum, $p$, as
\begin{equation}\label{eq:velocity}
    v = \frac{p}{\sqrt{m_0^2+\frac{p^2}{c^2}}}
\end{equation}
where the $p$ is given by
\begin{equation}\label{eq:moment}
    p = \frac{1}{c}\sqrt{E_k^2 + 2 E_k m_0 c^2}.
\end{equation}
Plugging (\ref{eq:moment}) into (\ref{eq:velocity}) into
(\ref{eq:simpletof}) and simplifying yields
\begin{subequations}
    \begin{equation}
        t = \frac{s}{c}\sqrt{\frac{(m_0 c^2 + E_k)^2}{E_k (E_k + 2 m_0
        c^2)}}\label{eq:toft}
    \end{equation}
    \begin{equation}
        E_k = \sqrt{m_0^2+\frac{m_0^2}{\left( t \frac{c}{s} \right)^2-1}} - m_0\label{eq:tofe}
    \end{equation}
\end{subequations}
where Eqs. (\ref{eq:toft}) and (\ref{eq:tofe}) are equivalent.

\begin{figure}[!htb]
  \centering
  \begin{minipage}{0.48\textwidth}
\includegraphics[width=\textwidth,bb=0 0 350 350]{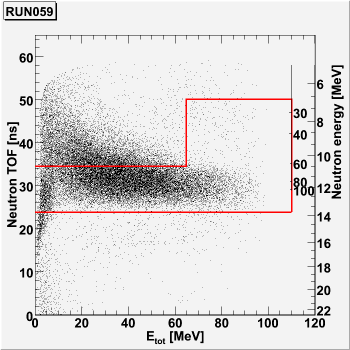}
  \end{minipage}
  \begin{minipage}{0.48\textwidth}
\includegraphics[width=\textwidth,bb=0 0 350 350]{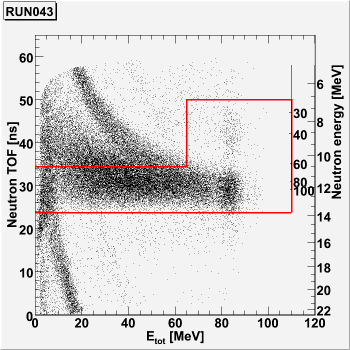}
  \end{minipage}
  \caption[Neutron TOF versus proton energy]{Neutron \ac{TOF} versus proton energy for calcium in the left panel, and CH$_2$ in the right panel, together with the \ac{TOF} cut. The the low energy tail in the neutron
spectra is clearly visible in the CH$_2$ plot. The extra \ac{TOF}
cut at high energies is motivated in section
\ref{sec:tofdiss}.}\label{fig:tofplots}
\end{figure}

But the time scale mentioned in section \ref{sec:timing} includes
both the neutrons \ac{TOF} as well as the charged particles \ac{TOF}
from the target to the telescope. As can be calculated from
(\ref{eq:toft}), if the particle is a 94.5 MeV neutron traveling a
distance of 3.74 m the \ac{TOF} would be about 29.9 ns. A 5 MeV
proton traveling 26.83 cm has a \ac{TOF} of 8.71 ns while a 90 MeV
proton would have a \ac{TOF} of only 2.19 ns. This makes quite a
large difference, especially at low energies, and is compensated for
by subtracting the charged particle \ac{TOF}, calculated from the
energy of $\Delta E_{1}$, $\Delta E_{2}$ and $E$ added together. The
final \ac{TOF} cuts are illustrated in figure~\ref{fig:tofplots}.

To choose a suitable sized cut, the elastic p(n,p) scattering is
chosen to define the width. This peak has a \ac{FWHM} of 6.6~ns,
quite in agreement with \citet{2000NIMPA.452..484D} where the
\ac{FWHM} is measured to be about 6-7 ns, and one of the large
contributions is the finite time of a beam pulse from the cyclotron.
The with of the cut is chosen as two standard deviations, and is a
compromise between statistics, minimizing mismatching errors, and a
high lower limit of the accepted neutron spectrum. For more details,
see section \ref{sec:mismatch}.

\subsection{Corrections}

\begin{figure}[!htb]
  \centering
  \begin{minipage}{0.45\textwidth}
\includegraphics[width=\textwidth,bb=0 0 142 142]{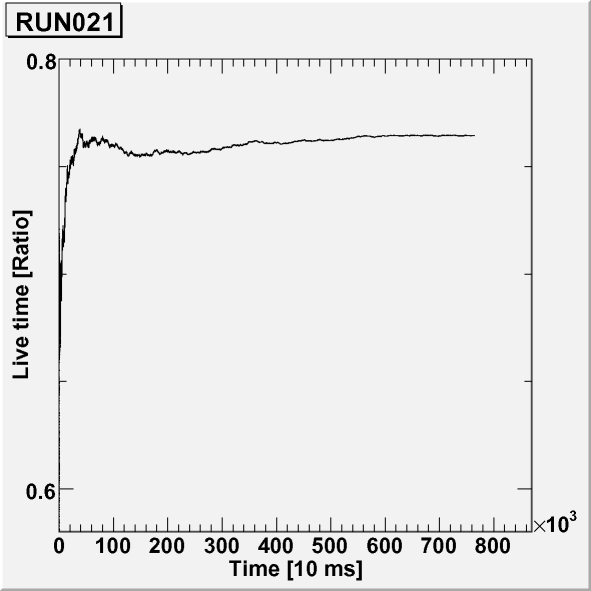}
    \caption[Typical live time of a run]{Typical live time of a run. The instabilities in the beginning are due to statistics.\vspace{10pt}}\label{fig:cLT}
  \end{minipage}
  \begin{minipage}{0.1\textwidth}
  \end{minipage}
  \begin{minipage}{0.45\textwidth}
\includegraphics[width=\textwidth,bb=0 0 567 567]{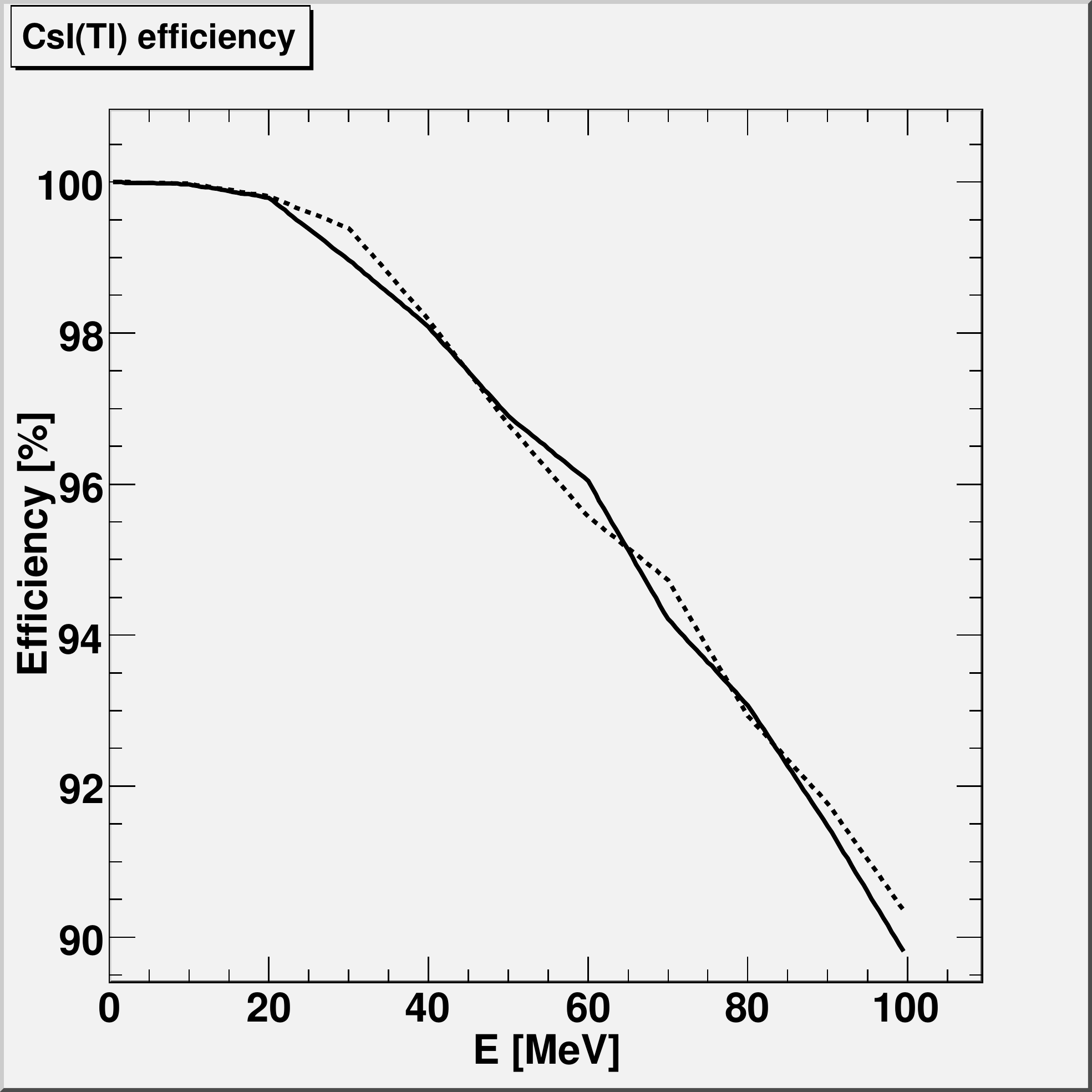}
    \caption[Efficiency of the CsI(Tl) detector]{Efficiency of the \ac{CsI(Tl)} detector. The dashed line is a simulation with MCNPX and the solid line is a simulation with GEANT \protect \citep{bildcomm}.}\label{fig:CsIeff}
  \end{minipage}
\end{figure}

For each run, the live time of the data acquisition system is
corrected for. As mentioned in section \ref{sec:readelectr}, events
that occur while an earlier event is treated are discarded. The
amount of discarded events is not constant, as seen in figure
\ref{fig:cLT}, but is assumed to be. The ratio of discarded events
is obtained by comparing the 100 Hz scaler and the 100 Hz computer
busy scaler. The finite efficiency of the \ac{CsI(Tl)} is also
corrected for. This has a non-linear dependence of energy deposited,
and this is simulated by both MCNPX and GEANT \citep{bildcomm}. The
results of these simulations are found in figure \ref{fig:CsIeff}.
Since they are in good agreement, one of them is choosen, in this
case the GEANT results.

\begin{figure}[!htb]
  \centering
  \begin{minipage}{0.48\textwidth}
\includegraphics[width=\textwidth,bb=0 0 284 284]{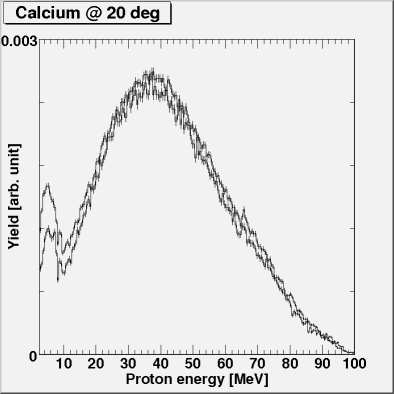}
  \end{minipage}
  \begin{minipage}{0.48\textwidth}
\includegraphics[width=\textwidth,bb=0 0 284 284]{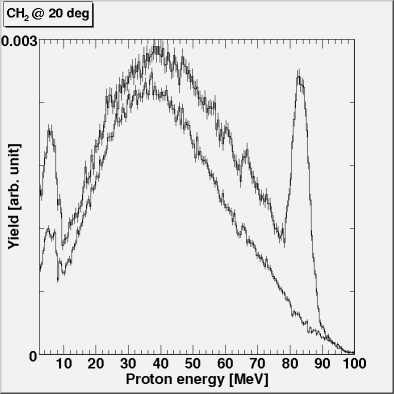}
  \end{minipage}\\
  \begin{minipage}{0.48\textwidth}
\includegraphics[width=\textwidth,bb=0 0 284 284]{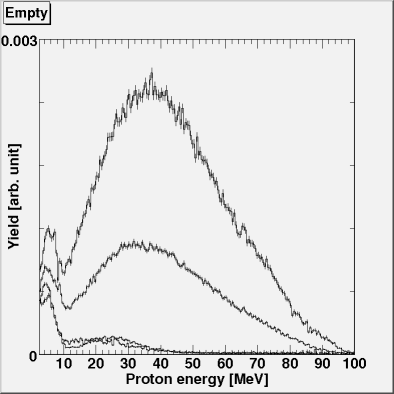}
  \end{minipage}
  \caption[Signal and background plots]{In the top panels, the signal and background for calcium, left, and polyethene, right, at an angle of 20 degrees. The bottom panel shows the background for the four angles analyzed: 140 degrees, 160 degrees, 40 degrees and 20 degrees in increasing order.}\label{fig:snratio}
\end{figure}

During the entire event-by-event analysis, the data sets from the
runs containing an empty frame were treated in parallel to the
calcium and the polyethene data. When switching to analysis of
histograms instead of events, the data from the background runs kan
be subtracted from the calcium and polyethene data. In that way one
can get pure signal histograms. A comparision between the calcium,
polyethene and background data can be found in figure
\ref{fig:snratio}. The background in MEDLEY has been analyzed
thoroughly by \citet{tippawan} for the old facility. Its main
components are particles produced by produced neutrons reacting in
the beam pipe and the reaction chamber, and the \ac{CsI(Tl)} and the
telescope housing acting as an active target for the beam halo.

\subsubsection{Target thickness}

\begin{figure}[!htb]
  \centering
  \begin{minipage}{0.48\textwidth}
\includegraphics[width=\textwidth,bb=0 0 567 567]{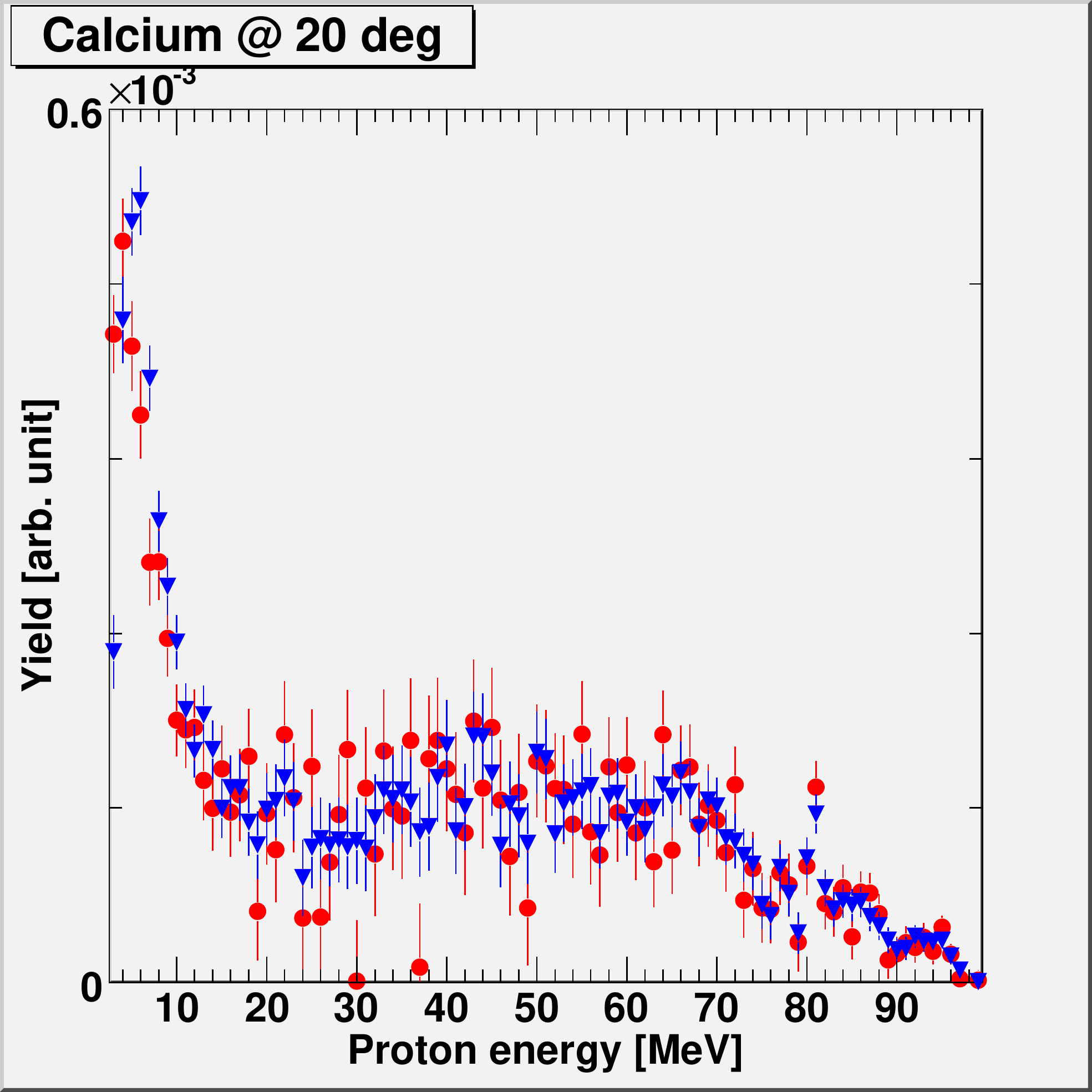}
  \end{minipage}
  \begin{minipage}{0.48\textwidth}
\includegraphics[width=\textwidth,bb=0 0 567 567]{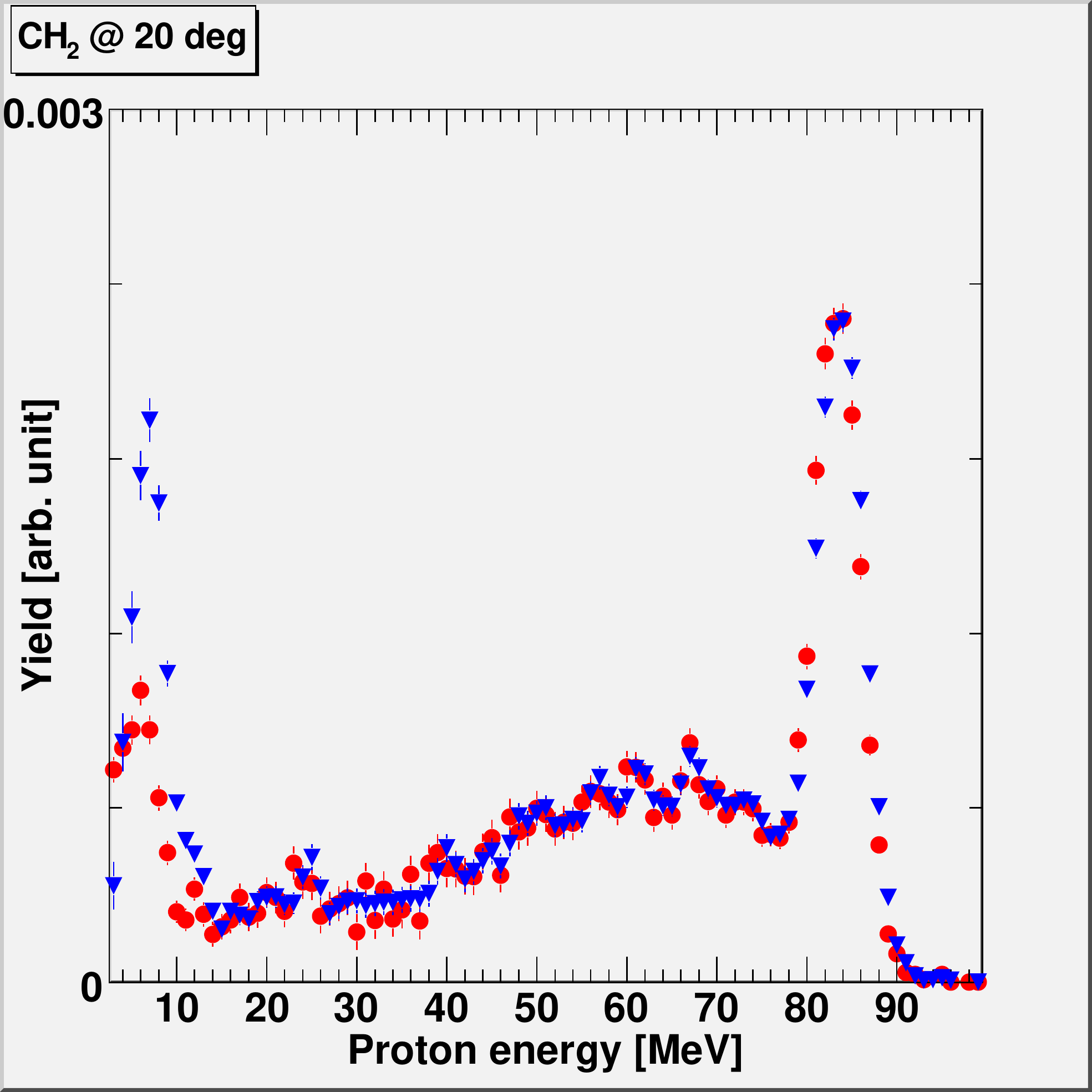}
  \end{minipage}
  \caption[Target correction results]{Target correction results for calcium in the left panel and polyethene in the right panel. Red circles represent the original data, while blue triangles show the corrected data, re-binned to 1 MeV for comparison. After three iterations the Kolmogorov test gave 100\% probability for the current iteration compared to the previous one.}\label{fig:corred}
\end{figure}

To achieve a count rate that is acceptable, some demands are put on
the target. Since the neutron beam is quite low in intensity and
only a small fraction of the neutrons interacts the target has to
have some finite thickness. For high-energy protons this is of
negligible importance. For low energy protons, on the other hand,
the energy loss can be quite significant. In table
\ref{tab:calstop}, \ac{SRIM} calculations for low energies are shown
to illustrate the problem. The target used in these measurements was
about 230 $\mu$m thick. These effects are compensated for through a
method implemented in a computer code, TCORR, by \citet{tcorr} and
is based on iterative calculations of response functions. The result
of the target correction procedure is found in figure
\ref{fig:corred}.

\begin{table}
  \centering
  \begin{tabular}{cccc}
    \hline
    Ion Energy [MeV] & $\textrm{d}E/\textrm{d}x$ [keV/$\mu$m] &  Range [$\mu$m] &
    Straggling [$\mu$m]
 \\
    \hline
    3.25  &  11.854 &  164.35  &    7.51   \\
    3.50  &  11.247 &  185.84  &    8.35   \\
    3.75  &  10.708 &  208.46  &    9.19   \\
    4.00  &  10.225 &  232.18  &   10.06   \\
    4.50  &  9.3934 &  282.85  &   12.85   \\
    5.00  &  8.7004 &  337.80  &   15.54   \\
    5.50  &  8.1137 &  396.92  &   18.20   \\
    6.00  &  7.6086 &  460.15  &   20.86   \\
    6.50  &  7.1698 &  527.41  &   23.55   \\
    \hline
  \end{tabular}
  \caption{Stopping of protons in calcium}\label{tab:calstop}
\end{table}

\subsection{Accepted neutron spectrum\label{sec:nspectrum}}

Even with the \ac{TOF} selection in section \ref{sec:tof}, not all
low energy neutrons were rejected. As can be seen in figure
\ref{fig:tofplots} the finite width of the \ac{TOF} cut results in
events induced by neutrons down to around 60~MeV to be accepted. In
the same figure one can also see the how the low energy p(n,p) tail
wraps around the time scale and allows events by about 10-14~MeV
neutrons inside the cut. To be able to subtract data originating
from these events the accepted neutron spectrum is deduced from the
CH$_2$ data. To get a clean neutron spectrum, the contribution from
carbon has to be subtracted from the CH$_2$ data. For 95 MeV
neutrons, the subtracted carbon cross section comes from existing
experimental data \citep{tippcomm}. This is not possible for lower
energy neutrons, so their contribution to the carbon spectrum is
taken from tabulated data in \citep{ICRU}.

To obtain a p(n,p) spectrum, the peak at 83 MeV is normalized to a
cross section obtained from the \ac{PWA} solution SP05 of
\citep{SAID} via a procedure described in appendix
\ref{sec:NN}\footnote{One should note that this is not the same
cross section as the one used for later normalization, but this
should pose no problem since this is only a relative measurement.}.
In this first normalization, the fact that also carbon contributes
slightly to the peak is ignored, but the peak is re-normalized to
the cross section from \citep{SAID} after the subtraction of
experimental carbon data.

\begin{figure}[!htb]
  \centering
  \begin{minipage}{0.48\textwidth}
\includegraphics[width=\textwidth,bb=0 0 567 546]{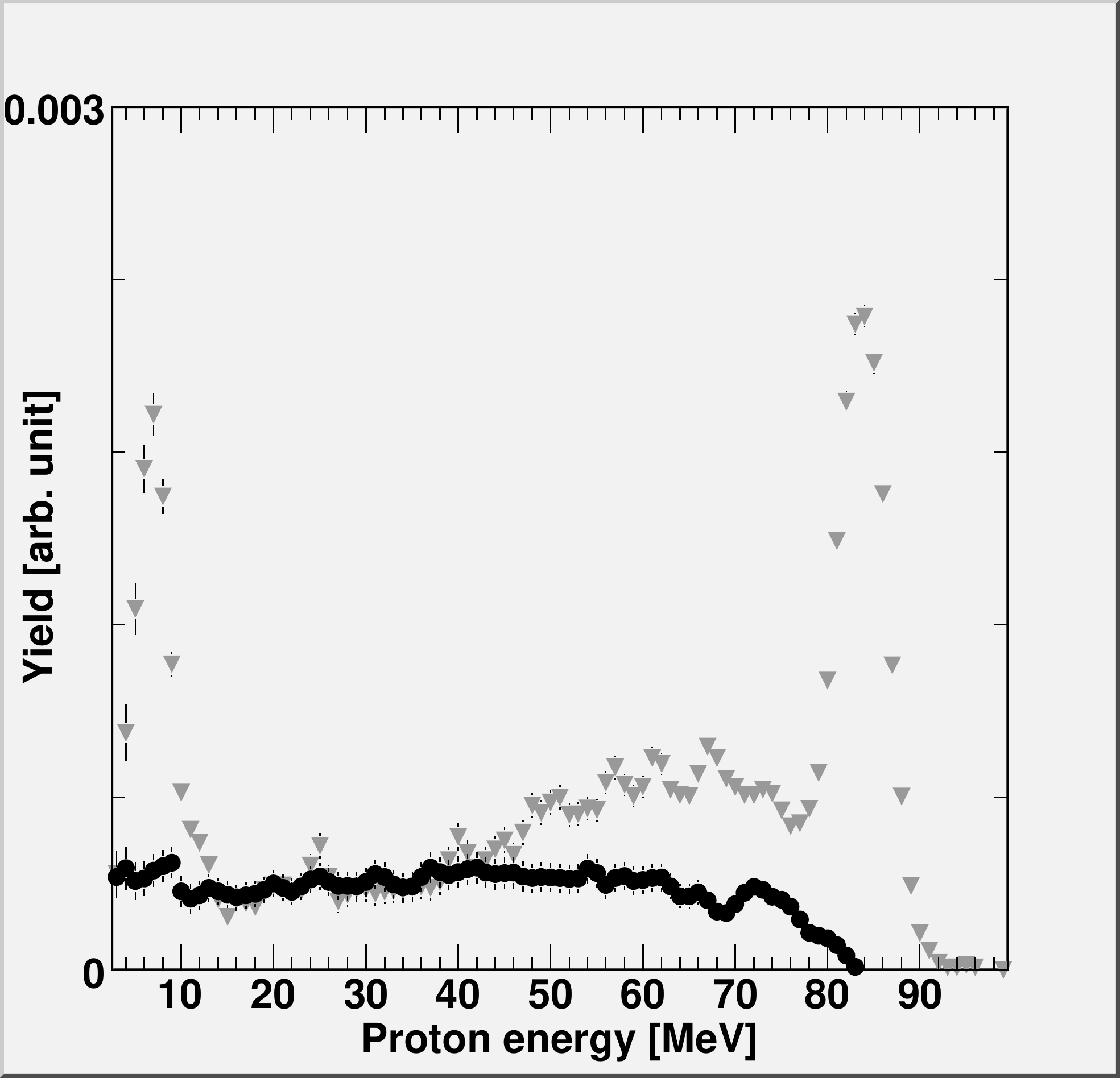}
  \end{minipage}
  \begin{minipage}{0.48\textwidth}
\includegraphics[width=\textwidth,bb=0 0 567 546]{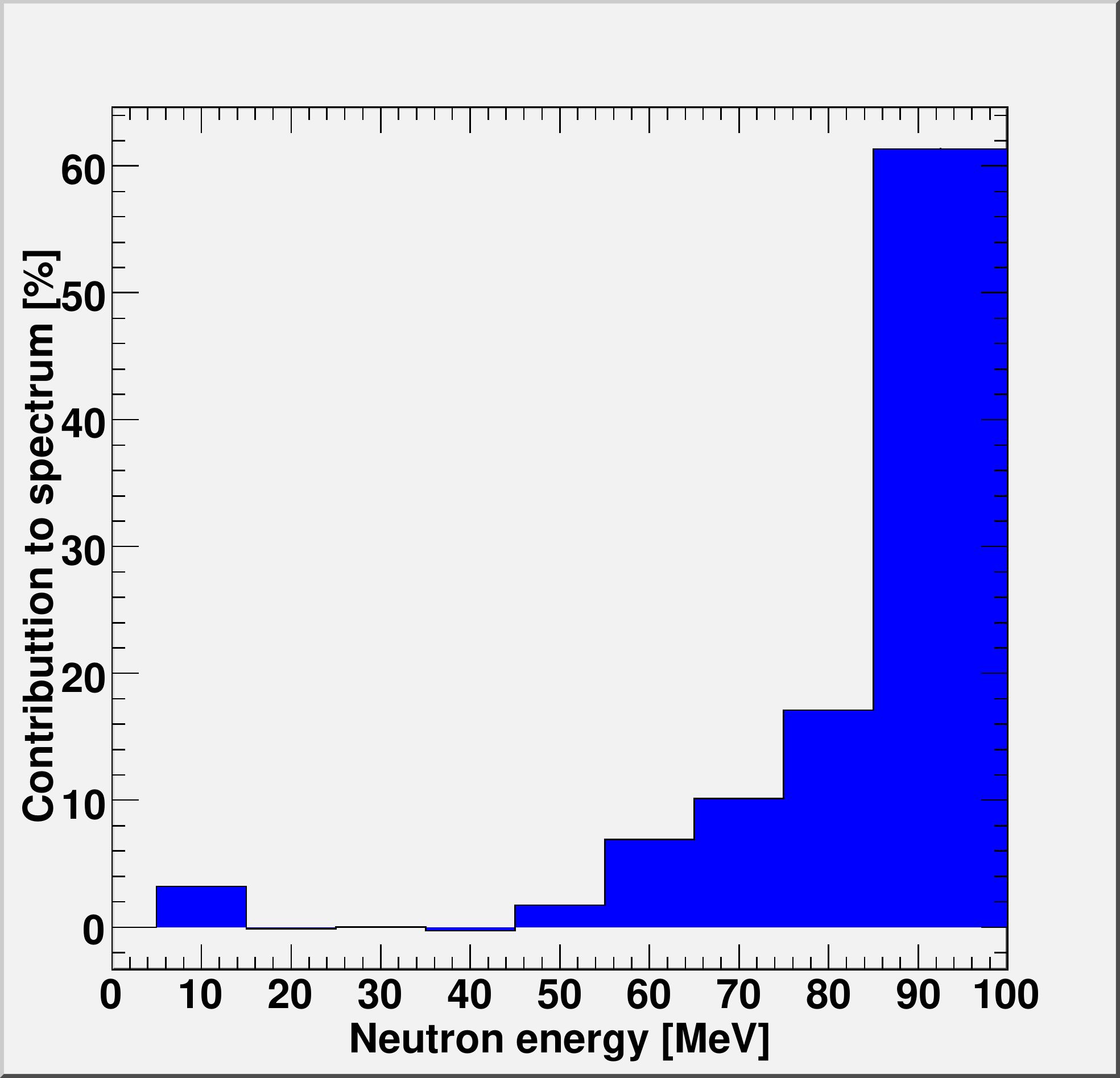}
  \end{minipage}
  \caption[Accepted neutron spectrum]{The deduced pure np spectrum in the left panel. Grey triangles are polyethene data and black circles are carbon data from the procedure in section \ref{sec:nspectrum}. In the right panel is, the fraction of accepted neutrons per bin is shown.}\label{fig:nspectra}
\end{figure}

Now the remaining CH$_2$ data are divided into bins corresponding to
10 MeV neutron energy each, translated to elastically scattered
proton energies as seen in table \ref{tab:neubins}. The bin right
below the peak is integrated and the number of events in the bin is
corrected for the variation in p(n,p) cross section, also listed in
table \ref{tab:neubins}. The cross section is assumed constant
within the bin and equal to the cross section in the center of the
bin. When doing this an assumption is made that the high Q value of
the C(n,p) reaction\footnote{The Q value for the C(n,p) reaction is
12.588 MeV \citep{physics}.} makes the contribution from the carbon
to the bin content negligible. Now, when one have the fraction of
accepted neutrons in that bin compared to the full energy peak, one
can subtract the corresponding carbon cross section obtained from
\citep{ICRU}. In parallel, the obtained bin content relative to the
peak is used to build up an accepted neutron spectrum. This process
is then iterated downwards in energy for each of the bins until a
pure p(n,p) spectrum and a corresponding neutron spectrum is
available, and the resulting spectrum is found in figure
\ref{fig:nspectra} and in table \ref{tab:nspectra}.

\begin{table}
  \centering
  \begin{tabular}{cc}
    \hline
    Bin [MeV] & Contribution [\%] \\
    \hline
    5-15 & 3.2 \\
    15-25 & 0 \\
    25-35 & 0 \\
    35-45 & 0 \\
    45-55 & 0 \\
    55-65 & 6.9 \\
    65-75 & 10 \\
    75-85 & 17 \\
    85-100 & 61 \\
    \hline
  \end{tabular}
  \caption[Accepted neutron spectrum]{Accepted neutron spectrum, as seen in figure \ref{fig:nspectra}.}\label{tab:nspectra}
\end{table}

\subsection{Normalization}

\begin{figure}[!htb]
  \centering
\includegraphics[width=0.48\textwidth,bb=0 0 567 546]{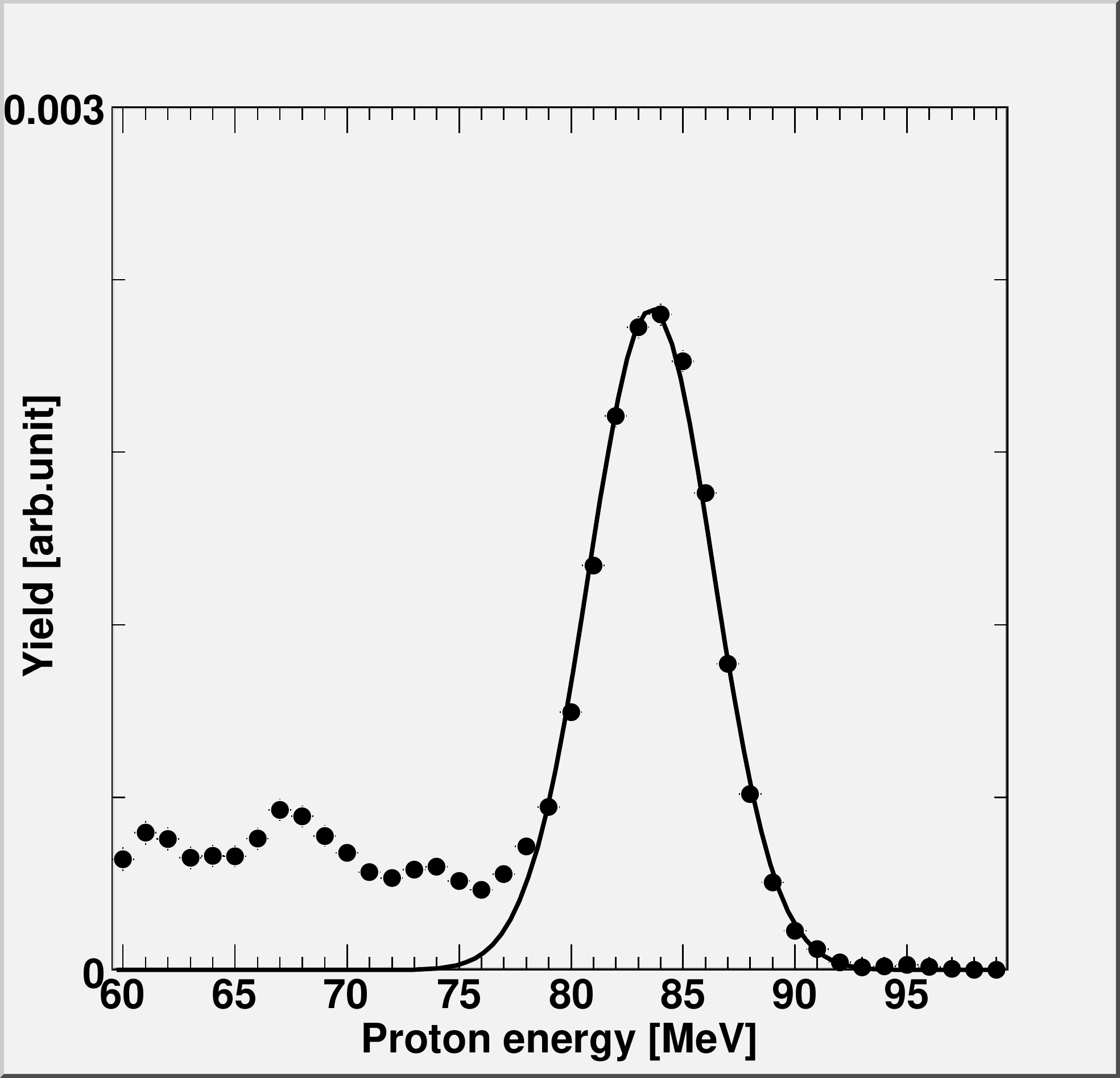}
  \caption[Elastic np peak]{The elastic np peak fitted with a gaussian. The data are CH$_2$ with carbon \protect \citep{tippcomm} subtracted.}\label{fig:np}
\end{figure}

To obtain a value for the cross section all data are normalized to
elastic p(n,p) scattering at 20 degrees. This since the p(n,p)
scattering cross section is well known and the peak is clear at this
angle. The peak together with a gaussian fit is seen in figure
\ref{fig:np}.

The \ac{FWHM} of the peak is found to be 6.4 MeV, in agreement with
the \ac{FWHM} in \citet{2000NIMPA.452..484D}, with the solid angle
of the telescope as the main contribution. The cross section,
$\sigma$, for each bin, $i$, in telescope $\textrm{T}X$ is now
calculated from
\begin{equation}\label{eq:xsecnorm}
    \frac{\sigma_{\textrm{Ca},i,\textrm{T}X}}{N_{\textrm{Ca},i,\textrm{T}X}} =
    \frac{\sigma_{\textrm{H}}}{N_{\textrm{H}}}
    \frac{2 m_{\textrm{CH}_2}}{M_{\textrm{CH}_2}}
    \frac{M_{\textrm{Ca}}}{m_{\textrm{Ca}}}
    \frac{\Omega_{\textrm{T}1,\textrm{T}8}}{\Omega_{\textrm{T}X}}
    \frac{\Phi_{\textrm{CH}_2}}{\Phi_{\textrm{Ca}}}
\end{equation}
where $N$ is the number of counts, $m$ is the target mass, $M$ is
the molecular mass, $\Phi$ is the relative neutron flux\footnote{The
normalization to neutron flux has actually already been done,
simultaneous to the life time correction, using the \ac{ICM}
monitor. But the factor should still be included in Eq.
(\ref{eq:xsecnorm}) for completeness.} and $\Omega$ is the solid
angle. Since the target distances for \ac{T1} and \ac{T8} are the
same, as seen in table \ref{tab:decset} the solid angles for these
telescopes are assumed to be equal. The other six telescopes, as
also seen in table \ref{tab:decset} however needs to be corrected
with a factor of 0.546, from a simple geometrical point of view. For
a more in-deep analysis of the solid angle in MEDLEY, see
\citep{flux}.

Since the polyethylene is the reference target that is regularly
used in MEDLEY experiments the weight, and number of protons, has
been measured with high precision to be $461.55 \pm 0.01$ mg and
$3.963 \cdot 10^{22}$ protons \citep{flux}. The weight of the
calcium target used was measured to be $235.9 \pm 0.1$ mg before the
runs and $237.3 \pm 0.1$ after the runs, which averages to 236.6 mg.

The cross section, $\sigma_{\textrm{H}}^\textrm{cm}$, is taken from
\citet{2001PhRvC..63d4001R} where the p(n,p) cross section at a
\ac{CM} angle of 139.0 degrees and at 96 MeV is measured to be 7.735
mb/sr. In order to extrapolate this to 94.5 MeV at a \ac{CM} angle
of 139.1 degrees, data from \ac{PWA} solution SP05 are used. From
\citep{SAID} the cross section for 94.5 MeV and 139.1 degrees is
8.041 mb/sr, while the cross section for 96 MeV and 139.0 degrees is
7.928 mb/sr. This gives an increase in cross section of about 1.43
\%. Via the procedure in section \ref{sec:NN} this is translated
into $\sigma_{\textrm{H}}^\textrm{lab} = 30.53$ mb/sr.

\clearpage
\setcounter{figure}{0} \setcounter{table}{0}
\setcounter{equation}{0}

\section{Results}

The results given here are those of the accepted neutron spectrum
obtained in section \ref{sec:nspectrum} and listed in table
\ref{tab:nspectra}, resulting in a mean value for the neutron energy
of 79.7 MeV. The total cross section for this spectrum is found in
table \ref{tab:restot} with a cutoff energy of 2.5 MeV.

\subsection{Double-differential cross sections}

\begin{figure}[!htb]
  \centering
  \begin{minipage}{0.48\textwidth}
\includegraphics[width=\textwidth,bb=0 0 567 567]{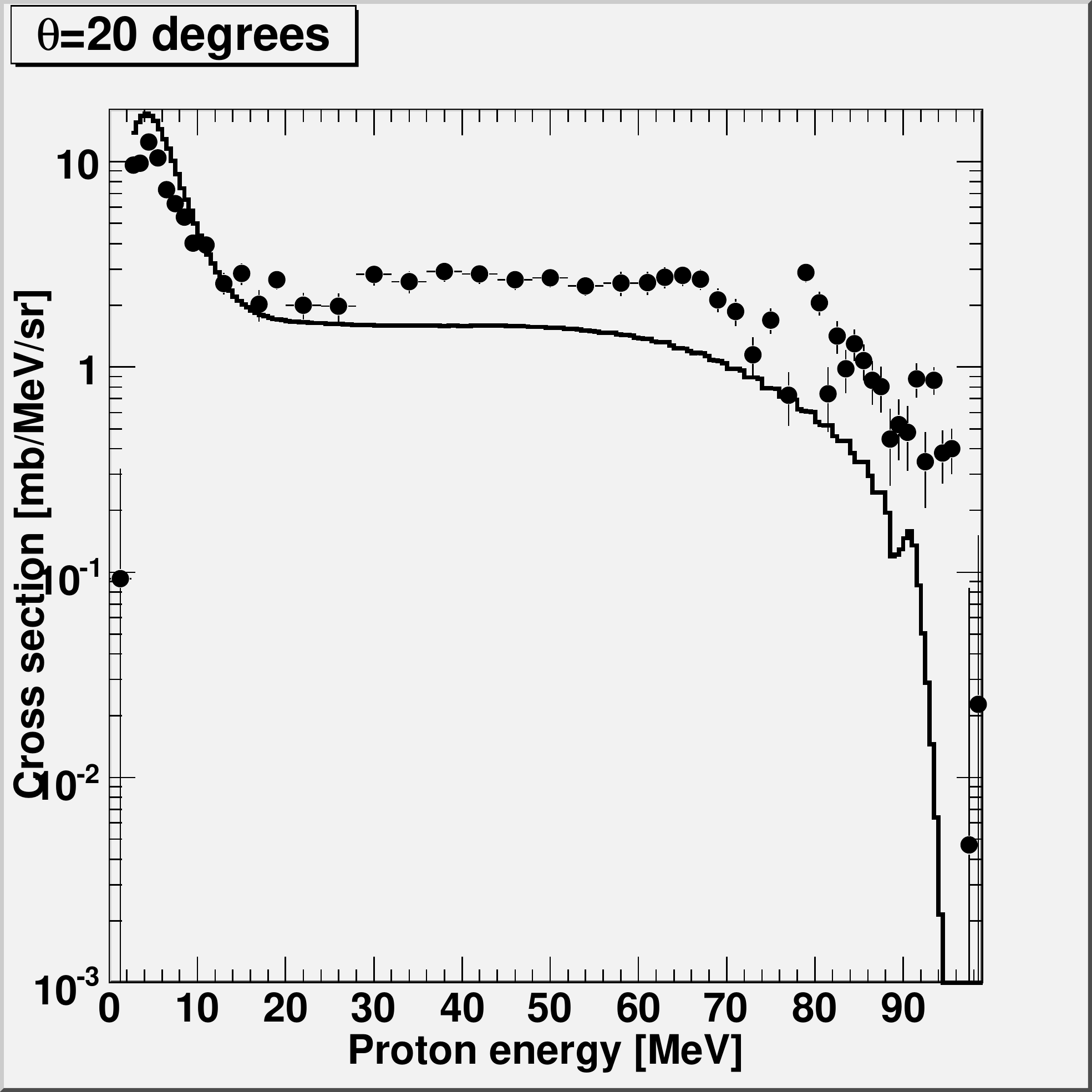}
  \end{minipage}
  \begin{minipage}{0.48\textwidth}
\includegraphics[width=\textwidth,bb=0 0 567 567]{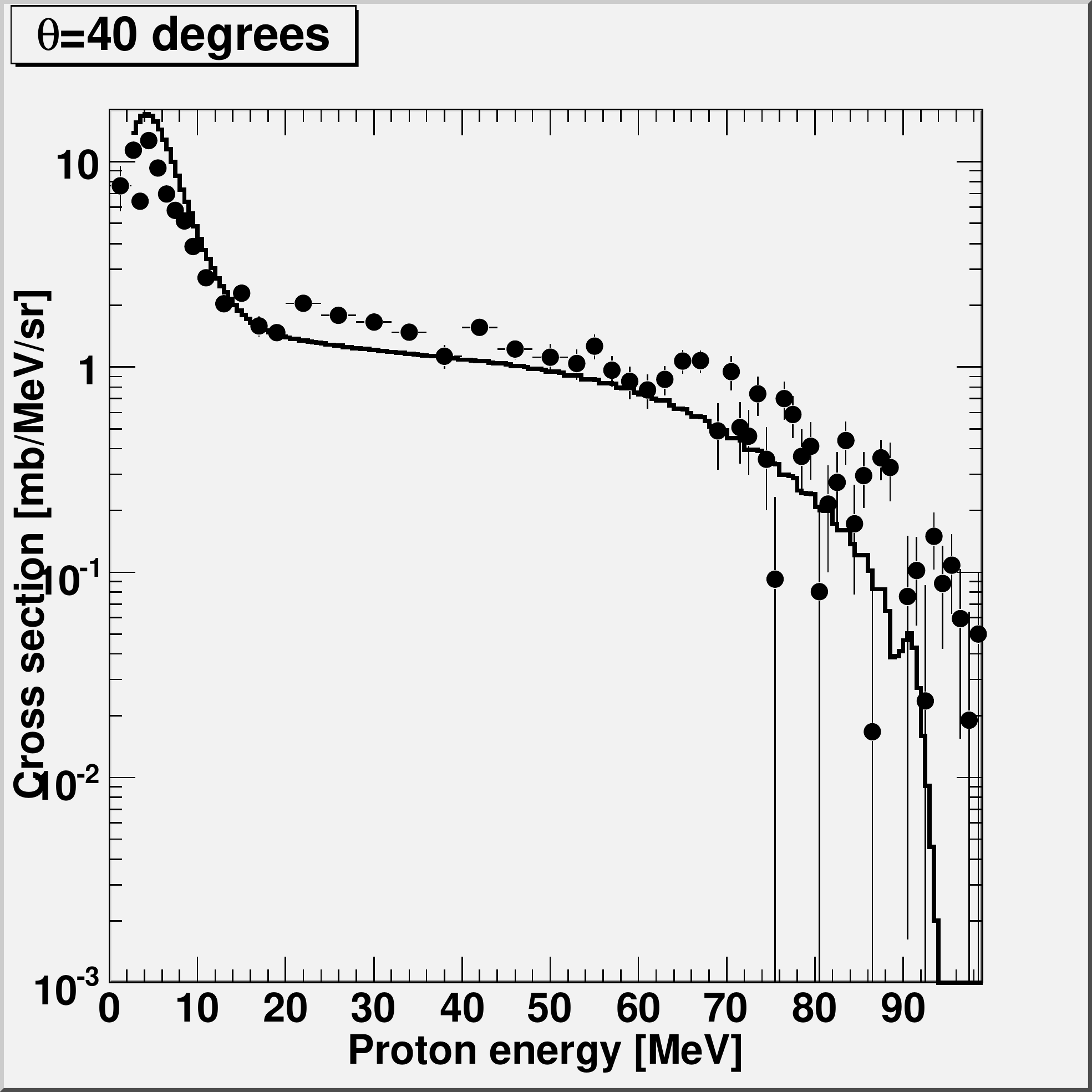}
  \end{minipage}\\
  \begin{minipage}{0.48\textwidth}
\includegraphics[width=\textwidth,bb=0 0 567 567]{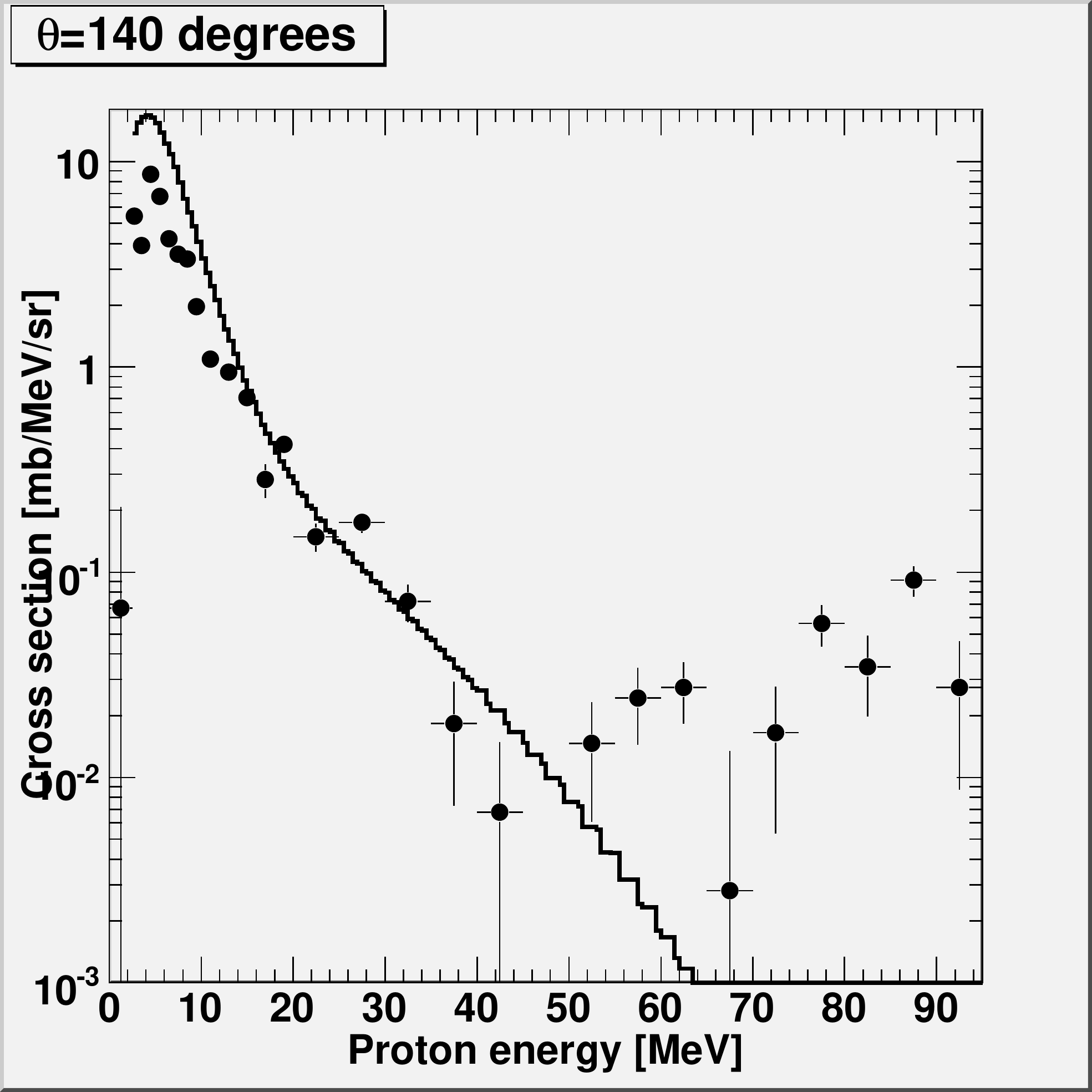}
  \end{minipage}
  \begin{minipage}{0.48\textwidth}
\includegraphics[width=\textwidth,bb=0 0 567 567]{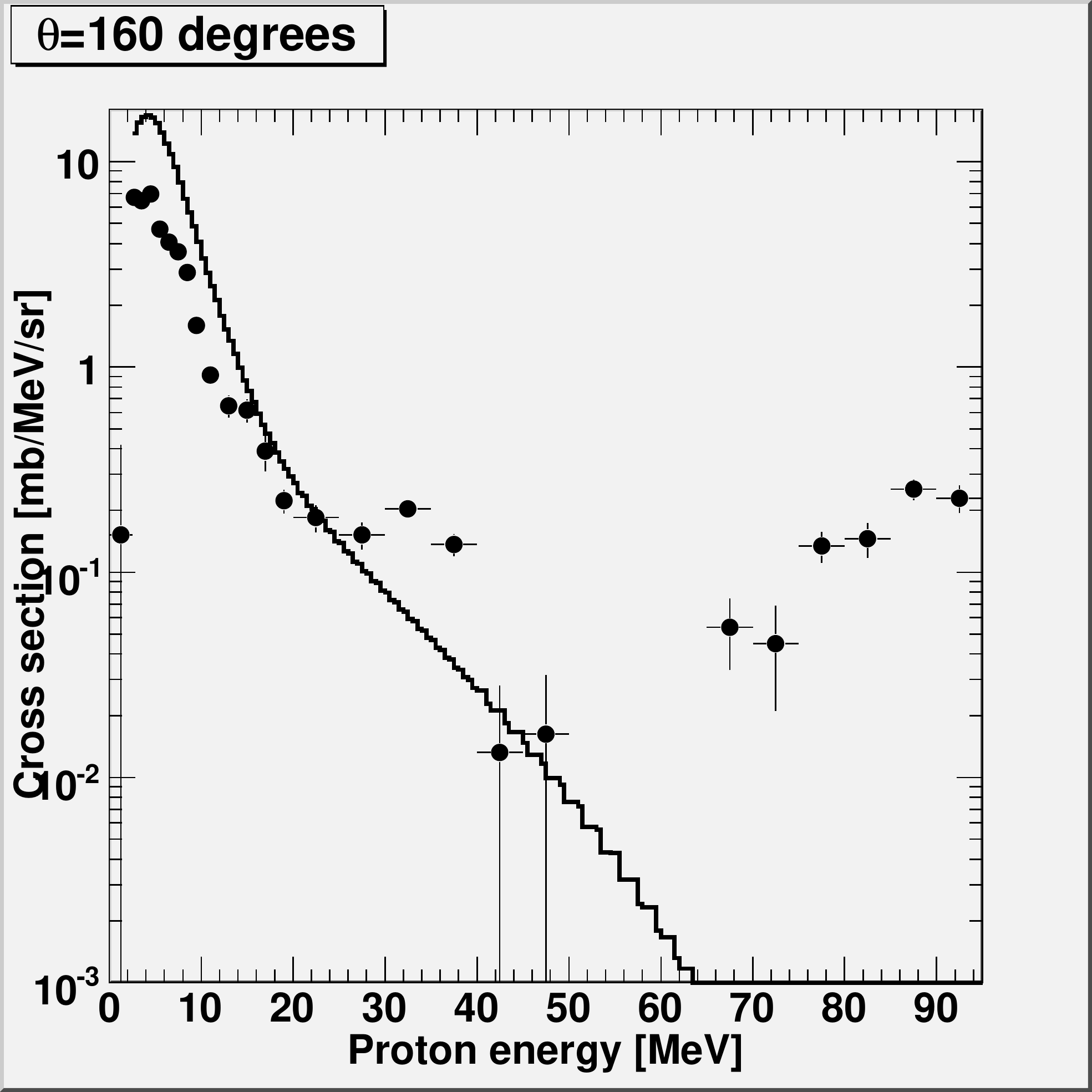}
  \end{minipage}
    \caption[Experimental double-differential cross sections]{Experimental double-differential cross sections, filled circles, for the neutron spectrum in table \ref{tab:nspectra} in the Ca(n,px) reaction. The solid histogram is TALYS calculations for this neutron spectrum. Cutoff energy is 2.5 MeV.\label{fig:dEdO_res}}
\end{figure}

Double-differential cross sections at four different angles have
been analyzed. The result is found in figure \ref{fig:dEdO_res} and
tables \ref{tab:res_deg20}-\ref{tab:res_deg160}, which also include
corrected data introduced in section \ref{sec:meth3}.

\subsection{Single-differential cross sections}

\subsubsection{Angular-differential cross section}

\begin{figure}[!htb]
  \centering
  \begin{minipage}{0.31\textwidth}
\includegraphics[width=\textwidth,bb=0 0 508 737]{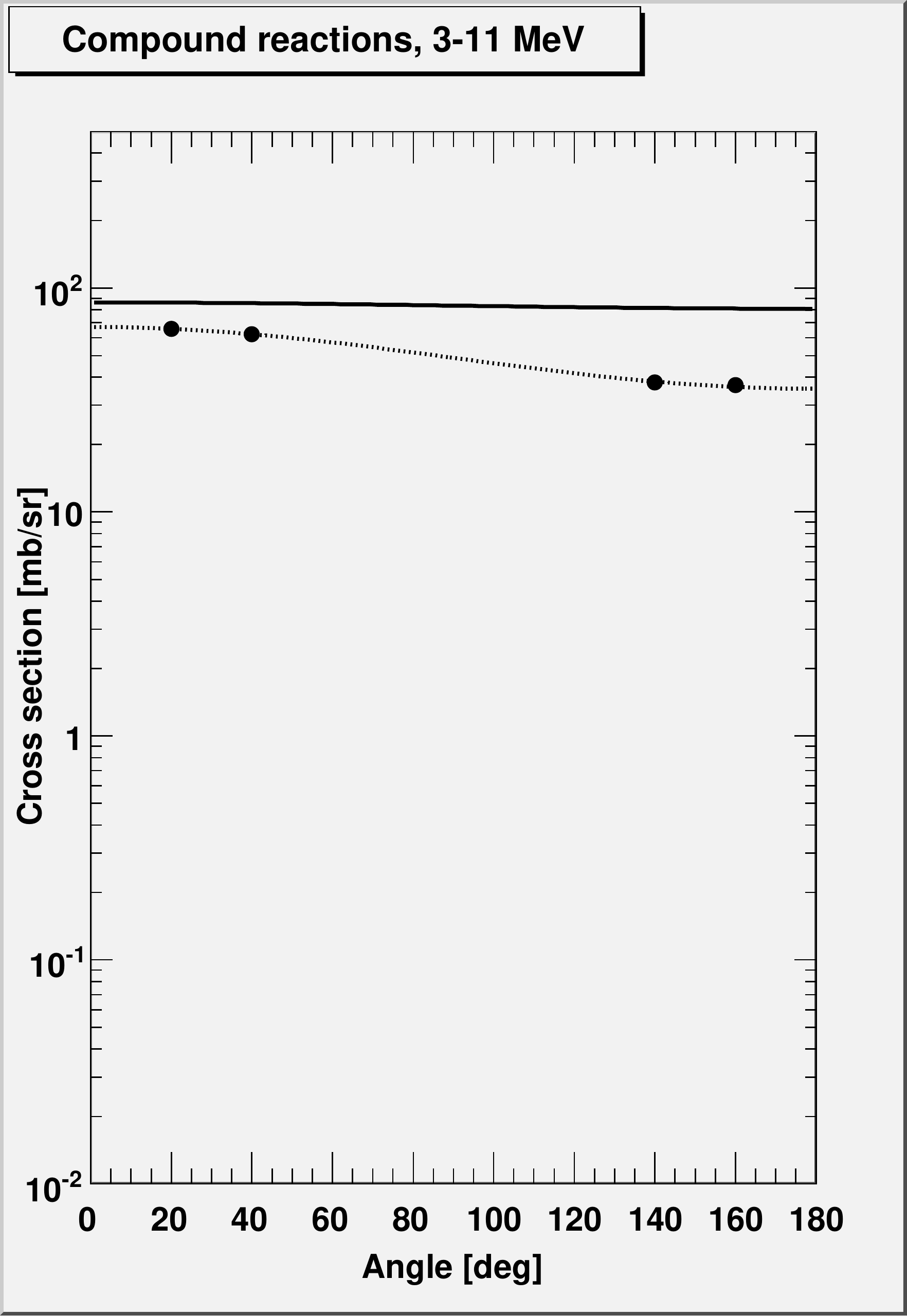}
  \end{minipage}
  \begin{minipage}{0.31\textwidth}
\includegraphics[width=\textwidth,bb=0 0 508 737]{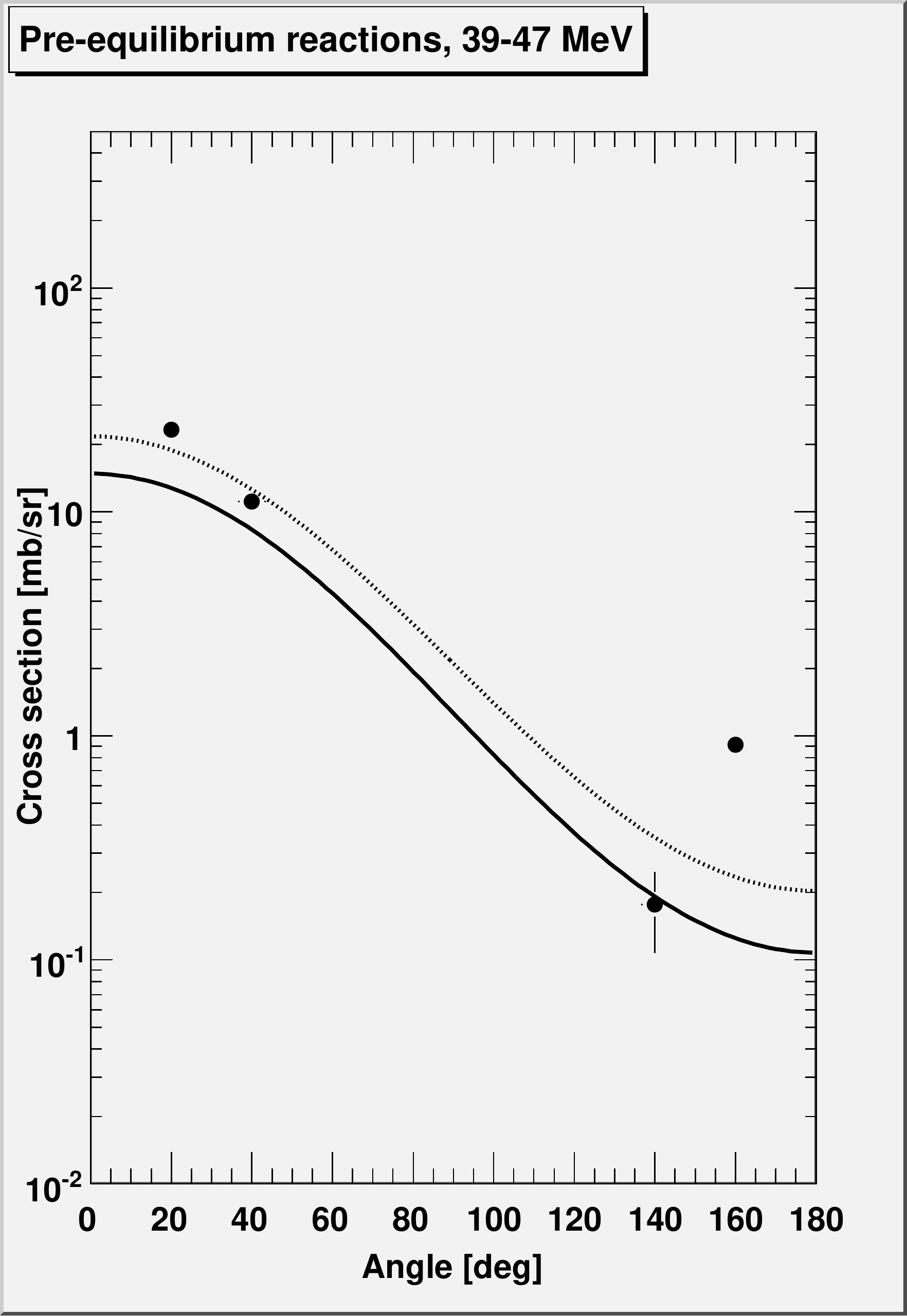}
  \end{minipage}
  \begin{minipage}{0.31\textwidth}
\includegraphics[width=\textwidth,bb=0 0 508 737]{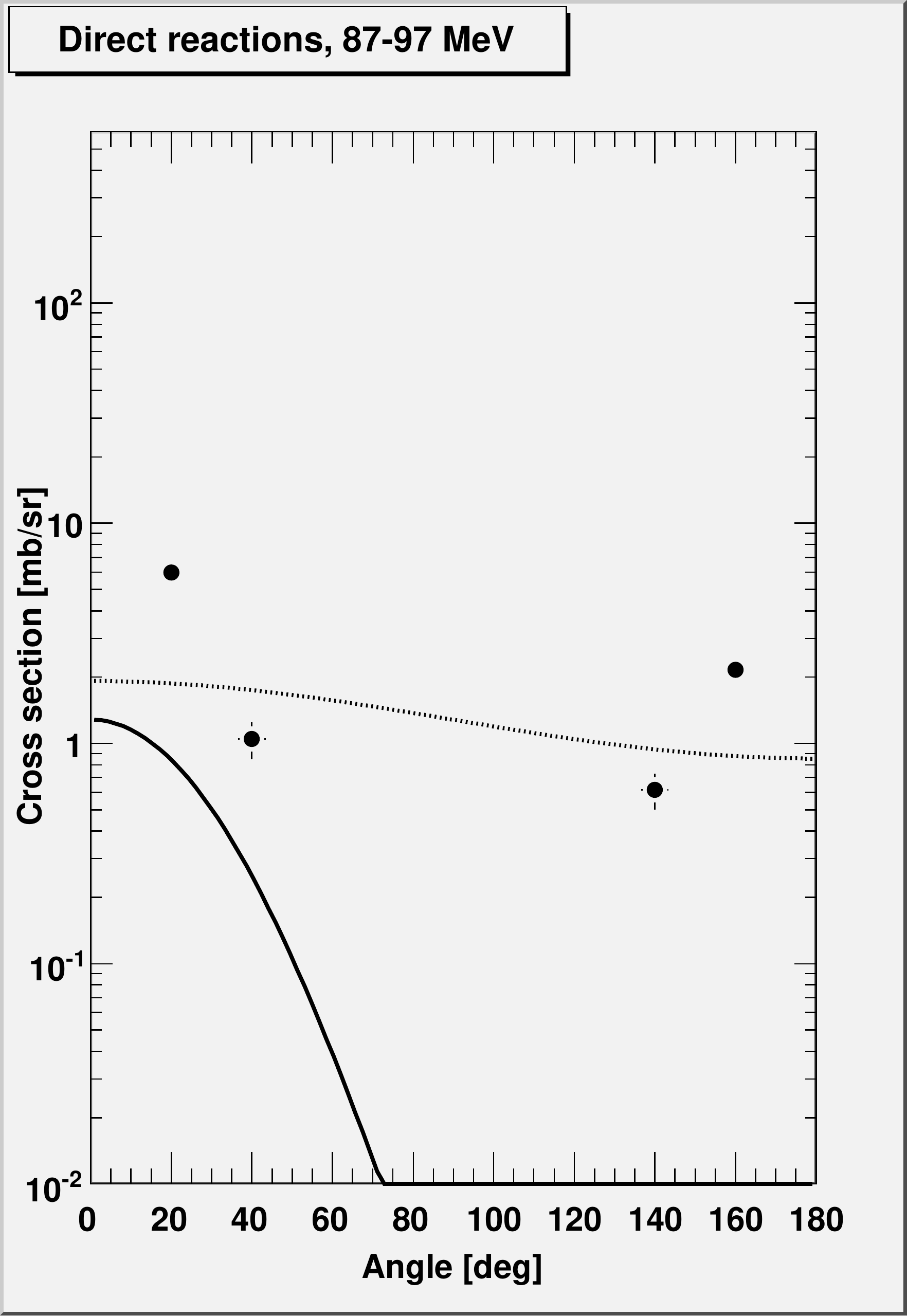}
  \end{minipage}
  \caption[Experimental angular-differential cross sections]{Experimental angular-differential cross sections for three different bins and the neutron spectra in table \ref{tab:nspectra} in the Ca(n,px) reaction. The chosen bins are assumed to represent the three specific reaction types. The solid line results from TALYS and the dotted line is a fit to the data according to Eq. (\ref{eq:gen_kalbach}). Cutoff energy is 2.5 MeV.}\label{fig:res_eint}
\end{figure}

To get angular-differential cross sections, the double-differential
spectra are summed over and weighted for the different bin widths.
The result is found in table \ref{tab:Eint_tot_all}. In figure
\ref{fig:res_eint} the angular-differential cross section for bins
corresponding to compound, pre-equilibrium and direct reactions is
found.

\subsubsection{Energy-differential cross section}

\begin{figure}[!htb]
  \centering
\includegraphics[width=\textwidth,bb=0 0 567 250]{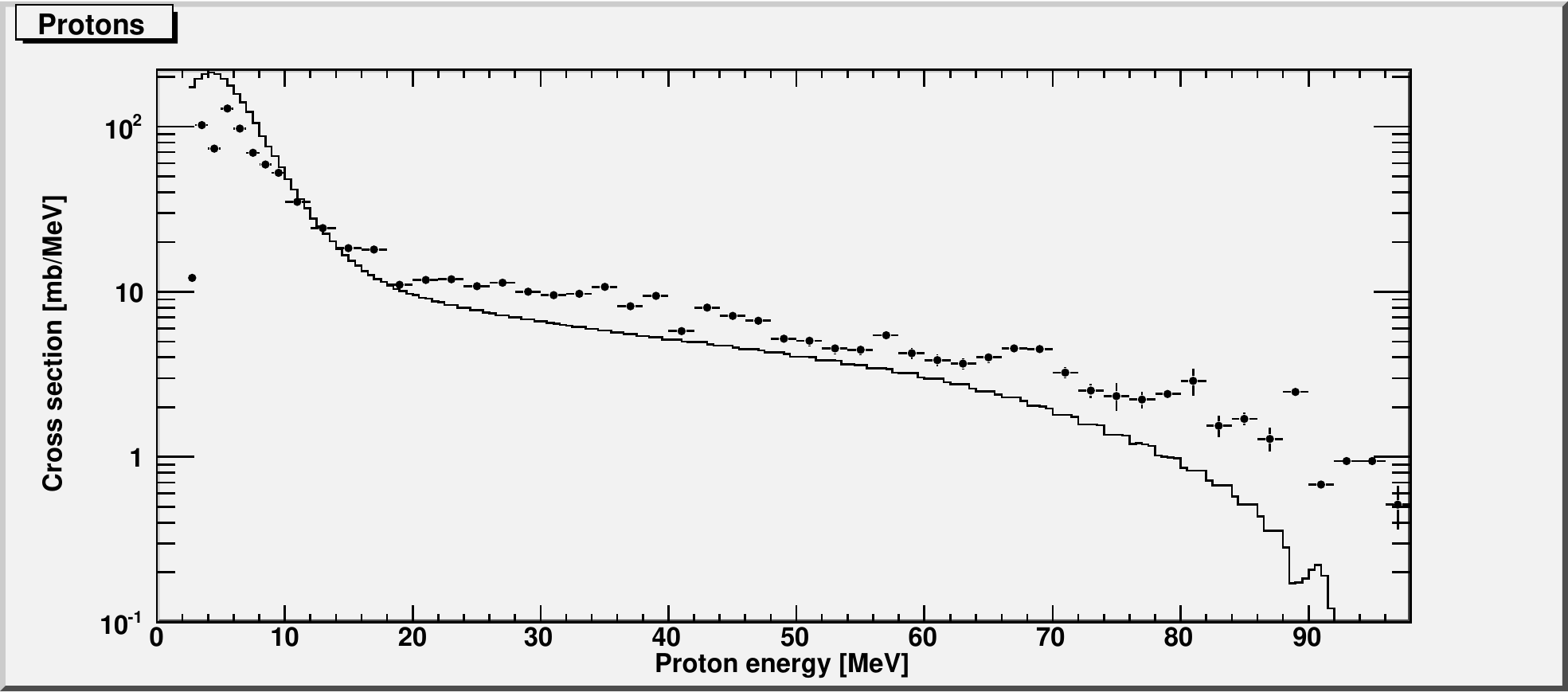}
  \caption[Experimental energy-differential cross sections]{Experimental energy-differential cross sections, filled circles, for the neutron spectrum in table \ref{tab:nspectra} in the Ca(n,px) reaction. The solid histogram is TALYS calculations for this neutron spectrum. Cutoff energy is 2.5 MeV.}\label{fig:res_angint}
\end{figure}

To get energy-differential cross sections, each of the energy bins
in the double-differential spectra are fitted with Eq.
(\ref{eq:gen_kalbach}). Using the results from this fit, the data
are interpolated to the angles 60, 80, 100 and 120 degrees, and also
extrapolated to 5 and 175 degrees. These data points are summed over
and weighted for the different bin as well as their angular
contribution. The result is found in figure \ref{fig:res_angint} and
in table \ref{tab:res_Aint}.

\clearpage
\setcounter{figure}{0} \setcounter{table}{0}
\setcounter{equation}{0}

\section{Discussion}

\subsection{Comparison with theory}

To be able to compare the result with theoretical calculations the
contribution from the low-energy neutrons need to be subtracted. To
do this three different methods, listed in sections
\ref{sec:meth1}-\ref{sec:meth3}, are presented. The difference in
total cross section between these methods is shown in table
\ref{tab:en_tot}.

\subsubsection{Method 1: Rescaling of experimental data\label{sec:meth1}}

In the first method for subtraction of low-energy neutron data, the
energy spectrum for proton emission is assumed to be independent of
the incoming neutron energy. The experimental cross section is
weighted with the amount of low energy neutrons in the accepted
spectrum from table \ref{tab:nspectra}, and rescaled to the peak in
the neutron spectrum. The result is found in figure
\ref{fig:dEdO_res_frac}.

The main advantage of this method is that one gets rid of the model
dependencies in the TALYS calculations, and the data used are pure
experimental. This means that the result at low energies, up to
about 40~MeV, should be quite accurate. The main disadvantage on the
other hand is the obvious fact that the cross section spectrum for
94~MeV neutrons and lower energy neutrons differ a lot at high
energies. This is especially relevant at energies above the low
energy bin under consideration.

\subsubsection{Method 2: Subtraction of TALYS data\label{sec:meth2}}

\begin{figure}[!htb]
  \centering
  \begin{minipage}{\textwidth}
    \begin{center}
        {\LARGE \sc Method 2}
    \end{center}
    \vspace{8pt}
  \end{minipage}\\
  \begin{minipage}{0.45\textwidth}
\includegraphics[width=\textwidth,bb=0 0 567 567]{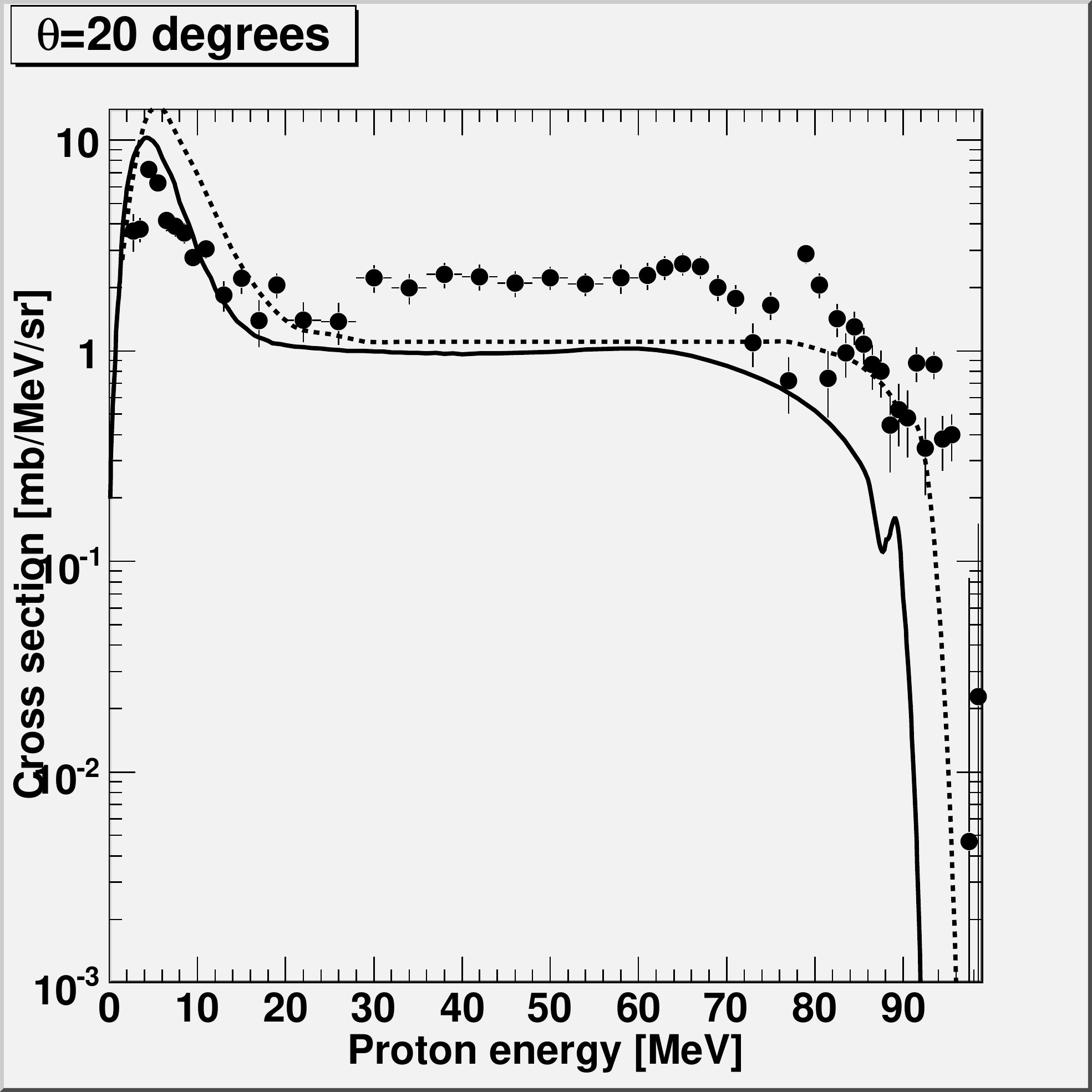}
  \end{minipage}
  \begin{minipage}{0.45\textwidth}
\includegraphics[width=\textwidth,bb=0 0 567 567]{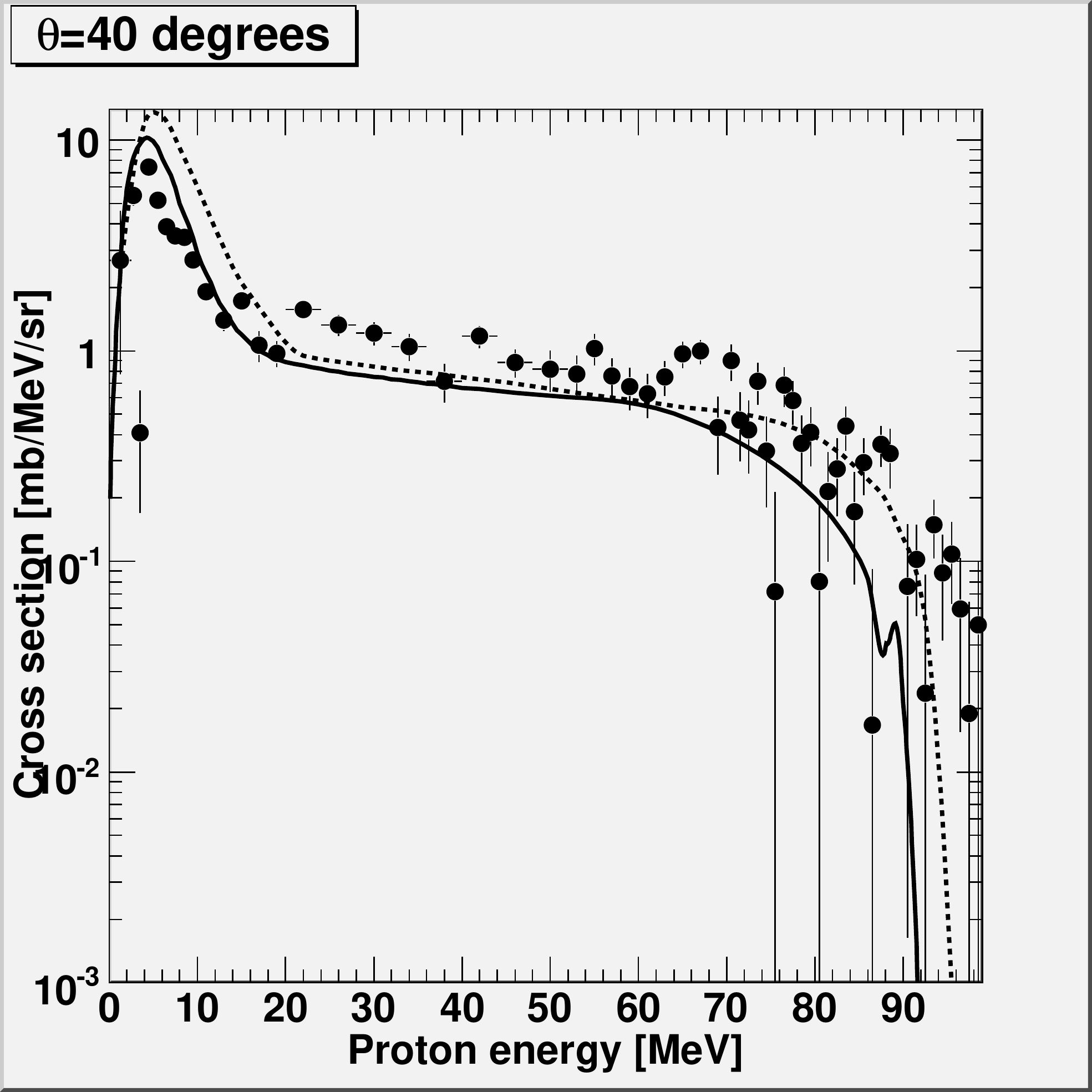}
  \end{minipage}\\
  \begin{minipage}{0.45\textwidth}
\includegraphics[width=\textwidth,bb=0 0 567 567]{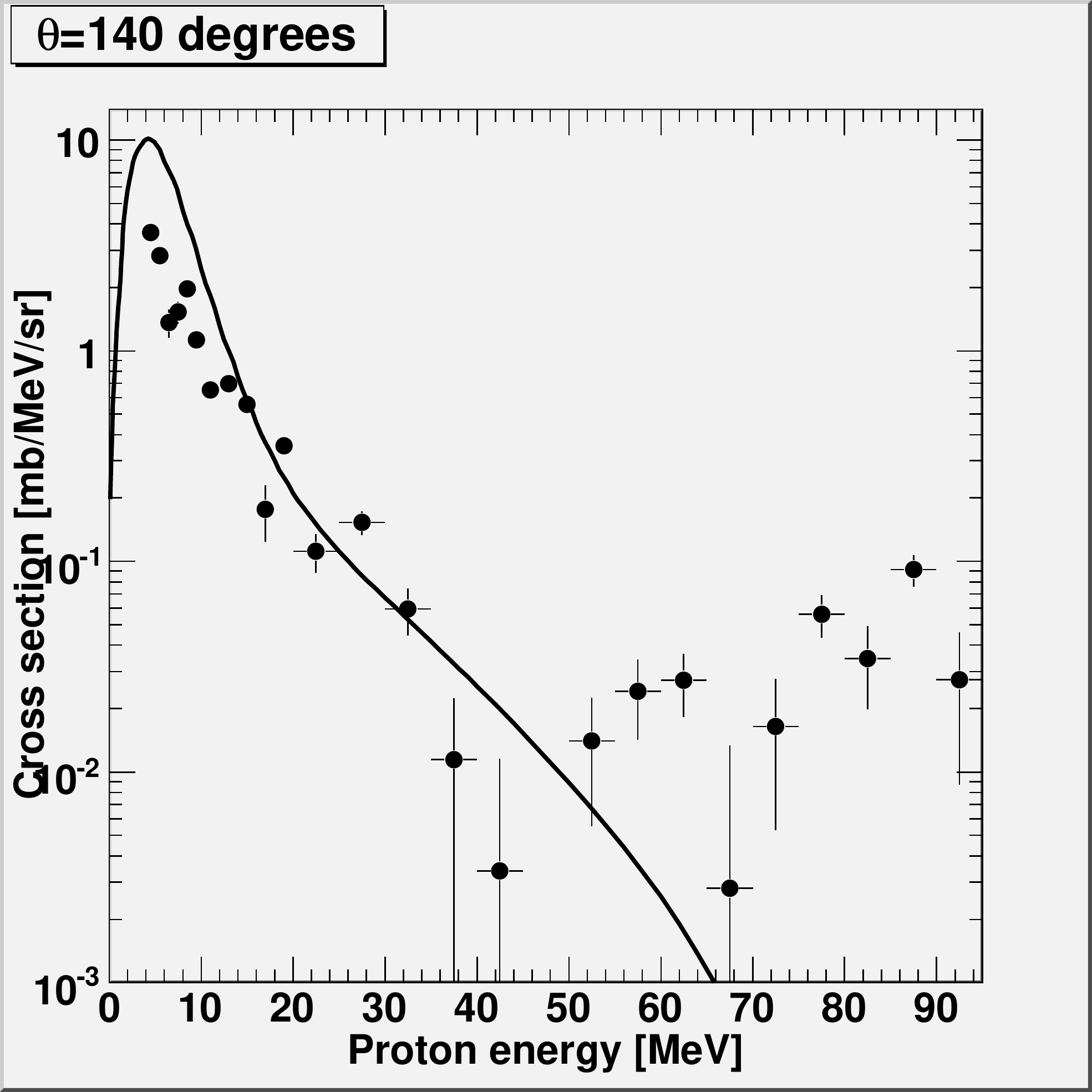}
  \end{minipage}
  \begin{minipage}{0.45\textwidth}
\includegraphics[width=\textwidth,bb=0 0 567 567]{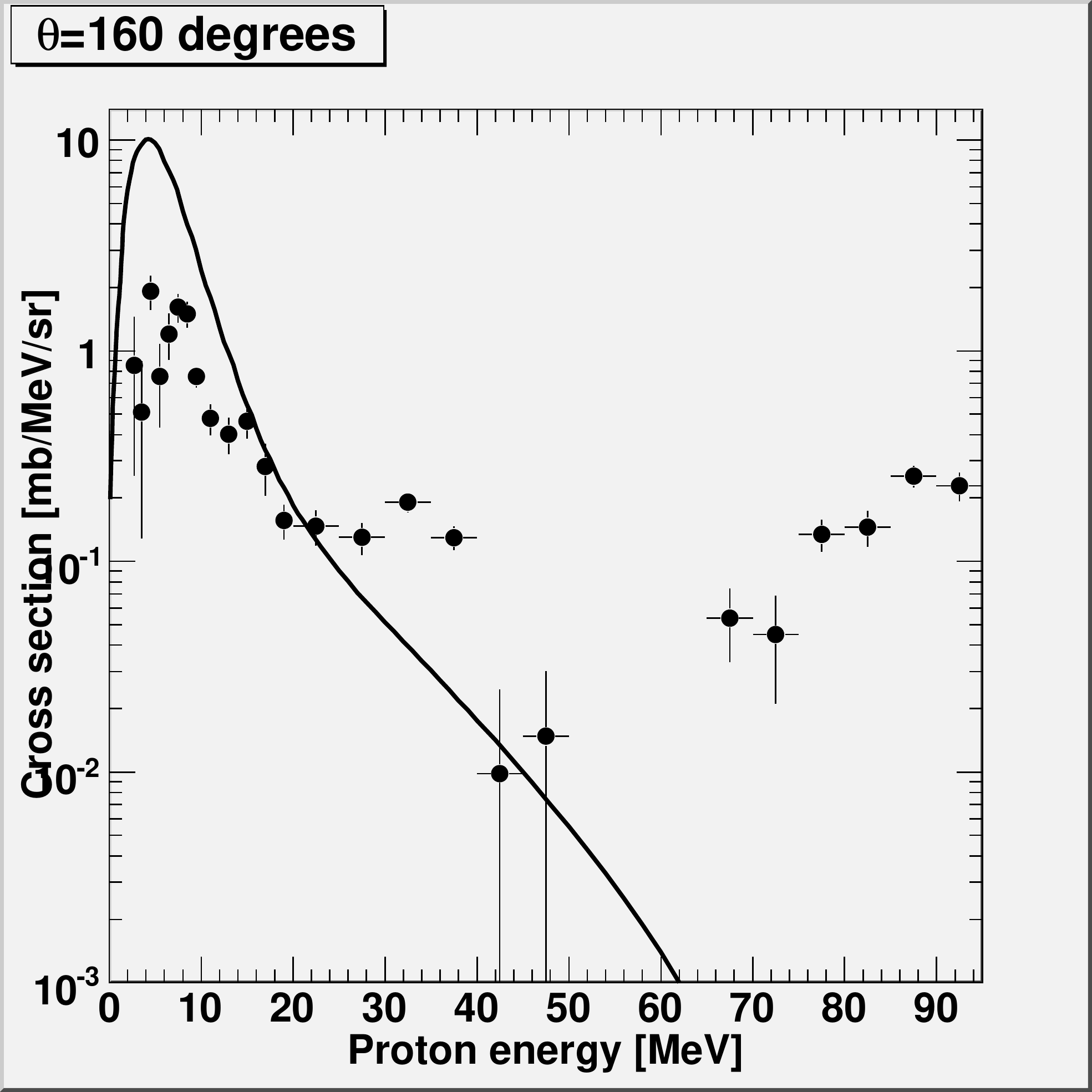}
  \end{minipage}\\
  \begin{minipage}{0.90\textwidth}
\includegraphics[width=\textwidth,bb=0 0 567 250]{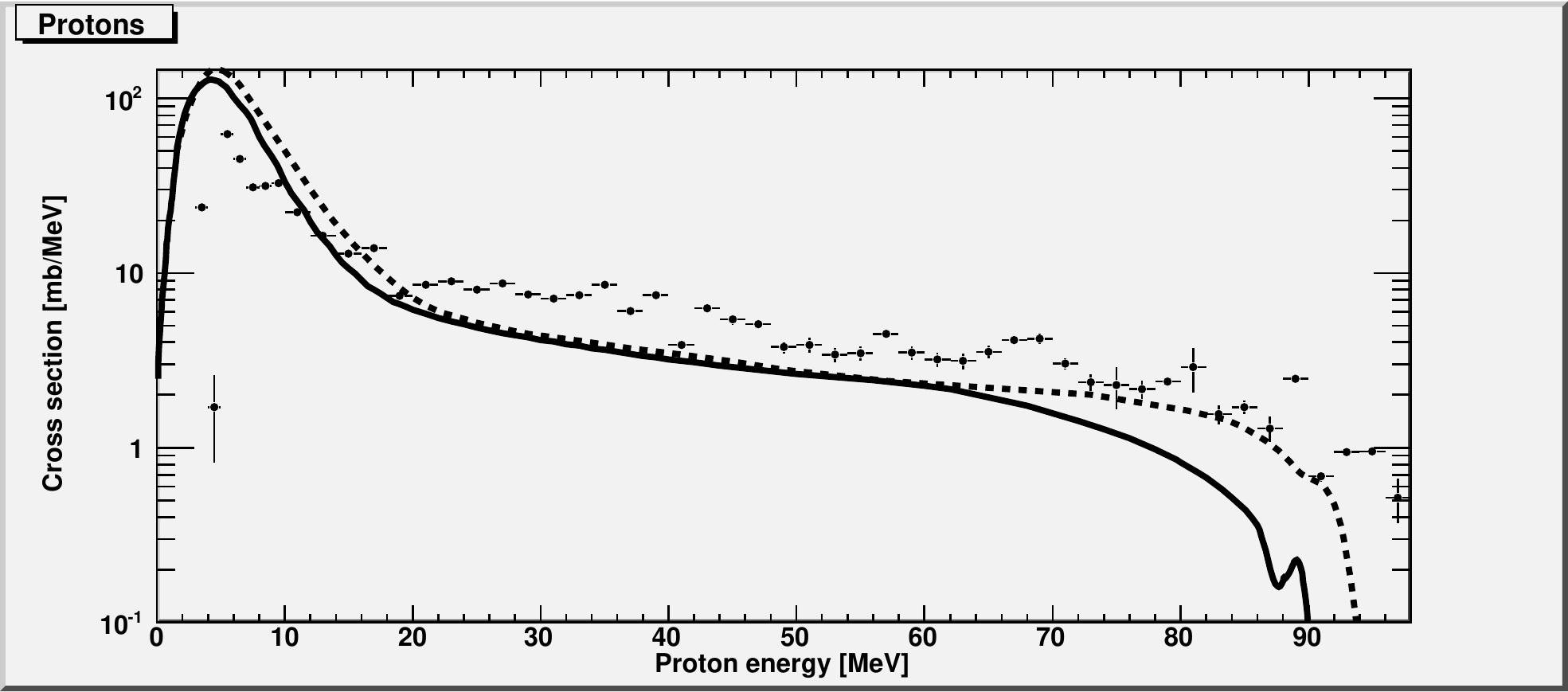}
  \end{minipage}
  \caption[Double- and energy-differential cross sections with TALYS data subtracted]{Double-differential cross sections in the four top panels, and angle integrated single-differential cross section in the bottom panel, with TALYS data (method 2) subtracted for 94 MeV neutrons in the Ca(n,px) reaction. Solid lines are predictions by TALYS, and dashed lines are predictions by GNASH \protect \citep{ICRU}. Cutoff energy is 2.5 MeV.\label{fig:dEdO_res_tal}}
\end{figure}

Another way to correct for the low-energy neutron contribution is by
using data from TALYS. For the accepted neutron spectrum in table
\ref{tab:nspectra}, TALYS is used to calculate cross sections for
each bin. The bins between 15~MeV and 55~MeV are left out, since the
contribution in these bins probably derive from statistics. The
result is found in figure \ref{fig:dEdO_res_tal}.

The advantage of using this method is that it makes the comparison
between experiment and theory straightforward. The difference one
can see in figure \ref{fig:dEdO_res_tal} between experiment and
theory is directly related to the TALYS calculations. So for code
verification purposes, this method works fine. But for obtaining
experimental data, this method is not so good, since the resulting
experimental data points will have a heavy model dependency on
calculations that are known to be incorrect.

\subsubsection{Method 3: Subtraction of modified TALYS data\label{sec:meth3}}

A third possible method is to try and modify the TALYS data to get
it more correct. From previous experiment, TALYS is known to heavily
overestimate the evaporation cross section in backward angles, and
slightly overestimate it at forward angles. It also appears to
underestimate pre-equilibrium cross sections in forward angles
\citep{2004PhRvC..69f4609T, 2006PhRvC..73c4611T}. Here, an
assumption is made that the over/under-estimations in the TALYS
calculations are truly systematic and do not vary with projectile
energy. From this assumption one can divide the experimental data
with the TALYS data in figure \ref{fig:dEdO_res} and get a
correction histogram. This correction histogram is assumed to
contain the true over/under-estimations in TALYS, and by multiplying
the low-energy TALYS calculations with this before subtracting them,
the true\footnote{Well, true under the condition that the assumption
is correct.} cross-section spectra are obtained. The result is found
in figure \ref{fig:dEdO_res_wtal} and, together with uncorrected
data, in tables \ref{tab:res_deg20}-\ref{tab:res_deg160}.

By using this method one can reduce the model dependencies compared
to unmodified TALYS data, by adjusting the theory to the experiment.
This method is the one that probably will give the result with
smallest systematic errors for the low-energy neutron correction.
However, it is first when one sees the results from similar
experiments at different energies that one can conclude if the
assumption made is too bold or not.

\begin{table}[!ht]
  \centering
  \begin{tabular}{lccccc}
    \hline
    Method      & $\frac{\textrm{d}\sigma}{\textrm{d}\Omega}(\theta=20)$ & $\frac{\textrm{d}\sigma}{\textrm{d}\Omega}(\theta=40)$ & $\frac{\textrm{d}\sigma}{\textrm{d}\Omega}(\theta=140)$ & $\frac{\textrm{d}\sigma}{\textrm{d}\Omega}(\theta=160)$ & $\sigma_{\textrm{prod}}(\textrm{n},\textrm{p}x)$\\
    \hline
    \hline
    1. Exp. Data   & $155 \pm 2.92$ & $99.4 \pm 1.54$ & $31.6 \pm 0.486$ & $32.1 \pm 0.700$ & $746 \pm 1.97$ \\
    2. TALYS       & $188 \pm 4.76$ & $107 \pm 2.51 $ & $17.6 \pm 0.792$ & $18.6 \pm 1.14$ & $679 \pm 3.74$ \\
    3. Mod. TALYS  & $175 \pm 6.69$ & $107 \pm 4.22$  & $33.7 \pm 1.73$ & $35.1 \pm 2.04$ & $788 \pm 5.83$ \\
    \hline
    TALYS & 131 & 106 & 62.4 & 62.4 & 1070 \\
    GNASH \protect \citep{ICRU} & 186 & 145 & - & - & 1300 \\
    \hline
  \end{tabular}
  \caption[Angular-differential and total cross sections with subtracted data]{Comparison of angular-differential and total cross sections with subtracted data. Differential cross sections are given in mb/sr and total cross sections in mb. The cutoff energy is 2.5 MeV, except for the GNASH calculations of $\sigma_{\textrm{prod}}(\textrm{n},\textrm{p}x)$, where no cut-off is was applied.}\label{tab:en_tot}
\end{table}

\subsection{Background and shielding}

As can be seen in figure \ref{fig:snratio} the background problem is
quite heavy. In the silicon data runs from the old facility, the
signal-to-background ratio was about 8 \citep{tippawan}. By moving
the spectrometer closer to the lithium target, and at the same time
reducing the shielding, the background increased dramatically. In
later data sets with energies of 180~MeV, the background problem
were so large that it was virtually impossible to get acceptable
results from the data \citep{terucomm}. To come to terms with the
background problems recent GEANT simulations at St. Petersburg
University show that the background for a 180 MeV neutron beam can
be reduced by a factor 20-50\footnote{Even up to a factor 100.}, by
exchanging the concrete wall to an iron wall. The background would
probably be reduced even more for a 94 MeV beam. Available iron
blocks from the old CELSIUS ring have actually been used when this
reconstruction was undertaken in January 2007. The first data with
the new shielding will probably be taken during February 2007.

\subsection{Special TOF cut\label{sec:tofdiss}}

\begin{figure}[!htb]
  \centering
  \begin{minipage}{0.48\textwidth}
\includegraphics[width=\textwidth,bb=0 0 350 350]{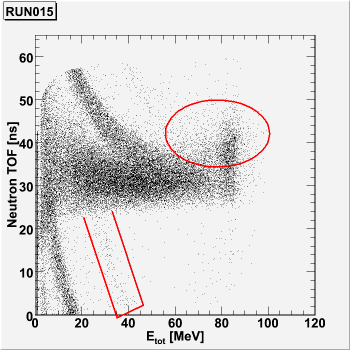}
  \end{minipage}
  \begin{minipage}{0.48\textwidth}
\includegraphics[width=\textwidth,bb=0 0 350 350]{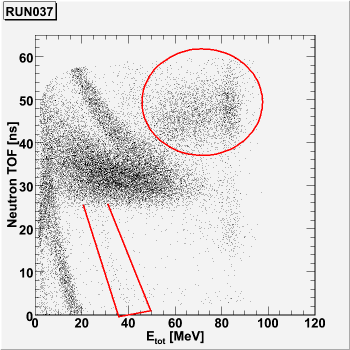}
  \end{minipage}
  \caption[TOF shift in T1 and run 37 in T8]{TOF shift in \ac{T1} in the left panel and run 37 in \ac{T8} in the right panel.}\label{fig:tofshift}
\end{figure}

Run 37 was itself unusable for the analysis, but still it may
contain some interesting information on the phenomena. As can be
seen in the right panel of figure~\ref{fig:tofshift} there exists a
really strong shift in the \ac{TOF} at high energies that does not
seem to be constant, but disappears around 60 MeV. This effect is
also visible in \ac{T1} as can be seen in the left panel of
figure~\ref{fig:tofshift}. This clear energy dependence makes a
correction for this effect quite difficult. Therefore, the \ac{TOF}
cut instead has been expanded to include these effects. This had to
be done for both \ac{T1} and \ac{T8} since the effect still was
present, but weak, after the threshold adjustment in \ac{T8}. The
reasons for this effect is unfortunately still unknown, but since it
was weakened by threshold adjustment\footnote{Before run 39 the
$\Delta E_2$ threshold in T8 was changed from 450 mV to 315 mV, and
the $\Delta E_2$ threshold in T7 was changed from 400 mV to 325 mV.
Before run 40 the $\Delta E_2$ threshold in T8 was further changed
to 295 mV, and the $\Delta E_2$ threshold in T3 was changed from 320
mV to 350 mV.} it is suspected to be correlated to the pulse height
in the $\Delta E_2$ detector. This suspicion is further strengthened
by the fact that an analysis of emitted deutrons shows no effect
like this.

One can also note the energy shift in the lower part of the plot,
that does not seem to be dependent on the threshold adjustment. If
the cause of these problems is because of instabilities in the
\ac{CFD} a possibility to get rid of this problem is to use a timer
based on the leading edge technique instead. This will require some
extra work with the \ac{TOF} selections and mean large
corrections\footnote{Of the order of 10~ns.}, since the timing of a
leading edge discriminator is dependent on the pulse height.

\subsection{Mismatch between T1 and T8\label{sec:mismatch}}

\begin{figure}[!htb]
  \centering
  \begin{minipage}{0.48\textwidth}
\includegraphics[width=\textwidth,bb=0 0 567 567]{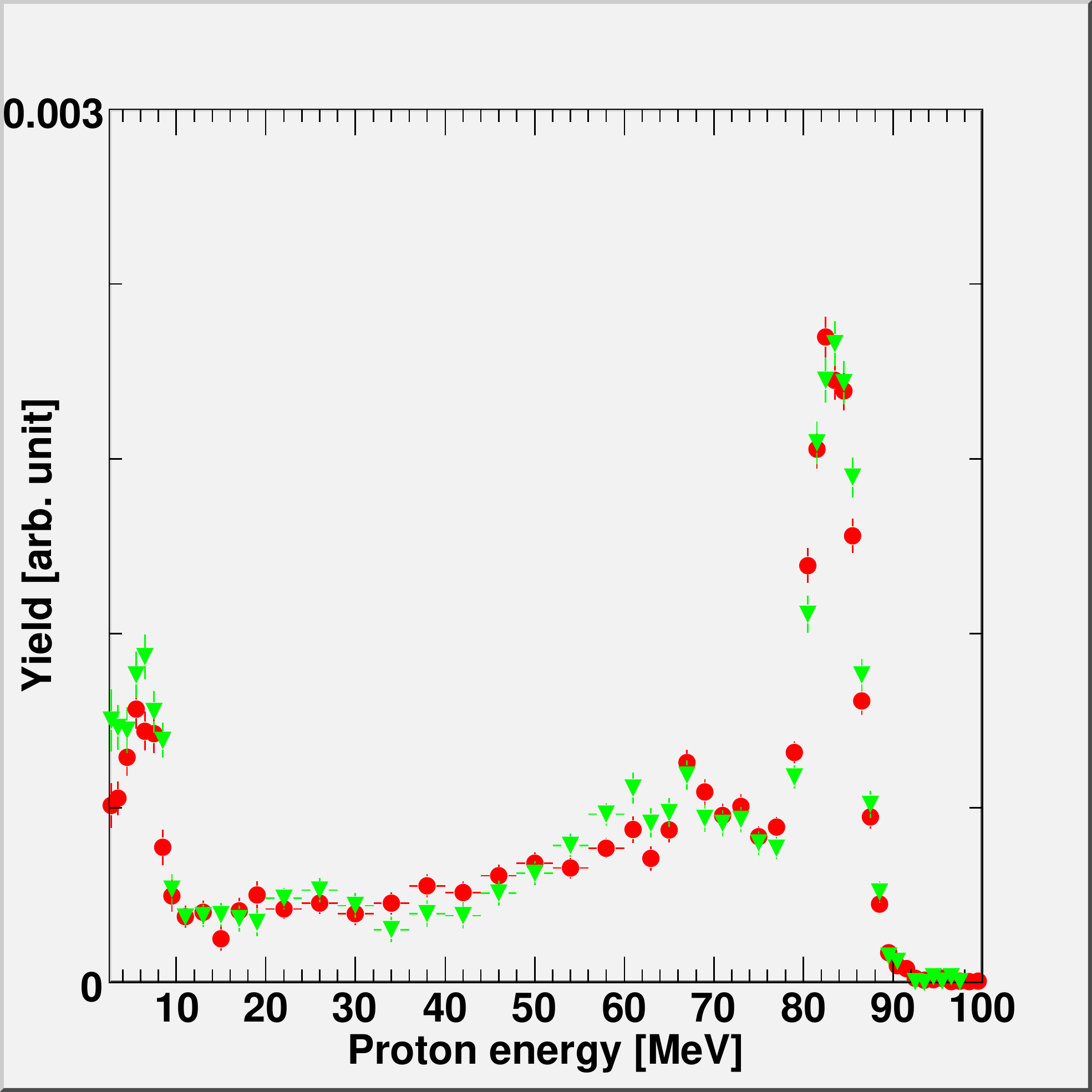}
  \end{minipage}
  \begin{minipage}{0.48\textwidth}
\includegraphics[width=\textwidth,bb=0 0 567 567]{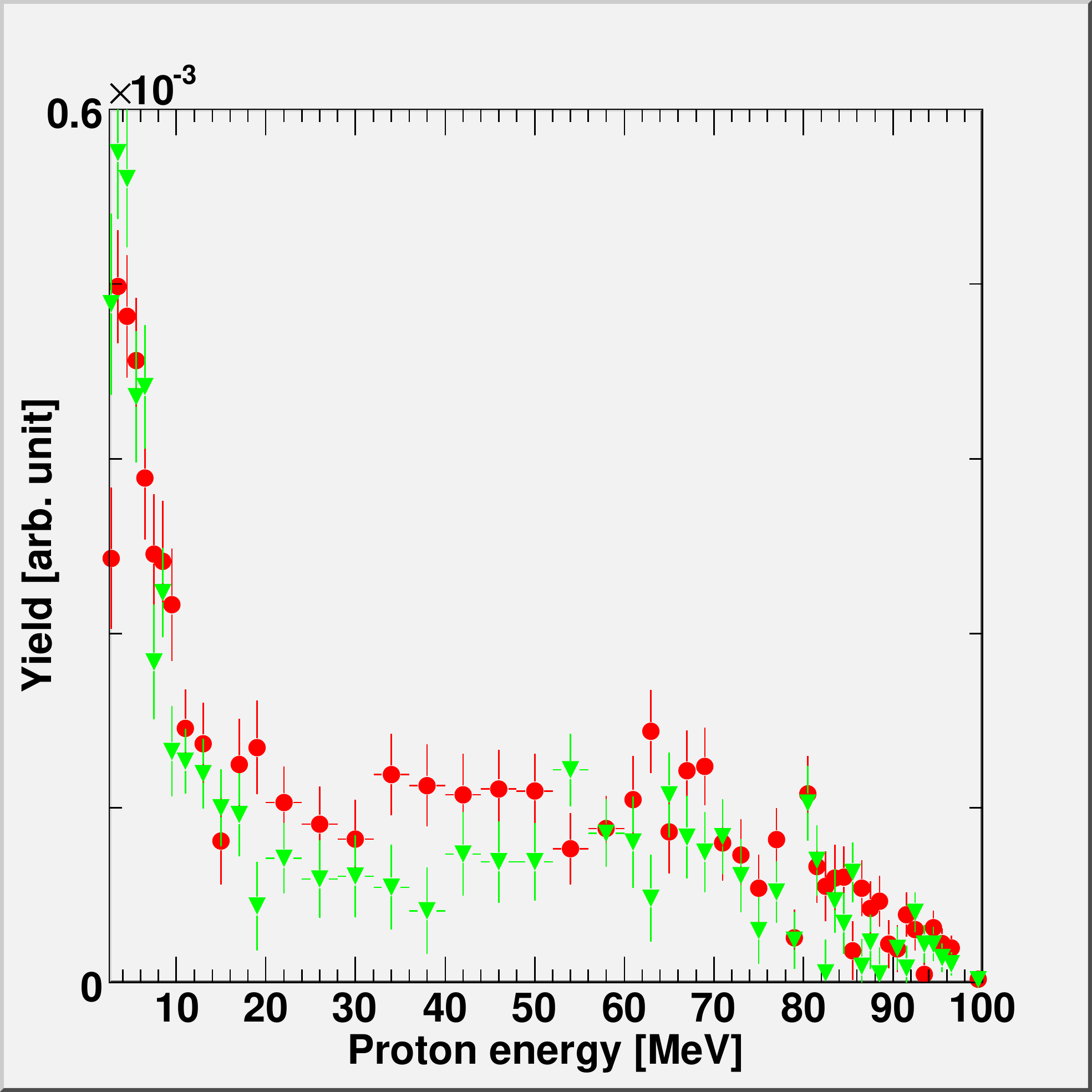}
  \end{minipage}
  \caption[Mismatch between T1 and T8]{Mismatch between \ac{T1} (filled circles) and \ac{T8} (filled triangles). The left panel shows the CH$_2$ runs and the right panel shows the calcium runs.}\label{fig:mismatch}
\end{figure}

As one can see in figure \ref{fig:mismatch} there are some
mismatches between \ac{T1} and \ac{T8}. The low energy mismatch has
been carefully examined, unfortunately without solving the problem.
The intermediate energy mismatch seen in the calcium data in figure
\ref{fig:mismatch} was however found to depend on the \ac{TOF} cut.
The mismatch was reduced when the upper part of the \ac{TOF} cut in
figure \ref{fig:tofplots} was increased. But increasing the \ac{TOF}
cut also increases the amount of accepted low energy neutrons that
need to be subtracted from the data. So the chosen \ac{TOF} cut of
two standard deviations from the peak, is a compromise between the
error from the telescope matching and the error from the low-energy
neutron correction.

\subsection{Reaction losses}

Due to nuclear interactions within the CsI(Tl) some protons may, for
example, produce gamma radiation that will escape the CsI(Tl)
undetected. Thus only a certain fraction of the incident protons is
detected at its true energy, and the rest will form a reaction tail
of apparently lower energy. This issue is currently under
investigation, but so far the results show that for low energies the lost
fraction is really small and correcting for these effects only gets
important at higher energies, as shown in figure \ref{fig:CsIeff}.

\subsection{Different approaches to angular distributions}

\begin{figure}[!htb]
  \centering
  \begin{minipage}{0.45\textwidth}
\includegraphics[width=\textwidth,bb=0 0 567 567]{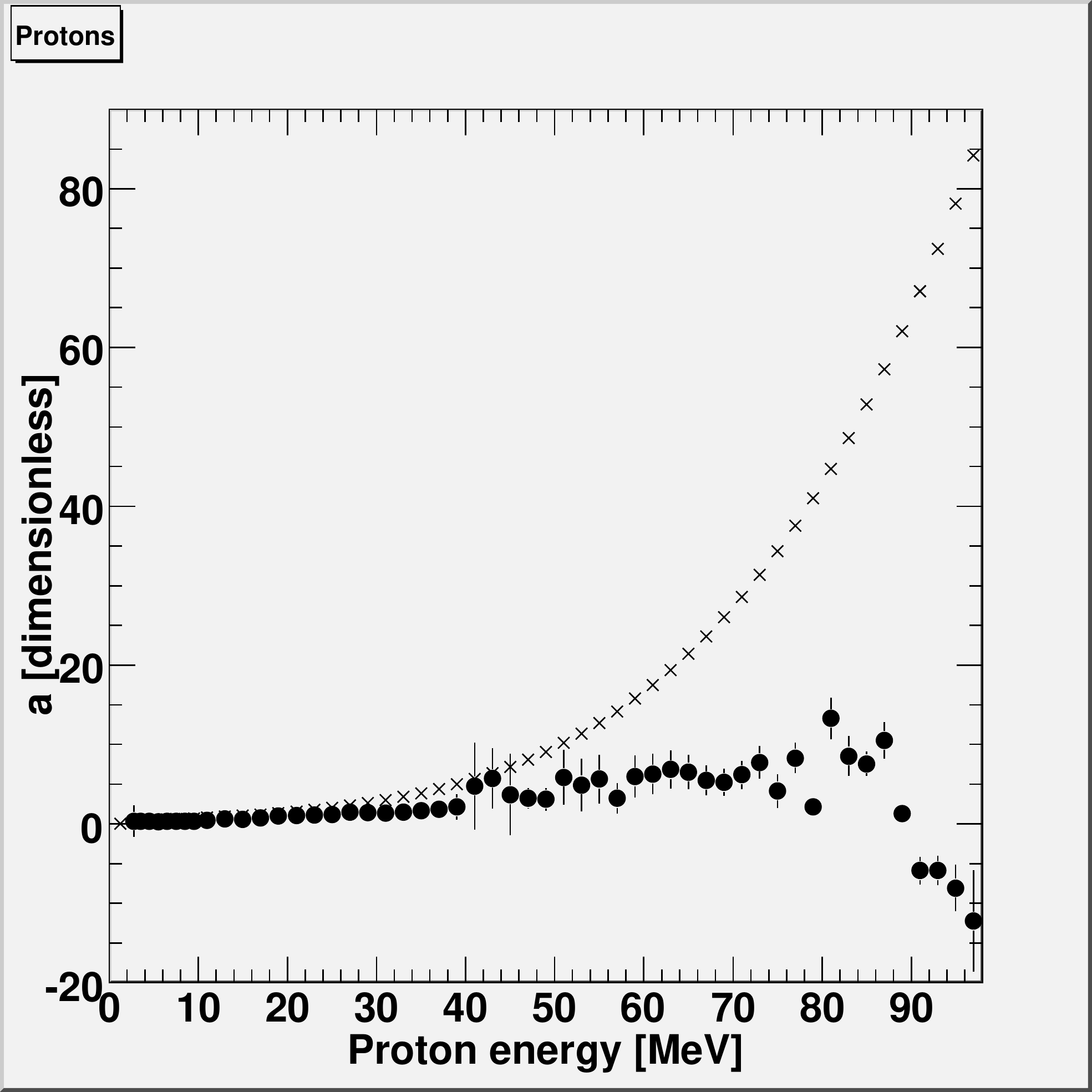}
    \caption[Theoretical and experimental values of angular forward peaking]{Theoretical and experimental values of angular forward peaking, $a$. The values of parameter $a$ are obtained from experimental fits (solid circles) and from Kalbach systematics (crosses).}\label{fig:ang_a}
  \end{minipage}
  \begin{minipage}{0.1\textwidth}
  \end{minipage}
  \begin{minipage}{0.45\textwidth}
\includegraphics[width=\textwidth,bb=0 0 567 567]{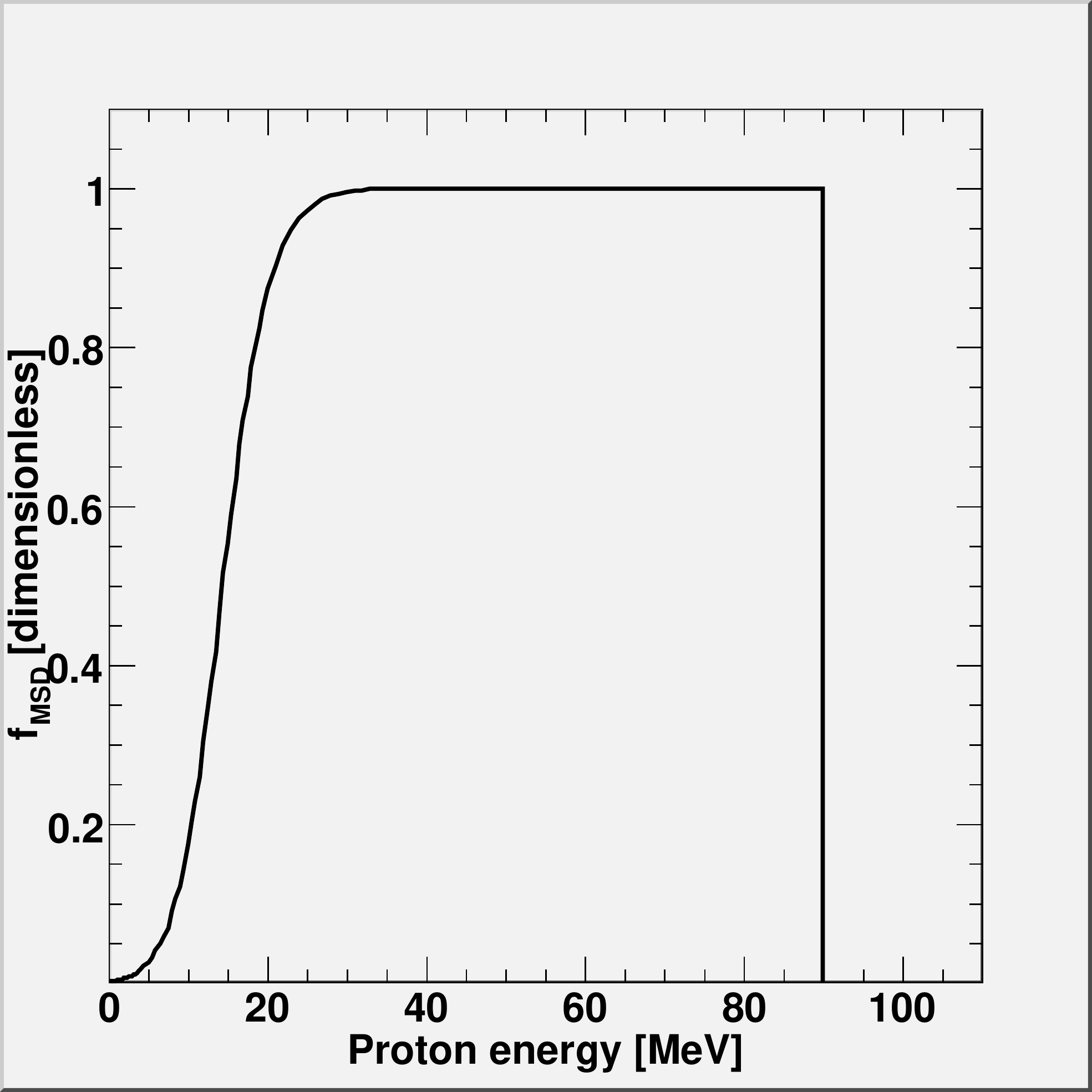}
    \caption[Theoretical pre-equilibrium ratio]{Theoretical pre-equilibrium ratio, $f_{\textrm{MSD}}$, as calculated by TALYS. \vspace{20pt}}\label{fig:ang_fmsd}
  \end{minipage}\\
  \begin{minipage}{0.96\textwidth}
\includegraphics[width=\textwidth,bb=0 0 567 250]{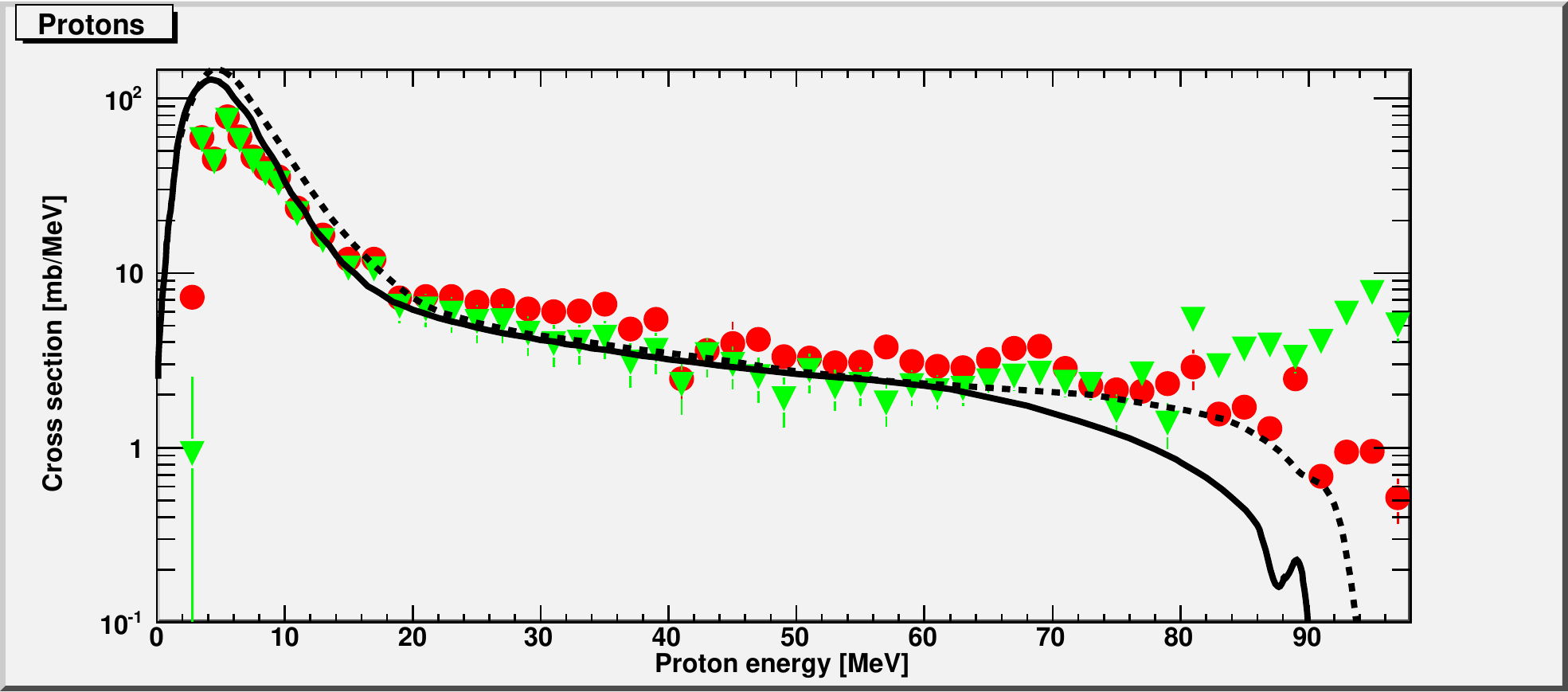}
    \caption[Theoretical and experimental approaches to angular distributions]{Theoretical and experimental approaches to angular distributions and the differences in energy-differential cross sections for the two approaches, where experimental angular distributions are filled circles and calculated angular distribution are filled triangles. The data used are the data with modified TALYS calculations subtracted.  The solid line is prediction by TALYS, and the dashed line is prediction by GNASH \protect \citep{ICRU}.}\label{fig:ang_appro}
  \end{minipage}
\end{figure}

In this work, the angular distributions were obtained by fitting
Eq.~(\ref{eq:gen_kalbach}) to the data. This is an approach coming
from experiment, and trying to fit some angular shape to the data,
that can be integrated to get the cross section. Another approach
one can use is to go from theory and calculate the predicted shape,
$a$, of the angular distributions and fitting Eq.~(\ref{eq:kalbach})
to the data under assumption that $f_{\textrm{MSD}}=1$. The actual
form of $f_{\textrm{MSD}}$, as calculated by TALYS, is found in
figure \ref{fig:ang_fmsd}. To calculate $a$, the parametrization
used \citep{1988PhRvC..37.2350K} is
\begin{equation}\label{eq:kalb_par1}
    a = 0.04\frac{E_1 {e_b}'}{{e_a}'}+1.8 \cdot 10^{-6}\left(\frac{E_1
    {e_b}'}{{e_a}'}\right)^{3}+6.7 \cdot 10^{-7}\left(\frac{E_3{e_b}'}{{e_a}'}\right)^{4}
\end{equation}
where $E_1 = \min({e_a}',130)$, $E_3 = \min({e_a}',41)$, ${e_a}' =
E_a + S_a$ and ${e_b}' = E_b + S_b$. Here $E_a$ is the incoming
neutron energy, $E_b$ is the outgoing proton energy, and $S_{a,b}$
is the separation energy between particle $a,b$ and the compound
nucleus. The separation energy, $S_{a,b}$, is calculated using the
Myers and Swiatecki mass formula \citep{talys}. As seen in figure
\ref{fig:ang_a} there exist some differences in angular
distributions between the theoretical and the experimental approach,
especially at higher energies, where the theory predicts the forward
peaking be larger. The negative values on $a$ at really high
energies, that would imply backward peaking, are probably
statistical artifacts from the background subtraction.

As can be seen in figure \ref{fig:ang_appro}, this parametrization
gives a result closer to the model at intermediate energies but
deviates more at high energies, while at low energies the difference
between the methods are small. This is quite expected from the
approximation of $f_{\textrm{MSD}}=1$. An investigation of the
parametrization with the calculated form of $f_{\textrm{MSD}}$, or
even experimental determination of $f_{\textrm{MSD}}$, is something
I think could be interesting in the future.

\subsection{Systematic uncertainties}

Throughout this work the error bars only represent the statistical
uncertainties in the experiment. To give a complete picture, the
systematic errors should be added to the result. These are listed in
table \ref{tab:sysunc}, where some values are for this particular
experiment, and some values are adapted from
\citet{2006PhRvC..73c4611T}.

\begin{table}
  \centering
  \begin{tabular}{lc}
    \hline
    Origin & Uncertainty \\
    \hline
    Target correction & 1-10 \% \\
    Solid angle & 1-5 \% \\
    Beam monitoring & 2-3 \% \\
    Number of Ca & 1 \% \\
    \ac{CsI(Tl)} efficiency & 1 \% \\
    Particle identification & 1 \% \\
    Dead time & $<0.1$ \% \\
    Absolute cross section & 5 \% \\
    \hline
  \end{tabular}
  \caption[Systematic uncertainties]{Systematic uncertainties}\label{tab:sysunc}
\end{table}

\subsection{Conclusions}

\begin{figure}[!htb]
  \centering
  \begin{minipage}{\textwidth}
    \begin{center}
        {\LARGE \sc Method 1}
    \end{center}
    \vspace{8pt}
  \end{minipage}\\
  \begin{minipage}{0.45\textwidth}
\includegraphics[width=\textwidth,bb=0 0 567 567]{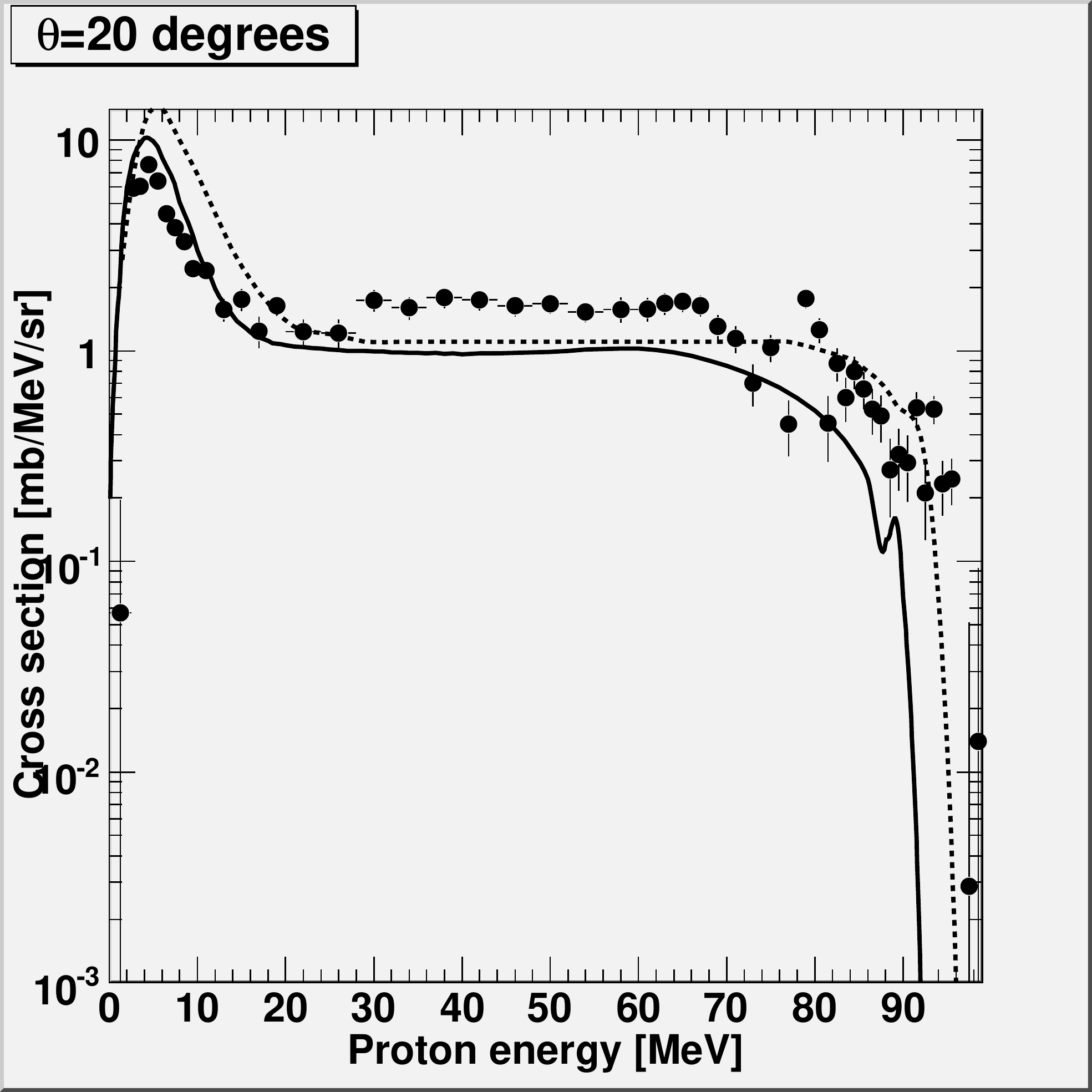}
  \end{minipage}
  \begin{minipage}{0.45\textwidth}
\includegraphics[width=\textwidth,bb=0 0 567 567]{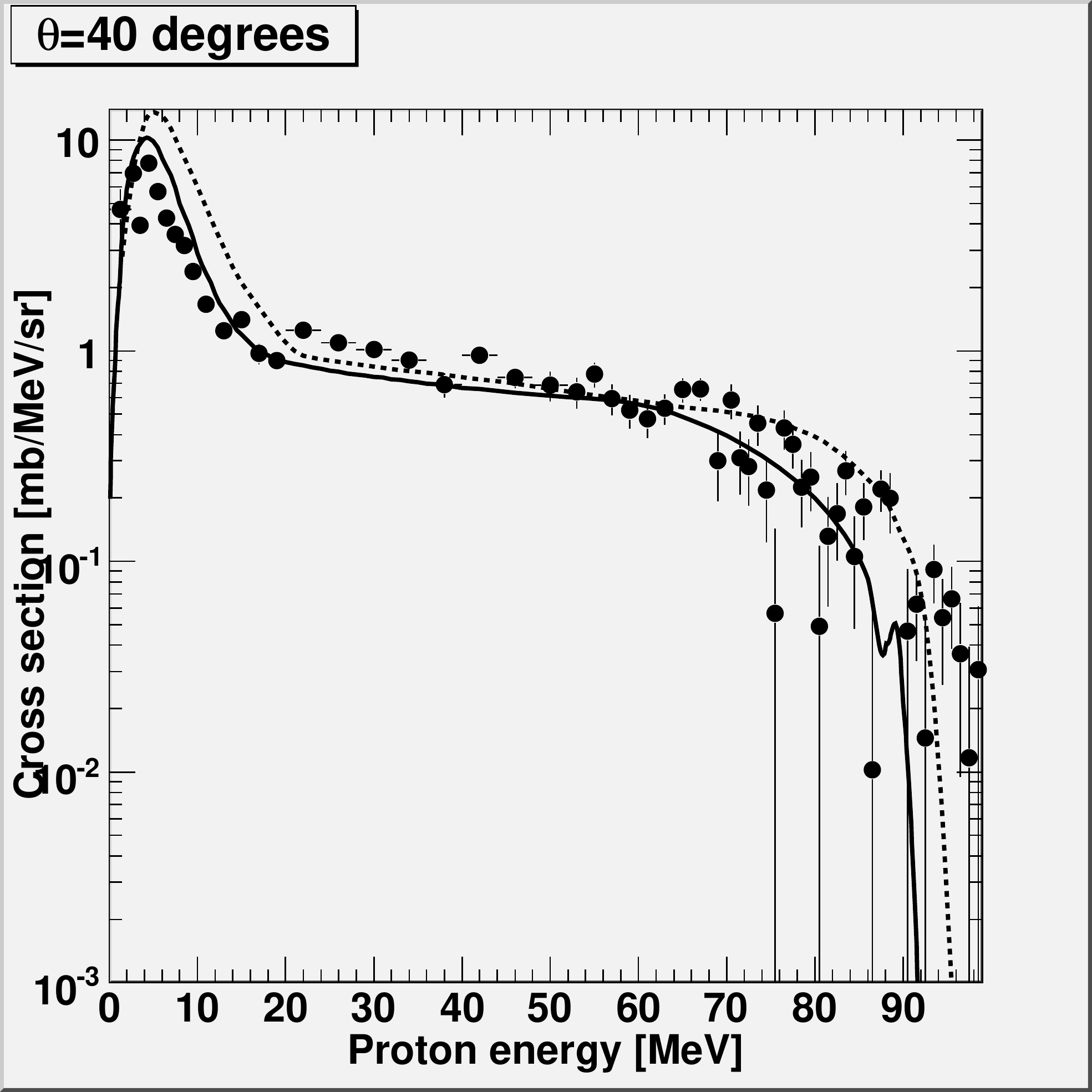}
  \end{minipage}\\
  \begin{minipage}{0.45\textwidth}
\includegraphics[width=\textwidth,bb=0 0 567 567]{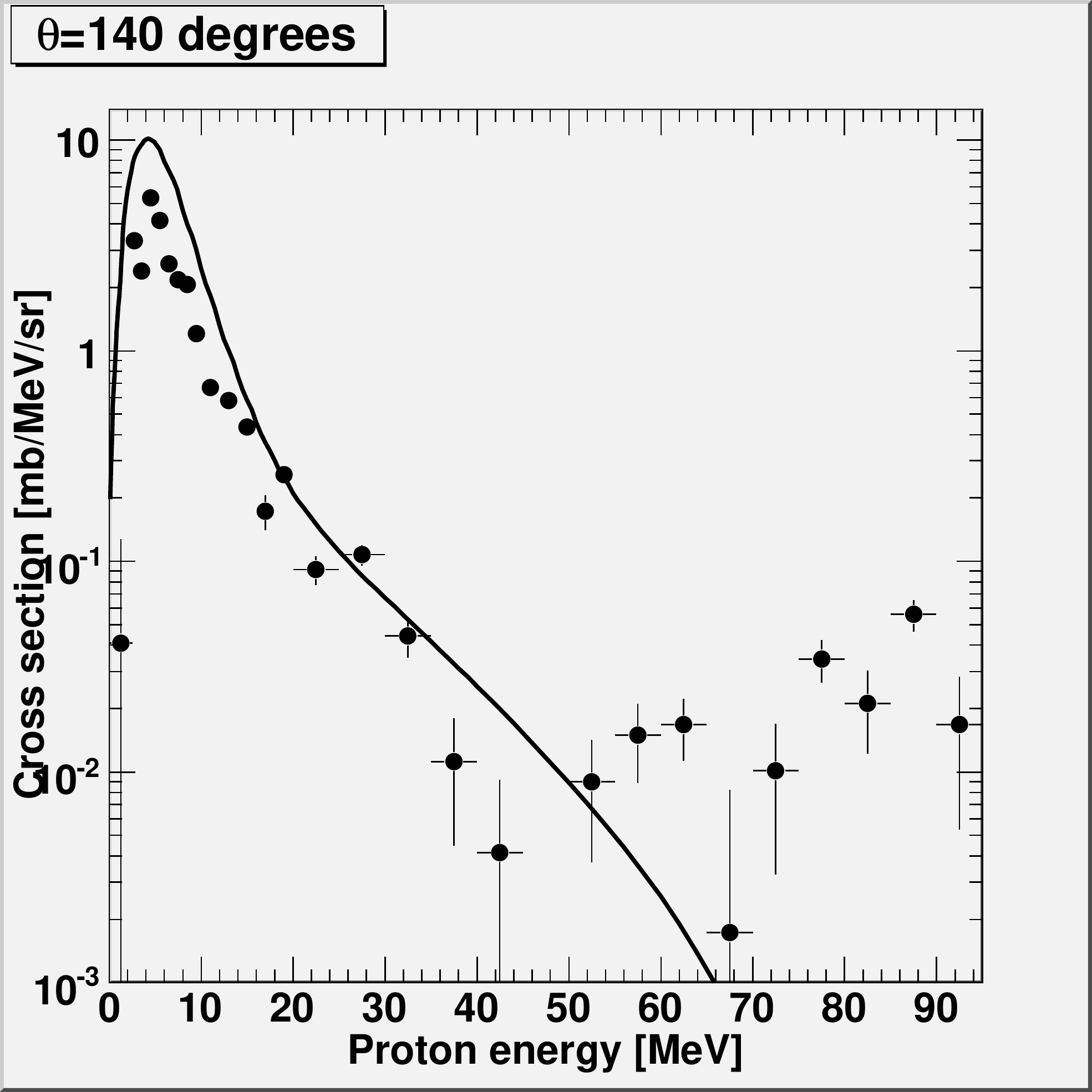}
  \end{minipage}
  \begin{minipage}{0.45\textwidth}
\includegraphics[width=\textwidth,bb=0 0 567 567]{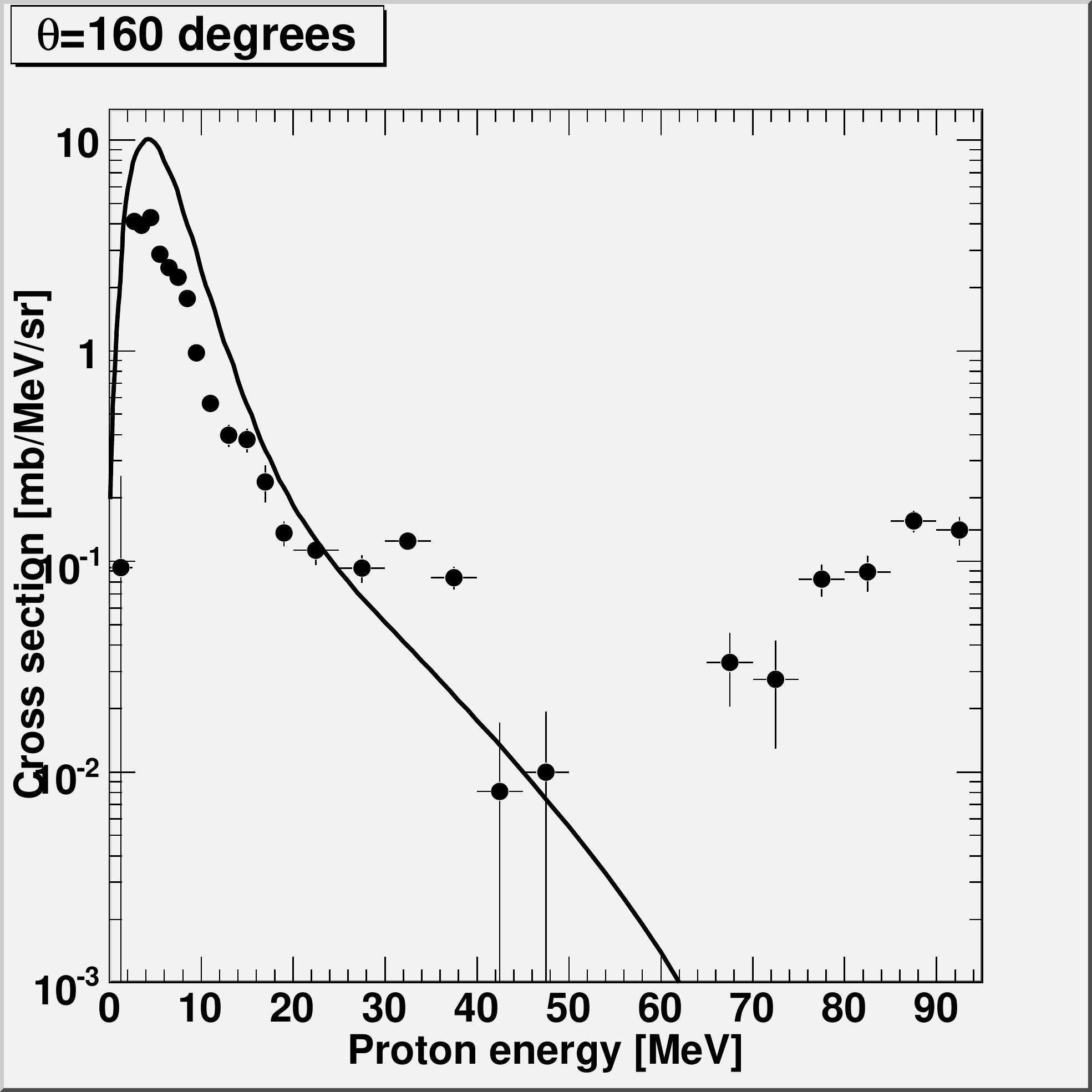}
  \end{minipage}\\
  \begin{minipage}{0.90\textwidth}
\includegraphics[width=\textwidth,bb=0 0 567 250]{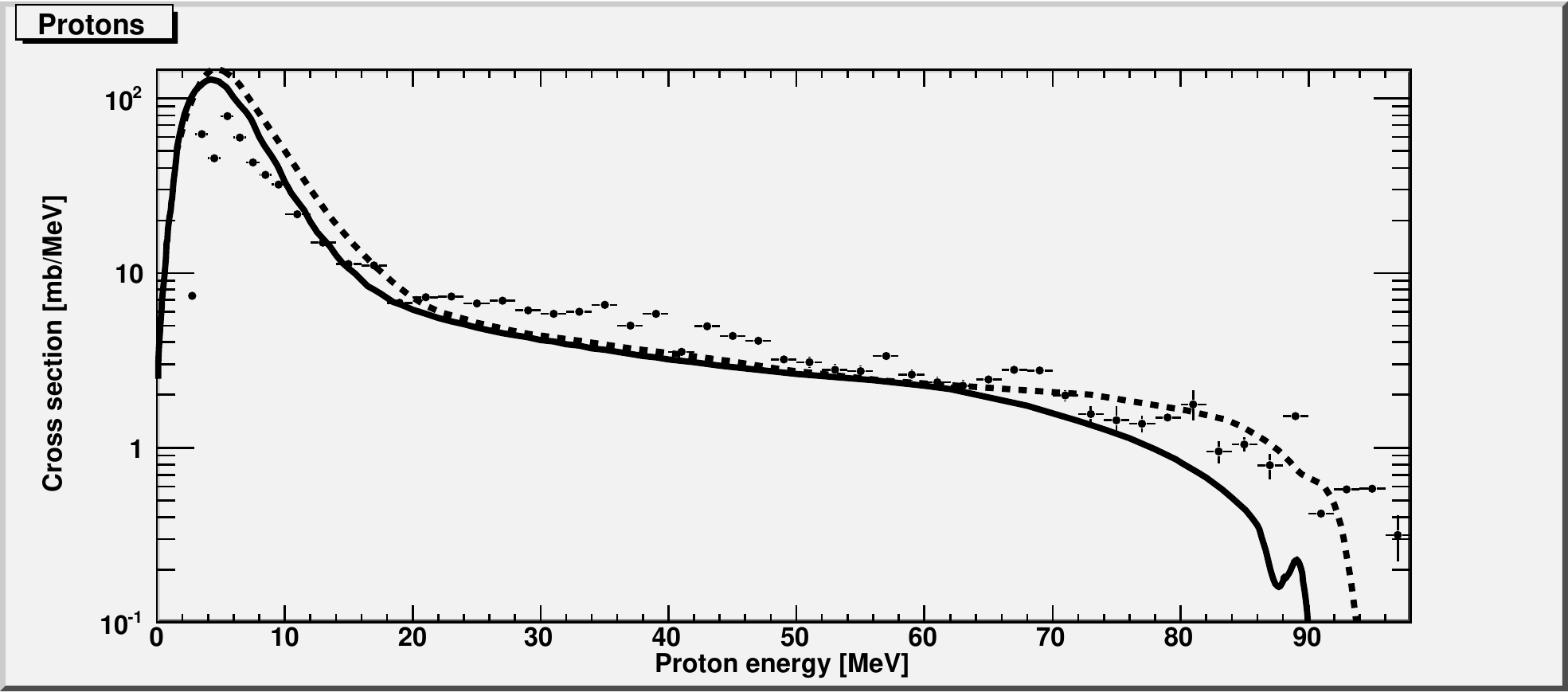}
  \end{minipage}
  \caption[Double- and energy-differential cross sections with experimental data subtracted]{Double-differential cross sections in the four top panels, and angle integrated energy-differential cross section in the bottom panel, with experimental data (method 1) subtracted for 94 MeV neutrons in the Ca(n,px) reaction. Solid lines are predictions by TALYS, and dashed lines are predictions by GNASH \protect \citep{ICRU}. Cutoff energy is 2.5 MeV.\label{fig:dEdO_res_frac}}
\end{figure}

\begin{figure}[!htb]
  \centering
  \begin{minipage}{\textwidth}
    \begin{center}
        {\LARGE \sc Method 3}
    \end{center}
    \vspace{8pt}
  \end{minipage}\\
  \begin{minipage}{0.45\textwidth}
\includegraphics[width=\textwidth,bb=0 0 567 567]{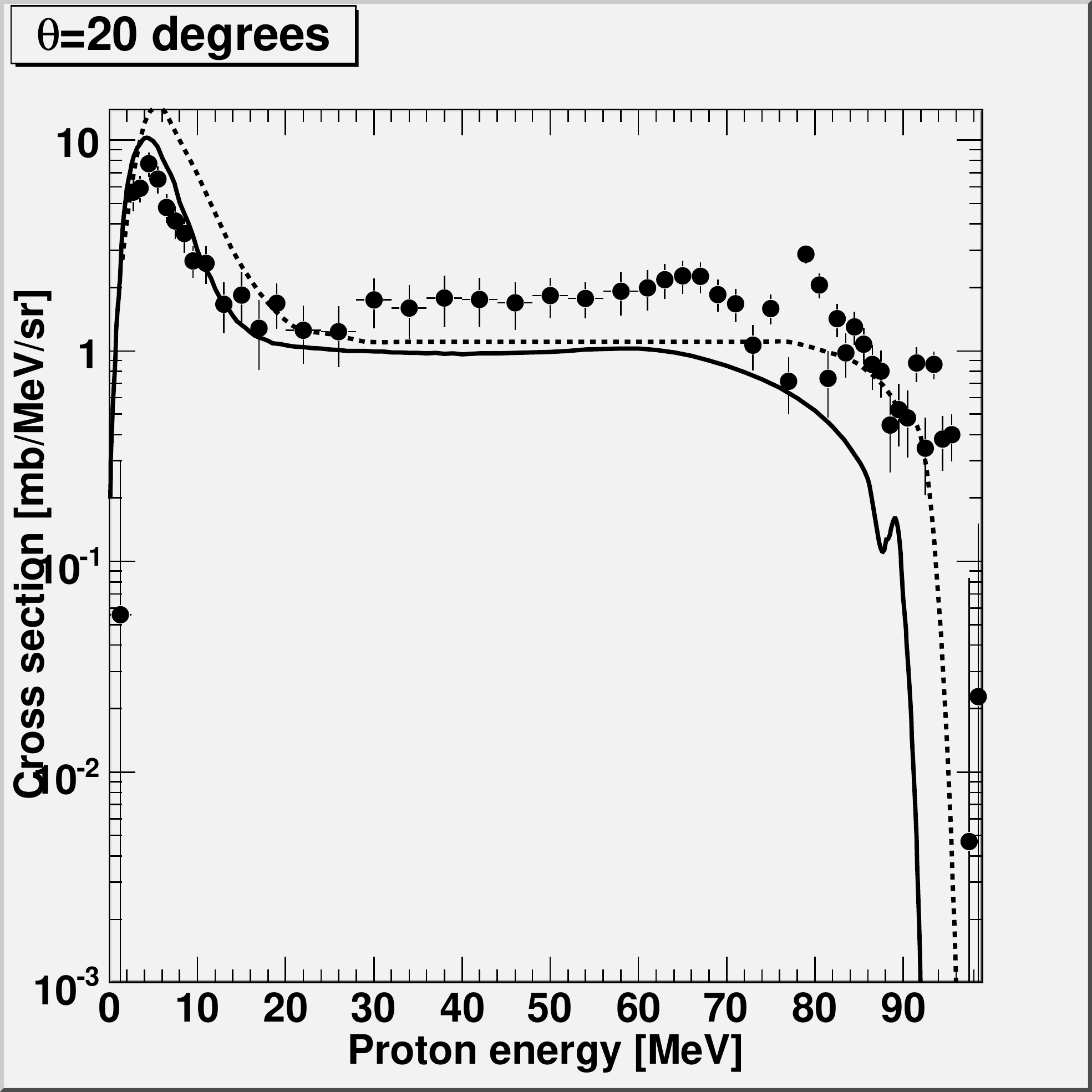}
  \end{minipage}
  \begin{minipage}{0.45\textwidth}
\includegraphics[width=\textwidth,bb=0 0 567 567]{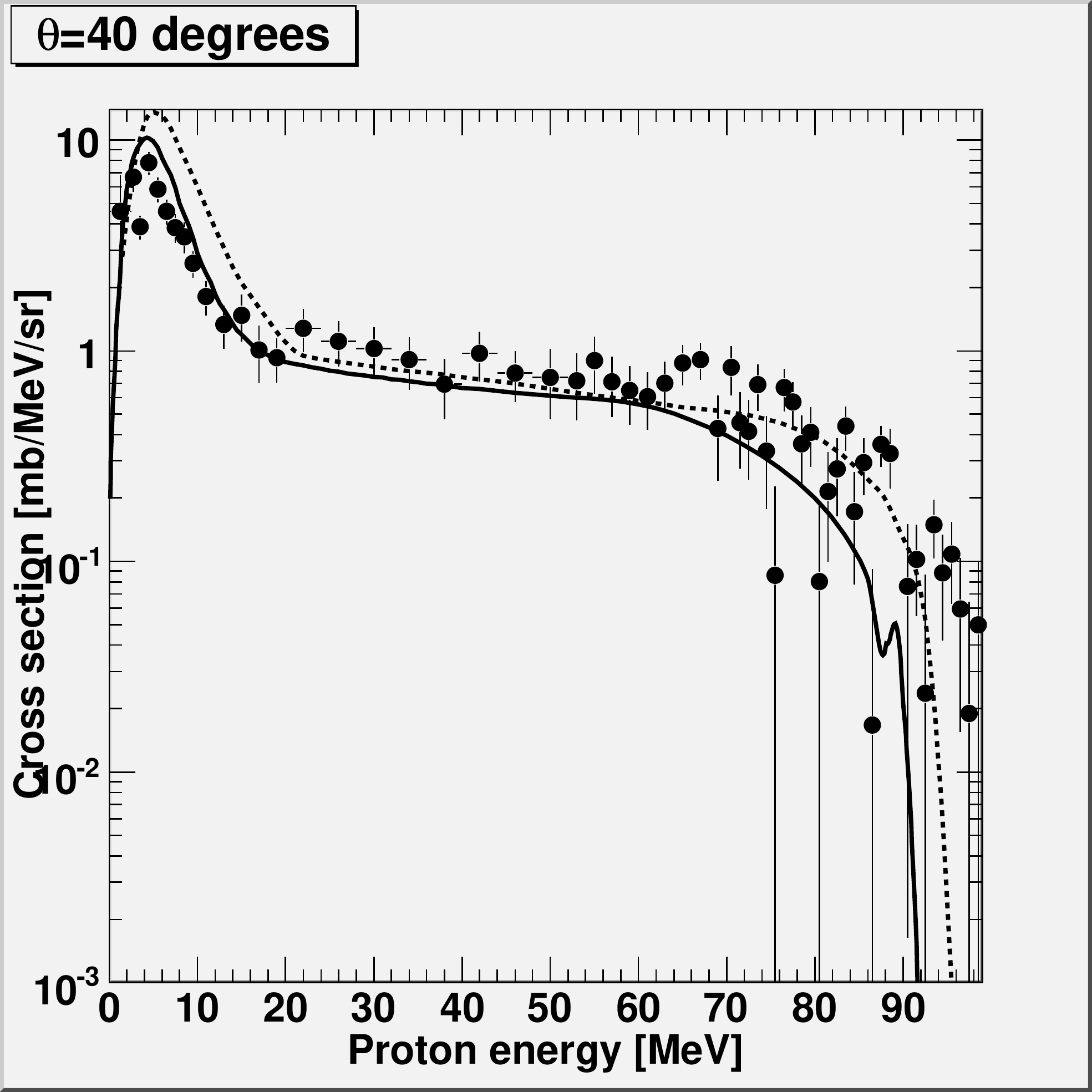}
  \end{minipage}\\
  \begin{minipage}{0.45\textwidth}
\includegraphics[width=\textwidth,bb=0 0 567 567]{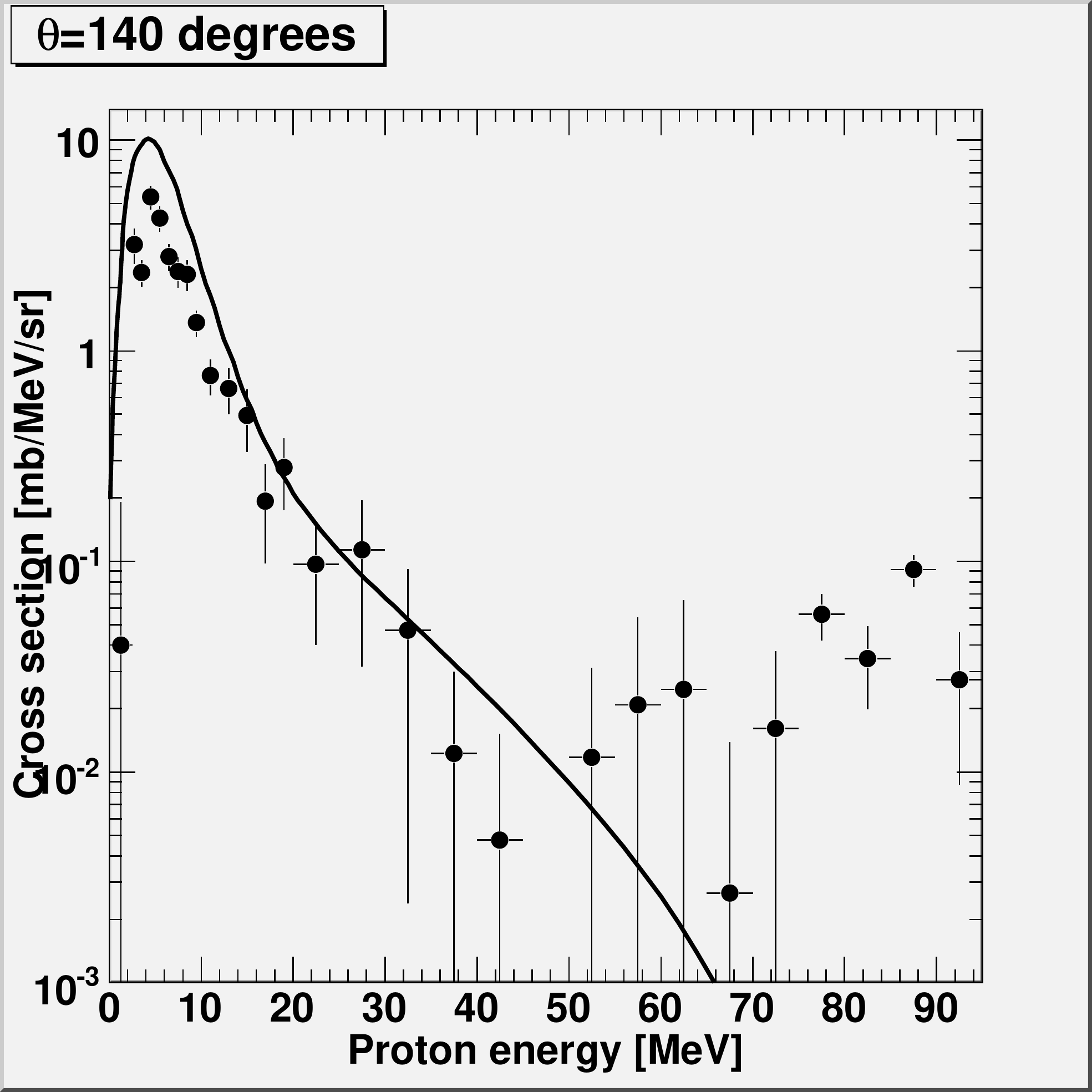}
  \end{minipage}
  \begin{minipage}{0.45\textwidth}
\includegraphics[width=\textwidth,bb=0 0 567 567]{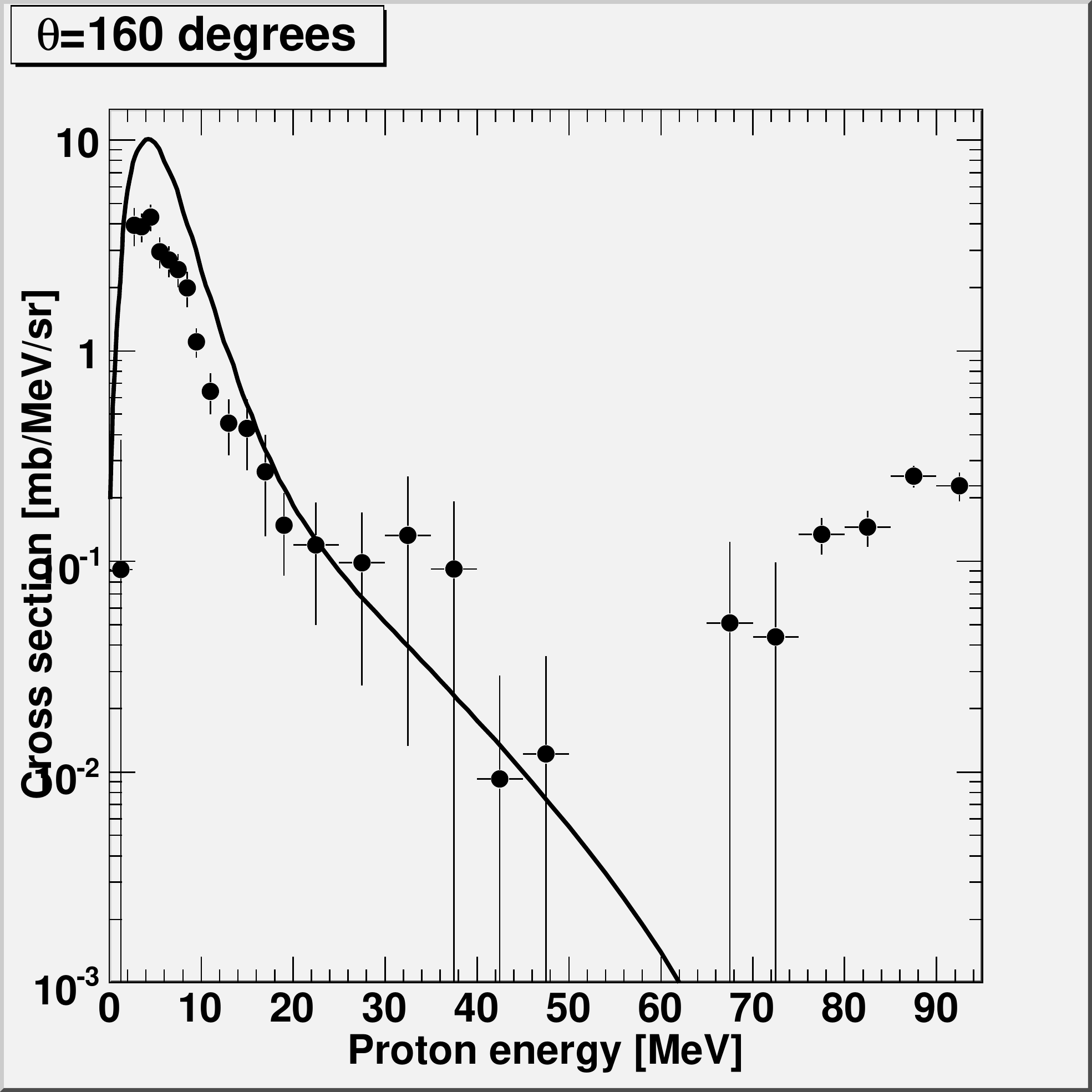}
  \end{minipage}\\
  \begin{minipage}{0.90\textwidth}
\includegraphics[width=\textwidth,bb=0 0 567 250]{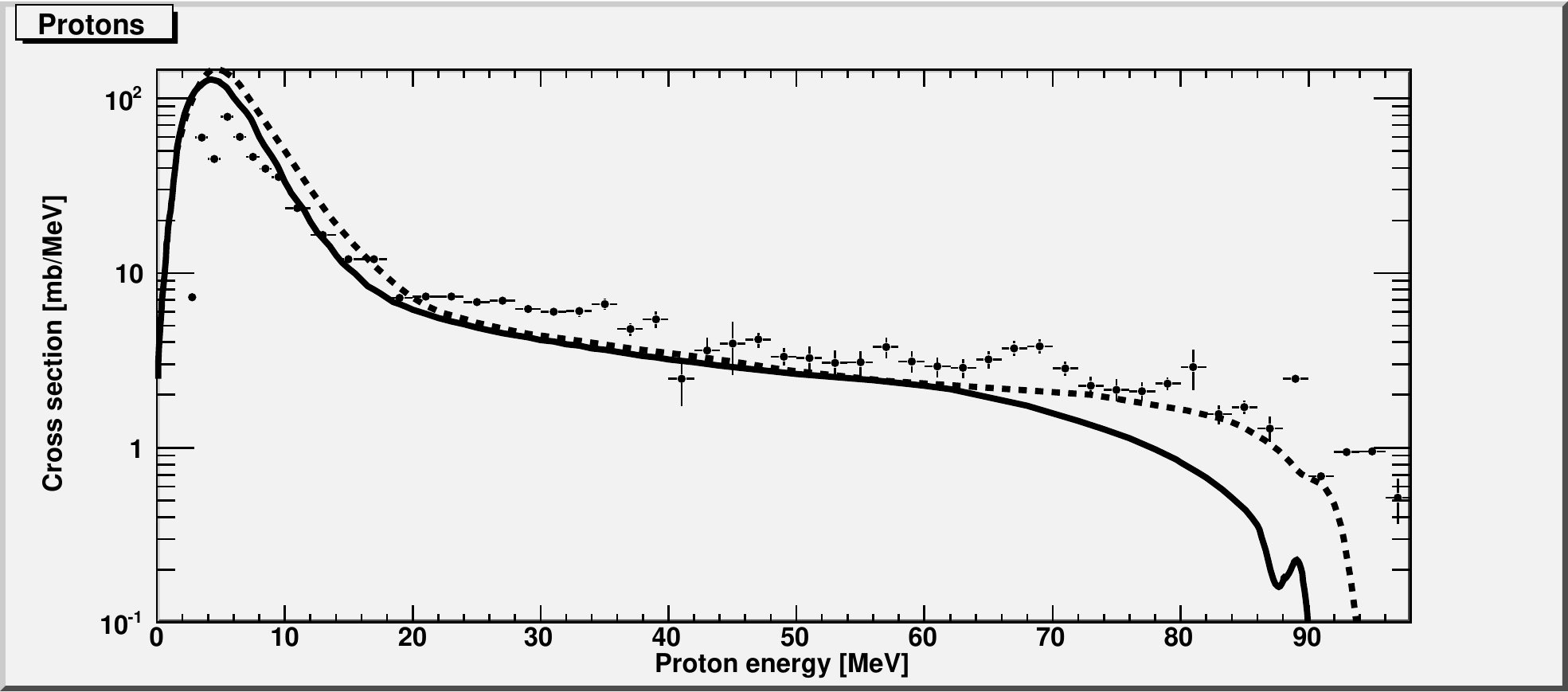}
  \end{minipage}
  \caption[Double- and energy-differential cross sections with modified TALYS data subtracted]{Double-differential cross sections in the four top panels, and angle integrated energy-differential cross section in the bottom panel, with modified TALYS data (method 3) subtracted for 94 MeV neutrons in the Ca(n,px) reaction. Solid lines are predictions by TALYS, and dashed lines are predictions by GNASH \protect \citep{ICRU}. Cutoff energy is 2.5 MeV.\label{fig:dEdO_res_wtal}}
\end{figure}

As presented in sections \ref{sec:meth1}-\ref{sec:meth3} and shown
in figures \ref{fig:dEdO_res_frac} - \ref{fig:dEdO_res_wtal},
comparisons between three different methods have been made. No
matter which method one chooses, the results all show the same trend
as the previous experiments \citep{2004PhRvC..69f4609T,
2006PhRvC..73c4611T}. The evaporation peak is overestimated by TALYS
and predicted to be almost isotropic, while in experiment it
significantly decreases with angle. The GNASH calculations
overestimate the peak even more. In the pre-equilibrium region,
TALYS instead underestimates the cross section, but the predictions
seem to be more accurate at larger angles.

The problem in the \ac{TOF} is an issue that needs to be
investigated, especially since similar effects have been noted in
other data sets as well \citep{tippcomm}. But since one of the
mismatches in section \ref{sec:mismatch} was found to be dependent
on the \ac{TOF} cut, a solution to the \ac{TOF} problem might shed
some light on the mismatch.

\subsection{Outlook}

The data sets analyzed in this work was the first, and at this
moment the only, data set available from the upgraded neutron
facility \citep{2005AIPC..769..780P} at \ac{TSL}. To further improve
the calcium results, experimental data from the angles 60, 80, 100
and 120 degrees should be analyzed. The reactions Ca(n,d), Ca(n,t),
Ca(n,{${^3}$He}) and Ca(n,$\alpha$) would also be interesting to
have a look at.

Apart from that, and as mentioned in section \ref{sec:setup}, lots
of work has previously been carried out at 96 MeV
\citep{2004PhRvC..69f4609T, 2006PhRvC..73c4611T,
2004PhRvC..70a4607B} and there still exists a never-ending selection
of isotopes to be measured. But since the nuclei under study in
these works are especially interesting from a biological and
technical point of view it is more interesting to use the
capabilities of the new neutron facility and run the experiments at
180 MeV. By obtaining more data sets, one extends the relevant
database for future applications. At the same time one makes it
possible to do a more quantitative analysis on the nuclei and get a
better systematic feedback for the development of TALYS.

\clearpage

\setcounter{figure}{0} \setcounter{table}{0}
\setcounter{equation}{0}

\section*{Acknowledgements}
\addcontentsline{toc}{section}{Acknowledgements}

\begin{itemize}
  \item \emph{S} -- Thank you for all your support and your infinite patience with me. And thank you for always being there for me in times of need.
  \item \emph{Jan ``Bumpen'' Blomgren} -- Always having something interesting to share from the world outside. And who, despite my probably very confused appearance when I first contacted him, understood what I wanted to do and introduced me to...
  \item \emph{Stephan Pomp} -- My guide and supervisor throughout this. Thank you for all the provided inspiration. And for always having time for me, and my questions and ideas of various quality and relevance.
  \item \emph{Masateru Hayashi} -- Who helped me a lot during the first part of my work, almost like an assisting supervisor. Thank you for discussing ROOT macros, C++ programming, detectors and electronics as well as life in Japan and Sweden with me.
  \item \emph{Johan Vegelius} -- For sharing pizza in the evenings and work during the days, and for simply being a great fellow diploma student.
\end{itemize}
I also would like to thank everyone at INF, and TSL, that has
contributed to a great working environment. And with these words, my
diploma thesis is completed. A special thanks to you for reading it.

\begin{flushright}
\includegraphics[clip=true, viewport=.0in .31in 1.6in 0.67in,bb=0 0 116 72]{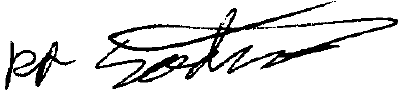}
\\P\"{a}r-Anders S\"{o}derstr\"{o}m
\end{flushright}

\clearpage

\appendix

\addcontentsline{toc}{section}{Appendices}

\setcounter{figure}{0} \setcounter{table}{0}
\setcounter{equation}{0}
\section*{Appendices}
\section{Figures and tables}


\begin{figure}[!hbt]
  \centering
\includegraphics[width=0.75\textwidth,bb=0 0 567 546]{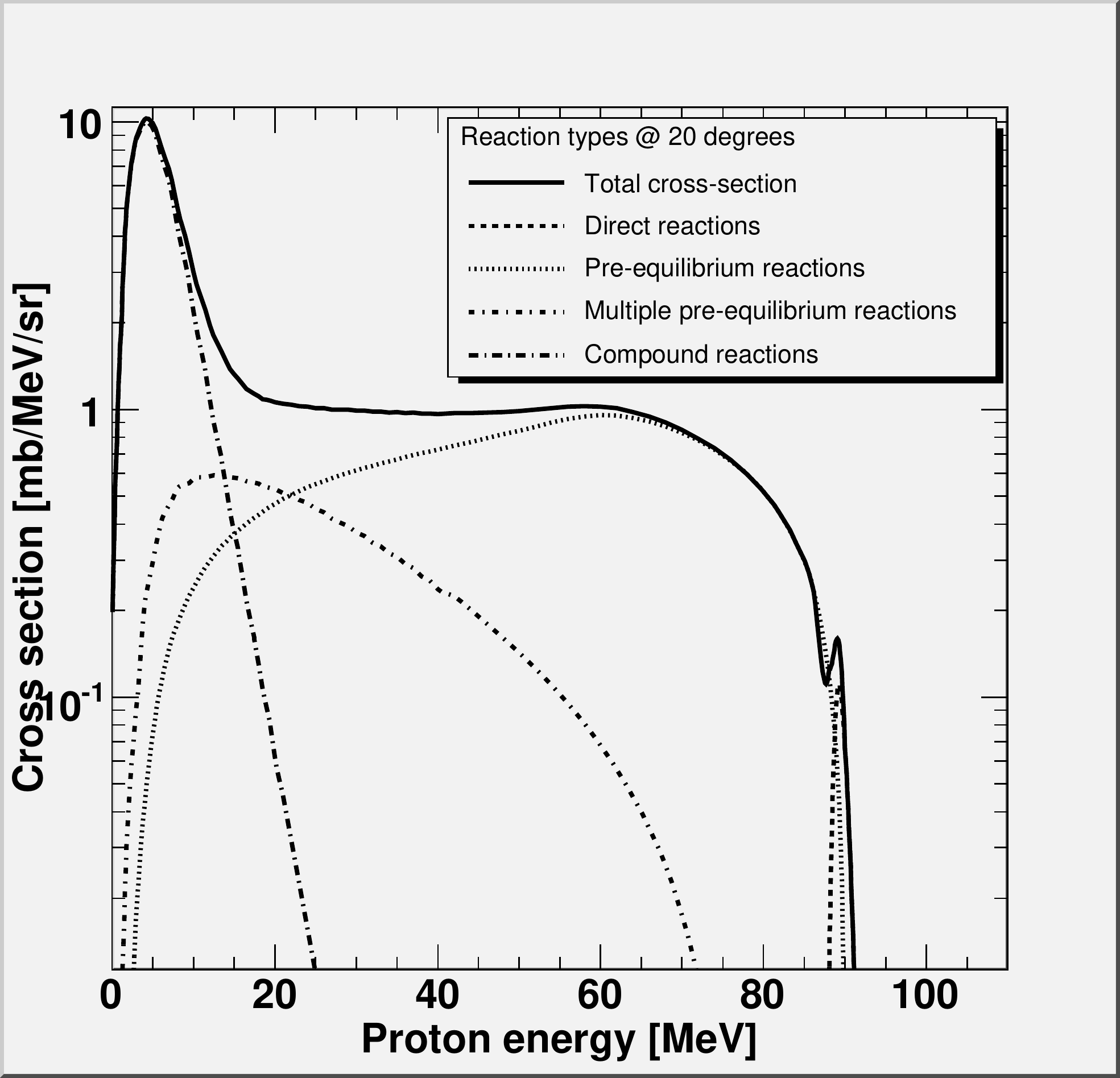}
  \caption[Contributions to total cross section from various reaction types]{Contribution to total cross section from various reaction types, according to TALYS calculations.}\label{fig:talplot}
\end{figure}

\begin{table}[!ht]
  \centering
  \begin{tabular}{ccccccc}
    \hline
    No & Si1 & Si2 & Distance [cm] & Angle [deg] & $\Delta x_1$ [$\mu$m]& $\Delta x_2$ [$\mu$m]\\
    \hline
    1 & 38-110B & 35-150B & 26.83 & 20 & 64.9 & 549\\
    2 & 37-198D & 37-051B & 19.83 & 40 & 60.5 & 538\\
    3 & 38-110D & 37-052A & 19.83 & 60 & 63.9 & 533\\
    4 & 37-198G & 37-052C & 19.83 & 80 & 61.7 & 526\\
    5 & 31-325C & 37-199A & 19.83 & 100& 50.4 & 439\\
    6 & 31-325D & 37-199E & 19.83 & 120& 50.1 & 432\\
    7 & 38-110G & 39-100B & 19.83 & 140& 61.6 & 550\\
    8 & 31-325H & 37-199G & 26.83 & 160& 52.9 & 424\\
    \hline
  \end{tabular}
  \caption{Detector setup}\label{tab:decset}
\end{table}

\clearpage
\begin{table}[!ht]
  \centering
  \begin{tabular}{ccccccc}
    \hline
    Run & Target & Beam  & Neutron fluence & Lifetime  & Runtime \\
        &        & [\textmu A] & [n/cm${^2}\cdot10^8$]& [\%] & [h]\\
    \hline \hline
    004 & Junk & & & & \\
    \hline
    005 & Empty & 1.2 & 3.1 & 84 & 0:57\\
    006 & Empty & 1.1 & 4.2 & 74 & 1:15\\
    007 & Empty & 1.2 & 5.2 & 72 & 1:29\\
    008 & Empty &     & 5.5 & 72 & 1:35\\
    009 & Empty &     &     &    & 1:27\\
    010 & Empty &     & 5.9 & 72 & 1:36\\
    011 & Empty & 1.2 & 3.5 & 72 & 0:57\\
    012 & Empty &     & 5.6 & 72 & 1:33\\
    013 & Empty &     & 6.9 & 72 & 1:55\\
    \hline
    014 & CH$_2$ & 1.2 & 6.5 & 71 & 1:45\\
    015 & CH$_2$ &     & 5.5 & 71 & 1:33\\
    016 & CH$_2$ &     & 4.3 & 71 & 1:09\\
    \hline
    017 & Empty & 1.2 & 4.7 & 72 & 1:22\\
    018 & Empty & 1.2 & 4.7 & 73 & 1:26\\
    \hline
    019 & Ca & 1.1 & 1.6 & 74 & 0:27\\
    020 & Ca &     & 5.9 & 74 & 1:50\\
    021 & Ca &     & 7.1 & 74 & 2:02\\
    022 & Ca & 1.1 & 7.1 & 74 & 2:11\\
    023 & Ca &     & 5.9 & 74 & 1:47\\
    024 & Ca &     & 5.6 & 74 & 1:37\\
    025 & Ca &     & 4.6 & 73 & 1:19\\
    \hline
    026 & CH$_2$ & & 5.1 & 72 & 1:28\\
    027 & CH$_2$ & & 4.6 & 72 & 1:19\\
    \hline
    028 & Ca &     & 7.1 & 73 & 2:01\\
    029 & Ca &     & 6.9 & 73 & 1:56\\
    030 & Ca &     & 5.5 & 73 & 2:00\\
    031 & Ca &     & 3.4 & 73 & 0:52\\
    032 & Ca &     & 5.7 & 73 & 1:37\\
    033 & Ca &     & 6.6 & 73 & 1:55\\
    034 & Ca &     & 7.0 & 73 & 2:01\\
    035 & Ca &     & 3.7 & 74 & 1:06\\
    036 & Ca &     & 1.6 & 74 & 0:35\\
    \hline
  \end{tabular}
  \caption[Runs with table at 180 degrees]{List of runs when the table was positioned at 180 degrees, that is with \ac{T1} in forward and \ac{T8} in backward direction.}\label{tab:runlist180}
\end{table}
\clearpage
\begin{table}[!ht]
  \centering
  \begin{tabular}{cccccccc}
    \hline
    Run & Target & Beam  & Neutron fluence & Lifetime  & Runtime \\
        &        & [\textmu A] & [n/cm${^2}\cdot10^8$]& [\%] & [h]\\
    \hline \hline
    001 & Junk &     & & & \\
    002 & Junk & 1.2 & & & \\
    003 & Junk & 1.2 & & & \\
    \hline
    037 & CH$_2$ & 1.1 &    & & 1:05 \\
    \hline
    038 & Junk &       & & &\\
    \hline
    039 & CH$_2$ & 1.1 & 0.96 & 71 & 0:15\\
    040 & CH$_2$ &     & 5.6 & 71 & 1:43\\
    041 & CH$_2$ & 1.1 & 5.5 & 71 & 1:35\\
    042 & CH$_2$ &     & 1.8 & 71 & 0:30\\
    043 & CH$_2$ &     & 6.9 & 71 & 2:00\\
    \hline
    044 & Empty &     & 5.9 & 72 & 1:44\\
    045 & Empty &     & 9.3 & 72 & 2:44\\
    046 & Empty &     & 7.2 & 72 & 2:07\\
    047 & Empty &     & 6.9 & 72 & 2:00\\
    048 & Empty &     & 7.0 & 72 & 2:01\\
    049 & Empty &     & 6.7 & 72 & 1:59\\
    050 & Empty &     & 6.8 & 72 & 2:00\\
    051 & Empty &     & 5.2 & 72 & 1:31\\
    \hline
    052 & Ca & 1.1 & 4.5 & 71 & 1:15\\
    053 & Ca & 1.1 & 5.3 & 72 & 1:37\\
    054 & Ca & 1.1 & 5.4 & 72 & 1:35\\
    055 & Ca &     & 5.7 & 72 & 1:45\\
    056 & Ca &     & 3.4 & 72 & 1:00\\
    057 & Ca &     & 4.7 & 72 & 1:25\\
    058 & Ca &     & 6.4 & 71 & 1:55\\
    059 & Ca &     & 6.9 & 71 & 2:02\\
    060 & Ca &     & 6.4 & 71 & 1:56\\
    061 & Ca &     & 6.6 & 71 & 2:01\\
    062 & Ca &     & 4.2 & 72 & 1:17\\
    063 & Ca & 1.1 & 5.4 & 72 & 1:39\\
    064 & Ca & 1.1 & 6.5 & 72 & 1:59\\
    065 & Ca &     & 6.5 & 72 & 1:59\\
    066 & Ca &     & 3.5 & 72 & 1:02\\
    \hline
  \end{tabular}
  \caption[Runs with table at 0 degrees]{List of runs when the table was positioned at 0 degrees, that is with \ac{T8} in forward and \ac{T1} in backward direction.}\label{tab:runlist0}
\end{table}
\clearpage

\begin{figure}[!ht]
  \centering
  \begin{tabular}{|c|c|c|c|c|c|c|c|c|c|}
    \hline
    \texttt{013C} & \textbf{\texttt{4264}} & \texttt{04FE} & \texttt{0000} & \texttt{0000} & \textbf{\texttt{4265}} & \texttt{05B2} & \texttt{0000} & \texttt{0000} & \textbf{\texttt{4266}} \\
    \hline
    \texttt{083E} & \texttt{0000} & \texttt{0000} & \textbf{\texttt{4267}} & \texttt{039B} & \texttt{0000} & \texttt{0000} & \textbf{\texttt{4268}} & \texttt{06B5} & \texttt{0000} \\
    \hline
    \texttt{0000} & \textbf{\texttt{4269}} & \texttt{06F4} & \texttt{0000} & \texttt{0000} & \textbf{\texttt{426A}} & \texttt{048B} & \texttt{0000} & \texttt{0000} & \textbf{\texttt{426B}}\\
    \hline
    \texttt{0325} & \texttt{0000} & \texttt{0000} & \textbf{\texttt{426C}}& \texttt{0A2C} & \texttt{0000} & \texttt{0000} & \textbf{\texttt{426D}} & \texttt{074E} & \texttt{0000} \\
    \hline
    \texttt{0000} & \textbf{\texttt{426E}} & \texttt{0C53} & \texttt{0000} & \texttt{0000} & \textbf{\texttt{426F}} & \texttt{000E} & \texttt{0000} & \texttt{0000} & \textbf{\texttt{4270}} \\
    \hline
    \texttt{02D9} & \texttt{0000} & \texttt{0000} & \textbf{\texttt{4271}} & \texttt{0000} & \texttt{0000} & \texttt{0000} & \textbf{\texttt{4272}} & \texttt{0000} & \texttt{0000} \\
    \hline
    \texttt{0000} & \textbf{\texttt{4273}} & \texttt{0000} & \texttt{0000} & \texttt{0000} & \textbf{\texttt{4274}} & \texttt{0000} & \texttt{0000} & \texttt{0000} & \textbf{\texttt{4275}} \\
    \hline
    \texttt{0000} & \texttt{0000} & \texttt{0000} & \textbf{\texttt{4276}} & \texttt{0000} & \texttt{0000} & \texttt{0000} & \textbf{\texttt{4277}} & \texttt{0000} & \texttt{0000} \\
    \hline
    \texttt{0000} & \textbf{\texttt{4501}} & \texttt{0066} & \texttt{0065} & \texttt{004F} & \texttt{0000} & \texttt{01F4} & \textbf{\texttt{4502}} & \texttt{004B} & \texttt{02CA} \\
    \hline
    \texttt{0137} & \texttt{0000} & \texttt{007E} & \textbf{\texttt{4503}} & \texttt{007A} & \texttt{0068} & \texttt{0040} & \texttt{0000} & \texttt{001C} & \textbf{\texttt{4504}} \\
    \hline
    \texttt{004B} & \texttt{0034} & \texttt{0043} & \texttt{0000} & \texttt{0206} & \textbf{\texttt{4505}} & \texttt{0064} & \texttt{005A} & \texttt{0036} & \texttt{0000} \\
    \hline
    \texttt{01F5} & \textbf{\texttt{4506}} & \texttt{0077} & \texttt{0049} & \texttt{003F} & \texttt{0000} & \texttt{01CF} & \textbf{\texttt{4507}} & \texttt{0062} & \texttt{006D} \\
    \hline
    \texttt{00D4} & \texttt{0000} & \texttt{0225} & \textbf{\texttt{4508}} & \texttt{0083} & \texttt{0065} & \texttt{004D} & \texttt{0000} & \texttt{01D3} & \textbf{\texttt{480A}} \\
    \hline
    \texttt{0060} & \texttt{0068} & \texttt{0067} & \texttt{004D} & \texttt{0069} & \texttt{0047} & \texttt{005A} & \texttt{0062} & \texttt{0000} & \texttt{4C09} \\
    \hline
    \texttt{0025} & \texttt{0038} & \texttt{004A} & \texttt{004E} & \texttt{0040} & \texttt{0039} & \texttt{002E} & \texttt{002F} & \texttt{0FDC} & \texttt{0FDC} \\
    \hline
    \texttt{0FDC} & \texttt{0FDC} & \texttt{0000} & \textbf{\texttt{4300}} & \textbf{\texttt{0000}} & \texttt{0083} & \texttt{018F} & \textbf{\texttt{FFFF}} &      & \\
    \hline
  \end{tabular}
  \caption[Typical sequence of hexadecimal raw data for an event]{Typical sequence of hexadecimal raw data for an event. Group identifiers are marked in bold. Groups 42XX are the scalers, 45XX are the readout from the telescopes and the group 4300 0000 is the identifier for the \ac{RF} signals. The event is ended by FFFF.}\label{tab:hexraw}
\end{figure}

\begin{table}[!ht]
  \centering
  \begin{tabular}{ccccc}
    \hline
      & T1 & T2 & T7 & T8 \\
    \hline
    \hline
    $k_1$ & 0.00334    & 0.00332 & 0.00309 & 0.00278 \\
    $m_1$ & -0.264  & -0.197  & -0.284  & -0.256 \\
    \hline
    $k_2$ & 0.00893 & 0.00863 & 0.00969  & 0.00904 \\
    $m_2$ & -0.851  & -0.667  &  -1.04   & -0.893 \\
    \hline
    $a$ & -2.09     & -0.734  &  -1.57   & -1.49 \\
    $b$ & 0.0259    & 0.0199  & 0.0226  & 0.0228 \\
    $c$ & 0.0032    & 0.0032  & 0.0032     & 0.0032 \\
    \hline
  \end{tabular}
  \caption[Calibration values]{Calibration values, as used in Eq. (\ref{eq:linear}) and Eq. (\ref{eq:energulight}).}\label{tab:fitparamet}
\end{table}

\begin{table}[!ht]
  \centering
  \begin{tabular}{ccc}
    \hline
    Neutron bin [MeV] & np bin [MeV] & $\sigma_{(\textrm{n},\textrm{p})}$ [mb]\\
    \hline
    5-15  & 4-13  & 292.60\\
    15-25 & 13-22 & 153.78\\
    25-35 & 22-31 & 100.51\\
    35-45 & 31-40 & 73.033\\
    45-55 & 40-48 & 57.336\\
    55-65 & 48-57 & 47.461\\
    65-75 & 57-66 & 40.841\\
    75-85 & 66-78 & 36.239\\
    \hline
  \end{tabular}
  \caption[Energy binning for carbon subtraction]{Energy binning for subtraction of carbon data from the CH$_2$ runs.}\label{tab:neubins}
\end{table}

\clearpage

\begin{table}[!ht]
  \centering
  \begin{tabular}{lcclcc}
    \hline
    Bin & Uncorrected & Corrected & Bin & Uncorrected & Corrected \\
    \multicolumn{1}{l}{[MeV]} & \multicolumn{2}{c}{[mb/sr/MeV]} & \multicolumn{1}{l}{[MeV]} & \multicolumn{2}{c}{[mb/sr/MeV]}\\
    \hline
2.5-3 & $0.0931 \pm 0.23$ & $0.0559 \pm 0.25$ & 68-70 & $2.67 \pm 0.30$ & $2.25 \pm 0.37$ \\
3-4 & $9.61 \pm 0.75$ & $5.66 \pm 1.1$ & 70-72 & $2.13 \pm 0.29$ & $1.85 \pm 0.32$ \\
4-5 & $9.81 \pm 0.50$ & $5.90 \pm 0.86$ & 72-74 & $1.87 \pm 0.28$ & $1.67 \pm 0.30$ \\
5-6 & $12.5 \pm 0.50$ & $7.71 \pm 1.00$ & 74-76 & $1.15 \pm 0.26$ & $1.07 \pm 0.26$ \\
6-7 & $10.4 \pm 0.50$ & $6.54 \pm 0.95$ & 76-78 & $1.69 \pm 0.25$ & $1.59 \pm 0.25$ \\
7-8 & $7.29 \pm 0.45$ & $4.79 \pm 0.75$ & 78-80 & $0.73 \pm 0.22$ & $0.716 \pm 0.22$ \\
8-9 & $6.25 \pm 0.44$ & $4.13 \pm 0.73$ & 80-81 & $2.89 \pm 0.30$ & $2.88 \pm 0.30$ \\
9-10 & $5.37 \pm 0.40$ & $3.61 \pm 0.68$ & 81-82 & $2.05 \pm 0.28$ & $2.05 \pm 0.28$ \\
10-12 & $4.00 \pm 0.28$ & $2.68 \pm 0.45$ & 82-83 & $0.739 \pm 0.26$ & $0.739 \pm 0.26$ \\
12-14 & $3.93 \pm 0.30$ & $2.60 \pm 0.53$ & 83-84 & $1.42 \pm 0.26$ & $1.42 \pm 0.26$ \\
14-16 & $2.56 \pm 0.31$ & $1.66 \pm 0.45$ & 84-85 & $0.979 \pm 0.23$ & $0.979 \pm 0.23$ \\
16-18 & $2.86 \pm 0.35$ & $1.83 \pm 0.53$ & 85-86 & $1.3 \pm 0.23$ & $1.30 \pm 0.23$ \\
18-20 & $2.02 \pm 0.35$ & $1.28 \pm 0.47$ & 86-87 & $1.07 \pm 0.21$ & $1.07 \pm 0.21$ \\
20-24 & $2.66 \pm 0.28$ & $1.68 \pm 0.41$ & 87-88 & $0.861 \pm 0.21$ & $0.861 \pm 0.21$ \\
24-28 & $2.00 \pm 0.30$ & $1.25 \pm 0.39$ & 88-89 & $0.801 \pm 0.20$ & $0.801 \pm 0.20$ \\
28-32 & $1.98 \pm 0.31$ & $1.23 \pm 0.39$ & 89-90 & $0.444 \pm 0.18$ & $0.444 \pm 0.18$ \\
32-36 & $2.83 \pm 0.33$ & $1.74 \pm 0.46$ & 90-91 & $0.524 \pm 0.17$ & $0.524 \pm 0.17$ \\
36-40 & $2.61 \pm 0.33$ & $1.59 \pm 0.46$ & 91-92 & $0.479 \pm 0.17$ & $0.479 \pm 0.17$ \\
40-44 & $2.92 \pm 0.32$ & $1.78 \pm 0.49$ & 92-93 & $0.874 \pm 0.17$ & $0.874 \pm 0.17$ \\
44-48 & $2.85 \pm 0.31$ & $1.75 \pm 0.47$ & 93-94 & $0.345 \pm 0.14$ & $0.345 \pm 0.14$ \\
48-52 & $2.66 \pm 0.29$ & $1.69 \pm 0.43$ & 94-95 & $0.862 \pm 0.13$ & $0.862 \pm 0.13$ \\
52-56 & $2.72 \pm 0.28$ & $1.83 \pm 0.40$ & 95-96 & $0.380 \pm 0.11$ & $0.380 \pm 0.11$ \\
56-60 & $2.49 \pm 0.26$ & $1.77 \pm 0.35$ & 96-97 & $0.400 \pm 0.10$ & $0.400 \pm 0.10$ \\
60-62 & $2.56 \pm 0.35$ & $1.92 \pm 0.45$ & 97-98 & $-0.0245 \pm 0.080$ & $-0.0245 \pm 0.080$ \\
62-64 & $2.57 \pm 0.33$ & $1.99 \pm 0.43$ & 98-99 & $0.00468 \pm 0.079$ & $0.00468 \pm 0.079$ \\
64-66 & $2.73 \pm 0.32$ & $2.18 \pm 0.41$ & 99-100 & $0.0227 \pm 0.13$ & $0.0227 \pm 0.13$ \\
66-68 & $2.80 \pm 0.316$ & $2.27 \pm 0.41$ & & & \\     \hline
  \end{tabular}
  \caption[Experimental double-differential cross sections at 20 degrees]{Experimental double-differential cross sections at 20 degrees  in the Ca(n,px) reaction for the neutron spectrum in table \ref{tab:nspectra} and for 94 MeV neutrons according to the subtraction method in section \ref{sec:meth3}.}\label{tab:res_deg20}
\end{table}

\clearpage

\begin{table}[!ht]
  \centering
  \begin{tabular}{lcclcc}
    \hline
    Bin & Uncorrected & Corrected & Bin & Uncorrected & Corrected \\
    \multicolumn{1}{l}{[MeV]} & \multicolumn{2}{c}{[mb/sr/MeV]} & \multicolumn{1}{l}{[MeV]} & \multicolumn{2}{c}{[mb/sr/MeV]}\\
    \hline
2.5-3 & $7.64 \pm 1.9$ & $4.59 \pm 2.2$ & 70-71 & $0.489 \pm 0.17$ & $0.428 \pm 0.19$ \\
3-4 & $11.3 \pm 0.57$ & $6.68 \pm 1.0$ & 71-72 & $0.950 \pm 0.18$ & $0.834 \pm 0.22$ \\
4-5 & $6.43 \pm 0.24$ & $3.87 \pm 0.51$ & 72-73 & $0.506 \pm 0.17$ & $0.456 \pm 0.18$ \\
5-6 & $12.6 \pm 0.34$ & $7.80 \pm 0.95$ & 73-74 & $0.459 \pm 0.16$ & $0.414 \pm 0.17$ \\
6-7 & $9.31 \pm 0.30$ & $5.84 \pm 0.78$ & 74-75 & $0.738 \pm 0.16$ & $0.689 \pm 0.17$ \\
7-8 & $6.97 \pm 0.28$ & $4.59 \pm 0.63$ & 75-76 & $0.355 \pm 0.15$ & $0.334 \pm 0.16$ \\
8-9 & $5.80 \pm 0.27$ & $3.84 \pm 0.60$ & 76-77 & $0.0924 \pm 0.14$ & $0.0861 \pm 0.14$ \\
9-10 & $5.16 \pm 0.25$ & $3.47 \pm 0.58$ & 77-78 & $0.700 \pm 0.15$ & $0.667 \pm 0.15$ \\
10-12 & $3.87 \pm 0.16$ & $2.60 \pm 0.38$ & 78-79 & $0.586 \pm 0.14$ & $0.571 \pm 0.14$ \\
12-14 & $2.72 \pm 0.15$ & $1.81 \pm 0.34$ & 79-80 & $0.366 \pm 0.13$ & $0.362 \pm 0.13$ \\
14-16 & $2.03 \pm 0.16$ & $1.33 \pm 0.31$ & 80-81 & $0.411 \pm 0.13$ & $0.409 \pm 0.13$ \\
16-18 & $2.29 \pm 0.17$ & $1.48 \pm 0.37$ & 81-82 & $0.0804 \pm 0.11$ & $0.0804 \pm 0.11$ \\
18-20 & $1.58 \pm 0.18$ & $1.01 \pm 0.31$ & 82-83 & $0.215 \pm 0.12$ & $0.215 \pm 0.12$ \\
20-24 & $1.47 \pm 0.14$ & $0.926 \pm 0.22$ & 83-84 & $0.274 \pm 0.11$ & $0.274 \pm 0.11$ \\
24-28 & $2.04 \pm 0.15$ & $1.28 \pm 0.29$ & 84-85 & $0.439 \pm 0.10$ & $0.439 \pm 0.10$ \\
28-32 & $1.78 \pm 0.15$ & $1.11 \pm 0.27$ & 85-86 & $0.172 \pm 0.094$ & $0.172 \pm 0.094$ \\
32-36 & $1.65 \pm 0.15$ & $1.02 \pm 0.27$ & 86-87 & $0.295 \pm 0.089$ & $0.295 \pm 0.089$ \\
36-40 & $1.48 \pm 0.15$ & $0.907 \pm 0.25$ & 87-88 & $0.0167 \pm 0.076$ & $0.0167 \pm 0.076$ \\
40-44 & $1.13 \pm 0.15$ & $0.694 \pm 0.22$ & 88-89 & $0.360 \pm 0.080$ & $0.36 \pm 0.080$ \\
44-48 & $1.55 \pm 0.14$ & $0.971 \pm 0.26$ & 89-90 & $0.325 \pm 0.10$ & $0.325 \pm 0.10$ \\
48-52 & $1.22 \pm 0.13$ & $0.785 \pm 0.21$ & 90-91 & $-0.202 \pm 0.00$ & $-0.202 \pm 0.00$ \\
52-54 & $1.12 \pm 0.18$ & $0.746 \pm 0.27$ & 91-92 & $0.076 \pm 0.074$ & $0.076 \pm 0.074$ \\
54-56 & $1.04 \pm 0.18$ & $0.720 \pm 0.25$ & 92-93 & $0.102 \pm 0.047$ & $0.102 \pm 0.047$ \\
56-58 & $1.26 \pm 0.17$ & $0.897 \pm 0.27$ & 93-94 & $0.0237 \pm 0.063$ & $0.0237 \pm 0.063$ \\
58-60 & $0.966 \pm 0.16$ & $0.712 \pm 0.23$ & 94-95 & $0.149 \pm 0.046$ & $0.149 \pm 0.046$ \\
60-62 & $0.851 \pm 0.16$ & $0.647 \pm 0.2$ & 95-96 & $0.0882 \pm 0.046$ & $0.0882 \pm 0.046$ \\
62-64 & $0.773 \pm 0.15$ & $0.605 \pm 0.18$ & 96-97 & $0.108 \pm 0.046$ & $0.108 \pm 0.046$ \\
64-66 & $0.871 \pm 0.14$ & $0.701 \pm 0.19$ & 97-98 & $0.0595 \pm 0.044$ & $0.0595 \pm 0.044$ \\
66-68 & $1.07 \pm 0.14$ & $0.876 \pm 0.19$ & 98-99 & $0.0190 \pm 0.045$ & $0.0190 \pm 0.045$ \\
68-70 & $1.07 \pm 0.13$ & $0.909 \pm 0.18$ & 99-100 & $0.0499 \pm 0.050$ & $0.0499 \pm 0.050$ \\
     \hline
  \end{tabular}
  \caption[Experimental double-differential cross sections at 40 degrees]{Experimental double-differential cross sections at 40 degrees  in the Ca(n,px) reaction for the neutron spectrum in table \ref{tab:nspectra} and for 94 MeV neutrons according to the subtraction method in section \ref{sec:meth3}.}\label{tab:res_deg40}
\end{table}

\clearpage

\begin{table}[!ht]
  \centering
  \begin{tabular}{lcclcc}
    \hline
    Bin & Uncorrected & Corrected & Bin & Uncorrected & Corrected \\
    \multicolumn{1}{l}{[MeV]} & \multicolumn{2}{c}{[mb/sr/MeV]} & \multicolumn{1}{l}{[MeV]} & \multicolumn{2}{c}{[mb/sr/MeV]}\\
    \hline
2.5-3 & $0.0668 \pm 0.14$ & $0.0401 \pm 0.15$ & 30-35 & $0.175 \pm 0.020$ & $0.114 \pm 0.082$ \\
3-4 & $5.41 \pm 0.42$ & $3.19 \pm 0.60$ & 35-40 & $0.0721 \pm 0.015$ & $0.0472 \pm 0.045$ \\
4-5 & $3.90 \pm 0.20$ & $2.35 \pm 0.35$ & 40-45 & $0.0183 \pm 0.011$ & $0.0123 \pm 0.018$ \\
5-6 & $8.68 \pm 0.30$ & $5.37 \pm 0.69$ & 45-50 & $0.00676 \pm 0.0082$ & $0.00474 \pm 0.010$ \\
6-7 & $6.76 \pm 0.27$ & $4.26 \pm 0.59$ & 50-55 & $-0.0278 \pm 0.0090$ & $-0.0208 \pm 0.031$ \\
7-8 & $4.21 \pm 0.21$ & $2.80 \pm 0.41$ & 55-60 & $0.0146 \pm 0.0085$ & $0.0117 \pm 0.020$ \\
8-9 & $3.55 \pm 0.18$ & $2.37 \pm 0.39$ & 60-65 & $0.0244 \pm 0.0099$ & $0.0209 \pm 0.033$ \\
9-10 & $3.37 \pm 0.16$ & $2.31 \pm 0.39$ & 65-70 & $0.0274 \pm 0.0090$ & $0.0246 \pm 0.041$ \\
10-12 & $1.97 \pm 0.055$ & $1.36 \pm 0.20$ & 70-75 & $0.00282 \pm 0.011$ & $0.00267 \pm 0.011$ \\
12-14 & $1.09 \pm 0.045$ & $0.762 \pm 0.15$ & 75-80 & $0.0165 \pm 0.011$ & $0.0161 \pm 0.021$ \\
14-16 & $0.946 \pm 0.046$ & $0.663 \pm 0.16$ & 80-85 & $0.0562 \pm 0.013$ & $0.0561 \pm 0.014$ \\
16-18 & $0.709 \pm 0.052$ & $0.493 \pm 0.16$ & 85-90 & $0.0346 \pm 0.015$ & $0.0346 \pm 0.015$ \\
18-20 & $0.283 \pm 0.053$ & $0.193 \pm 0.096$ & 90-95 & $0.0914 \pm 0.016$ & $0.0914 \pm 0.016$ \\
20-25 & $0.420 \pm 0.024$ & $0.280 \pm 0.11$ & 95-100 & $0.0274 \pm 0.019$ & $0.0274 \pm 0.019$ \\
25-30 & $0.149 \pm 0.0233$ & $0.0968 \pm 0.057$ & & & \\
\hline
  \end{tabular}
  \caption[Experimental double-differential cross sections at 140 degrees]{Experimental double-differential cross sections at 140 degrees  in the Ca(n,px) reaction for the neutron spectrum in table \ref{tab:nspectra} and for 94 MeV neutrons according to the subtraction method in section \ref{sec:meth3}.}\label{tab:res_deg140}
\end{table}

\begin{table}[!ht]
  \centering
  \begin{tabular}{lcclcc}
    \hline
    Bin & Uncorrected & Corrected & Bin & Un-corrected
     & Corrected \\
    \multicolumn{1}{l}{[MeV]} & \multicolumn{2}{c}{[mb/sr/MeV]} & \multicolumn{1}{l}{[MeV]} & \multicolumn{2}{c}{[mb/sr/MeV]}\\
    \hline
2.5-3 & $0.152 \pm 0.26$ & $0.0915 \pm 0.28$ & 30-35 & $0.152 \pm 0.023$ & $0.0984 \pm 0.073$ \\
3-4 & $6.70 \pm 0.60$ & $3.95 \pm 0.81$ & 35-40 & $0.204 \pm 0.020$ & $0.133 \pm 0.12$ \\
4-5 & $6.44 \pm 0.38$ & $3.87 \pm 0.61$ & 40-45 & $0.137 \pm 0.017$ & $0.0918 \pm 0.10$ \\
5-6 & $6.97 \pm 0.36$ & $4.31 \pm 0.62$ & 45-50 & $0.0132 \pm 0.015$ & $0.00927 \pm 0.019$ \\
6-7 & $4.68 \pm 0.32$ & $2.95 \pm 0.50$ & 50-55 & $0.0163 \pm 0.015$ & $0.0122 \pm 0.023$ \\
7-8 & $4.05 \pm 0.30$ & $2.69 \pm 0.46$ & 55-60 & $-0.00418 \pm 0.017$ & $-0.00335 \pm 0.018$ \\
8-9 & $3.63 \pm 0.25$ & $2.43 \pm 0.44$ & 60-65 & $-0.00651 \pm 0.019$ & $-0.00558 \pm 0.021$ \\
9-10 & $2.89 \pm 0.21$ & $1.98 \pm 0.38$ & 65-70 & $-0.0545 \pm 0.019$ & $-0.0491 \pm 0.081$ \\
10-12 & $1.59 \pm 0.083$ & $1.10 \pm 0.17$ & 70-75 & $0.0539 \pm 0.021$ & $0.0511 \pm 0.073$ \\
12-14 & $0.915 \pm 0.079$ & $0.640 \pm 0.14$ & 75-80 & $0.0449 \pm 0.024$ & $0.0439 \pm 0.055$ \\
14-16 & $0.647 \pm 0.079$ & $0.454 \pm 0.14$ & 80-85 & $0.134 \pm 0.023$ & $0.134 \pm 0.026$ \\
16-18 & $0.616 \pm 0.080$ & $0.428 \pm 0.16$ & 85-90 & $0.146 \pm 0.028$ & $0.146 \pm 0.028$ \\
18-20 & $0.388 \pm 0.078$ & $0.266 \pm 0.13$ & 90-95 & $0.254 \pm 0.030$ & $0.254 \pm 0.030$ \\
20-25 & $0.223 \pm 0.030$ & $0.148 \pm 0.063$ & 95-100 & $0.229 \pm 0.036$ & $0.229 \pm 0.036$ \\
25-30 & $0.184 \pm 0.0281$ & $0.120 \pm 0.070$ & & & \\     \hline
  \end{tabular}
  \caption[Experimental double-differential cross sections at 160 degrees]{Experimental double-differential cross sections at 160 degrees  in the Ca(n,px) reaction for the neutron spectrum in table \ref{tab:nspectra} and for 94 MeV neutrons according to the subtraction method in section \ref{sec:meth3}.}\label{tab:res_deg160}
\end{table}

\clearpage

\begin{table}[!ht]
  \centering
  \begin{tabular}{lcclcc}
    \hline
    Bin & Uncorrected & Corrected & Bin & Uncorrected & Corrected \\
    \multicolumn{1}{l}{[MeV]} & \multicolumn{2}{c}{[mb/MeV]} & \multicolumn{1}{l}{[MeV]} & \multicolumn{2}{c}{[mb/MeV]}\\
    \hline
0-2.5 & - & - & 46-48 & $6.69 \pm 0.29$ & $4.15 \pm 0.41$ \\
2.5-3 & $12.1 \pm 0.32$ & $7.24 \pm 0.36$ & 48-50 & $5.23 \pm 0.28$ & $3.32 \pm 0.37$ \\
3-4 & $102 \pm 0.52$ & $59.7 \pm 0.77$ & 50-52 & $5.04 \pm 0.35$ & $3.26 \pm 0.55$ \\
4-5 & $73.7 \pm 0.26$ & $45.0 \pm 0.47$ & 52-54 & $4.54 \pm 0.33$ & $3.05 \pm 0.56$ \\
5-6 & $129 \pm 0.33$ & $78.7 \pm 0.73$ & 54-56 & $4.46 \pm 0.33$ & $3.07 \pm 0.49$ \\
6-7 & $97.3 \pm 0.30$ & $60.4 \pm 0.62$ & 56-58 & $5.44 \pm 0.29$ & $3.77 \pm 0.53$ \\
7-8 & $69.7 \pm 0.26$ & $46.1 \pm 0.49$ & 58-60 & $4.27 \pm 0.32$ & $3.12 \pm 0.44$ \\
8-9 & $59.2 \pm 0.24$ & $39.5 \pm 0.47$ & 60-62 & $3.85 \pm 0.31$ & $2.91 \pm 0.39$ \\
9-10 & $52.5 \pm 0.21$ & $35.4 \pm 0.45$ & 62-64 & $3.68 \pm 0.30$ & $2.86 \pm 0.37$ \\
10-12 & $35.2 \pm 0.11$ & $23.6 \pm 0.26$ & 64-66 & $4.00 \pm 0.27$ & $3.21 \pm 0.36$ \\
12-14 & $24.2 \pm 0.10$ & $16.4 \pm 0.23$ & 66-68 & $4.54 \pm 0.29$ & $3.70 \pm 0.37$ \\
14-16 & $18.3 \pm 0.10$ & $12.0 \pm 0.22$ & 68-70 & $4.50 \pm 0.28$ & $3.79 \pm 0.35$ \\
16-18 & $18.0 \pm 0.11$ & $12.0 \pm 0.26$ & 70-72 & $3.25 \pm 0.24$ & $2.84 \pm 0.28$ \\
18-20 & $11.0 \pm 0.13$ & $7.19 \pm 0.22$ & 72-74 & $2.53 \pm 0.27$ & $2.27 \pm 0.28$ \\
20-22 & $11.8 \pm 0.11$ & $7.34 \pm 0.24$ & 74-76 & $2.34 \pm 0.45$ & $2.15 \pm 0.32$ \\
22-24 & $12.0 \pm 0.11$ & $7.35 \pm 0.27$ & 76-78 & $2.23 \pm 0.25$ & $2.11 \pm 0.25$ \\
24-26 & $10.9 \pm 0.12$ & $6.82 \pm 0.28$ & 78-80 & $2.42 \pm 0.12$ & $2.33 \pm 0.20$ \\
26-28 & $11.3 \pm 0.14$ & $6.95 \pm 0.30$ & 80-82 & $2.89 \pm 0.56$ & $2.89 \pm 0.74$ \\
28-30 & $10.0 \pm 0.14$ & $6.25 \pm 0.31$ & 82-84 & $1.55 \pm 0.23$ & $1.55 \pm 0.19$ \\
30-32 & $9.53 \pm 0.12$ & $5.99 \pm 0.31$ & 84-86 & $1.71 \pm 0.15$ & $1.71 \pm 0.15$ \\
32-34 & $9.74 \pm 0.12$ & $6.04 \pm 0.39$ & 86-88 & $1.29 \pm 0.21$ & $1.29 \pm 0.21$ \\
34-36 & $10.7 \pm 0.13$ & $6.64 \pm 0.46$ & 88-90 & $2.47 \pm 0.070$ & $2.47 \pm 0.070$ \\
36-38 & $8.17 \pm 0.11$ & $4.76 \pm 0.39$ & 90-92 & $0.684 \pm 0.045$ & $0.684 \pm 0.041$ \\
38-40 & $9.51 \pm 0.12$ & $5.42 \pm 0.61$ & 92-94 & $0.942 \pm 0.044$ & $0.942 \pm 0.044$ \\
40-42 & $5.76 \pm 0.13$ & $2.49 \pm 0.77$ & 94-96 & $0.949 \pm 0.060$ & $0.949 \pm 0.060$ \\
42-44 & $8.05 \pm 0.21$ & $3.59 \pm 0.68$ & 96-98 & $0.516 \pm 0.15$ & $0.516 \pm 0.15$ \\
44-46 & $7.12 \pm 0.32$ & $3.94 \pm 1.3$ & 98-100 & $0.315 \pm 0.15$ & $0.315 \pm 0.15$ \\
44-46 & $7.12 \pm 0.31$ & $3.94 \pm 0.41$ & & & \\     \hline
  \end{tabular}
  \caption[Experimental energy-differential cross sections]{Experimental energy-differential cross sections in the Ca(n,px) reaction for the neutron spectrum in table \ref{tab:nspectra} and for 94 MeV neutrons according to the subtraction method in section \ref{sec:meth3}.}\label{tab:res_Aint}
\end{table}

\begin{table}[!ht]
  \centering
  \begin{tabular}{lcc}
    \hline
    Angle & Uncorrected & Corrected \\
    \multicolumn{1}{l}{[deg]} & \multicolumn{2}{c}{[mb/sr]} \\
    \hline
    20 & $253 \pm 4.76$ & $175 \pm 6.69$\\
    40 & $162 \pm 2.51$ & $107 \pm 4.22$\\
    140 & $51.5 \pm 0.792$ & $33.7 \pm 1.73$\\
    160 & $52.4 \pm 1.14$ & $35.1 \pm 2.04$\\
    \hline
  \end{tabular}
  \caption[Experimental angular-differential cross sections]{Experimental angular-differential cross sections in the Ca(n,px) reaction for the neutron spectrum in table \ref{tab:nspectra} and for 94 MeV neutrons according to the subtraction method in section \ref{sec:meth3}. Cutoff energy is 2.5 MeV.}\label{tab:Eint_tot_all}
\end{table}

\begin{table}[!ht]
  \centering
  \begin{tabular}{cc}
    \hline
    Uncorrected & Corrected \\
    \multicolumn{2}{c}{[mb]} \\
    \hline
    $1220 \pm 3.21$ & $788 \pm 5.83$\\
    \hline
  \end{tabular}
  \caption[Experimental total cross sections]{Experimental total cross sections in the Ca(n,px) reaction for the neutron spectrum in table \ref{tab:nspectra} and for 94 MeV neutrons according to the subtraction method in section \ref{sec:meth3}. Cutoff energy is 2.5 MeV.}\label{tab:restot}
\end{table}

\clearpage

\section{Notes about extracting information from NN databases\label{sec:NN}}

\begin{figure}[!hbt]
  \centering
\includegraphics[width=0.6\textwidth,bb=0 0 825 267]{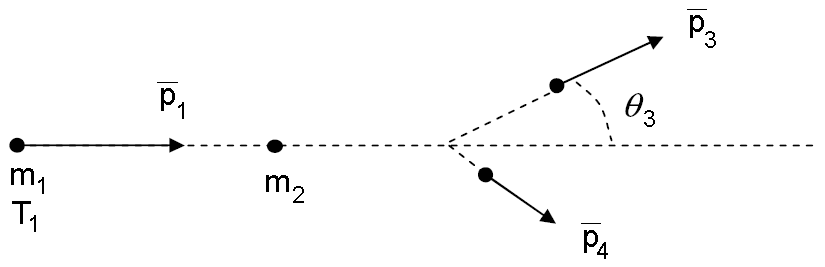}
  \caption[Scattering in the laboratory frame]{Scattering in the laboratory frame, adapted from \protect \citep{lecturenotes}.}\label{fig:kinema}
\end{figure}

When using databases to obtain data on elastic \ac{NN} scattering it
is not always intuitive to get what one wants from it. The problem
is that \ac{NN} databases often gives the cross section for a given
neutron scattering angle, but one most often knows the proton angle.
To get the cross section one wants, some transformations between
different frames and calculations of ratio-factors are needed
\citep{flux}. This is an eight-step process to get through this
without (too many) tears. First some necessary kinematic relations
for a general scattering reaction are given in \citep{lecturenotes},
\begin{subequations}
    \begin{equation}
        E_1 = T_1 + m_1\label{eq:kine1}
    \end{equation}
    \begin{equation}
        p_1 = \sqrt{T_1^2 + 2T_1m_1}\label{eq:kine2}
    \end{equation}
    \begin{equation}
        s = m_1^2+m_2^2+2E_1m_2\label{eq:kine3}
    \end{equation}
    \begin{equation}
        E_3^\textrm{cm} = \frac{1}{2\sqrt{s}} \left( s+m_3^2-m_4^2 \right)\label{eq:kine4}
    \end{equation}
    \begin{equation}
        v_\textrm{cm} = \frac{p_1}{E_1+m_2}\label{eq:kine5}
    \end{equation}
    \begin{equation}
        p_3^\textrm{cm} = \sqrt{(E_3^\textrm{cm})^2-m_3^2}\label{eq:kine6}
    \end{equation}
    \begin{equation}
        \tan\theta_3 = \frac{\sin\theta_3^\textrm{cm} \sqrt{1-v_\textrm{cm}^2}}{\cos\theta_3^\textrm{cm} +
        \frac{v_\textrm{cm}E_3^\textrm{cm}}{p_3^\textrm{cm}}}\label{eq:kine7}.
    \end{equation}
\end{subequations}
The relevant quantities are defined in figure \ref{fig:kinema},
where $m_i$ is the mass, $T_i$ is the kinetic energy and $p_i$ is
the momentum of particle $i$. Quantities in the \ac{CM} system are
labeled cm, unlabeled quantities are in the laboratory frame.

For a \ac{NN} reaction with a neutron on a proton target we switch
the indices in the notations as $1,4 \rightarrow n$ and $2,3
\rightarrow p$. Known parameters are assumed to be $m_p$, $m_n$,
$T_n$ and $\theta_\textrm{lab}^p$. An unlabeled $\theta$ is always
the scattering angle of the neutron.
\begin{enumerate}
  \item Using the relations (\ref{eq:kine1}) to (\ref{eq:kine7}) it
  is straightforward, although (\ref{eq:kine7}) might make it analytically tricky, to calculate $\theta^\textrm{cm}_p$.
  \item The easy part: $\theta^\textrm{cm}_n = 180 -
  \theta^\textrm{cm}_p$.
  \item Calculate $\theta^\textrm{lab}_n$. This is preferably done
  via some computer code for relativistic kinematics, for example
  the code RELKIN circulating at \ac{INF} \citep{pompcomm}.
  \item From the \ac{NN} database of choice, extract the cross section in the \ac{CM} frame for the
  \emph{opposite} case, that is when the neutron is scattered in the
  desired proton angle. That is, the cross section $\left( \frac{\textrm{d}\sigma}{\textrm{d}\Omega}
  \right)^\textrm{cm}_{\theta_n =
  \theta^\textrm{cm}_p}$.\label{list:steppcm}
  \item In the same way as in step \ref{list:steppcm}, use a \ac{NN} database to extract the cross section in the \ac{CM} frame for the
  desired case. That is, the cross section $\left( \frac{\textrm{d}\sigma}{\textrm{d}\Omega}
  \right)^\textrm{cm}_{\theta_n = \theta^\textrm{cm}_n}$.\label{list:stepncm}
  \item Do step \ref{list:steppcm}, but in the laboratory frame, to get $\left( \frac{\textrm{d}\sigma}{\textrm{d}\Omega}
  \right)^\textrm{lab}_{\theta_n = \theta^\textrm{lab}_p}$ from the
  \ac{NN} database.\label{list:stepplab}
  \item Now to calculate the ratio factor, $R$, introduced in
  \citep{flux} for transformations between lab frame cross section and \ac{CM} frame cross section. Since the $R$ given in the reference only is valid for particles of equal masses the results from step \ref{list:steppcm} and step \ref{list:stepplab} is used to calculate $R$ explicitly from the relation $\left( \frac{\textrm{d}\sigma}{\textrm{d}\Omega}
  \right)^\textrm{cm}_{\theta_n = \theta^\textrm{cm}_p} = R \left( \frac{\textrm{d}\sigma}{\textrm{d}\Omega}
  \right)^\textrm{lab}_{\theta_n = \theta^\textrm{lab}_p}$.\label{list:stepR}
  \item Using $R$ obtained in step \ref{list:stepR} it is now possible to transform the cross section
   from step \ref{list:stepncm} into the lab frame, by the relation previously used and obtained from \citep{flux}, $\left( \frac{\textrm{d}\sigma}{\textrm{d}\Omega}
  \right)^\textrm{lab}_{\theta_n = \theta^\textrm{lab}_n} = \left( \frac{\textrm{d}\sigma}{\textrm{d}\Omega}
  \right)^\textrm{cm}_{\theta_n = \theta^\textrm{cm}_n}\frac{1}{R}$.
\end{enumerate}

Finally we have the cross section
$\left(\frac{\textrm{d}\sigma}{\textrm{d}\Omega}\right)^\textrm{lab}_{\theta
= \theta^\textrm{lab}_n}$ which is just what we wanted, since the
cross section of scattering a neutron at an angle of
$\theta^\textrm{lab}_n$ is equivalent to the cross section of
scattering of a proton at an angle $\theta^\textrm{lab}_p$.

\begin{example}
Let us find the cross section for np scattering at
$\theta_\textrm{lab}^p = 20$ degrees with an incoming neutron energy
of $T_n=50$ MeV. Using the mass values $m_n = 939.57$ MeV/c$^2$ and
$m_n = 938.27$ MeV/c$^2$ the relations (\ref{eq:kine1}) to
(\ref{eq:kine7}) will give $\theta^\textrm{cm}_p \approx 40.5$
degrees and, consequentially, $\theta^\textrm{cm}_n \approx 139.5$
degrees. Using RELKIN \citep{pompcomm}, a code for relativistic
kinematics, $\theta^\textrm{lab}_p$ is calculated to be
$\theta^\textrm{lab}_p \approx 69.4$ degrees. From the SAID \ac{NN}
database \citep{SAID} we get
\begin{subequations}
\begin{equation}
    \left( \frac{\textrm{d}\sigma}{\textrm{d}\Omega} \right)^\textrm{cm}_{\theta_n = 40.5} \approx
    14.1 \frac{\textrm{mb}}{\textrm{sr}}\label{eq:ex_xsecpcm}
\end{equation}
\begin{equation}
    \left( \frac{\textrm{d}\sigma}{\textrm{d}\Omega} \right)^\textrm{cm}_{\theta_n = 139.5} \approx
    15.0 \frac{\textrm{mb}}{\textrm{sr}}\label{eq:ex_xsecncm}
\end{equation}
\begin{equation}
    \left( \frac{\textrm{d}\sigma}{\textrm{d}\Omega} \right)^\textrm{lab}_{\theta_n = 20.0} \approx
    53.9. \frac{\textrm{mb}}{\textrm{sr}}\label{eq:ex_xsecplab}
\end{equation}
\end{subequations}
Note the different observable keywords \texttt{DSG} for cross
section in the center of mass frame, and \texttt{DSGL} for the cross
section in the laboratory frame. The ratio factor, $R$, can now be
calculated using the information in (\ref{eq:ex_xsecpcm}),
(\ref{eq:ex_xsecplab}) and the definition in step \ref{list:stepR}
\begin{equation}
    R \equiv \frac{\left( \frac{\textrm{d}\sigma}{\textrm{d}\Omega}
  \right)^\textrm{cm}_{\theta_n = \theta^\textrm{cm}_p}}{ \left( \frac{\textrm{d}\sigma}{\textrm{d}\Omega}
  \right)^\textrm{lab}_{\theta_n = \theta^\textrm{lab}_p}} =
  \frac{14.1}{53.9} \approx 0.261.\label{eq:ex_ratio}
\end{equation}
The ratio factor from (\ref{eq:ex_ratio}) and the cross section
(\ref{eq:ex_xsecncm}) can now be used to obtain
\begin{equation}
\left(\frac{\textrm{d}\sigma}{\textrm{d}\Omega}
  \right)^\textrm{lab}_{\theta_n = 69.4} = \left( \frac{\textrm{d}\sigma}{\textrm{d}\Omega}
  \right)^\textrm{cm}_{\theta_n = 139.5}\frac{1}{R} = 15.0 \cdot \frac{1}{0.261}
  \approx 57.5 \frac{\textrm{mb}}{\textrm{sr}}
\end{equation}
which is the sought cross section.
\end{example}

\end{document}